\documentclass[cernpreprint,atlasdraft=false,texlive=2016,UKenglish,texmf]{atlasdoc}
 
\usepackage[subcaption]{atlaspackage}
\usepackage{multirow}
\usepackage{atlasbiblatex}
 
\usepackage[BSM=true,jetetmiss=true,process=true]{atlasphysics}
 
\graphicspath{{logos/}{figures/}}
 
\usepackage{ANA-EXOT-2018-43-PAPER-defs}

\AtlasTitle{Search for dark matter produced in association with a single top quark in $\sqrt{s}=13$~TeV \(pp\) collisions with the ATLAS detector}
 
\PreprintIdNumber{CERN-EP-2020-184}
\AtlasAbstract{
This paper presents a search for dark matter in the context of a two-Higgs-doublet model together with an additional pseudoscalar mediator, $a$,
which decays into the dark-matter particles.
Processes where the pseudoscalar mediator is produced in association with a single top quark in the 2HDM+$a$ model are explored for the first time at the LHC.
Several final states which include either one or two charged leptons (electrons or muons) and a
significant amount of missing transverse momentum are considered.
The analysis is based on proton--proton collision data collected with the ATLAS experiment at
$\sqrt{s} = 13$~TeV during LHC Run~2 (2015--2018), corresponding to an integrated luminosity of 139~fb$^{-1}$.
No significant excess above the Standard Model predictions is found.
The results are expressed as 95\% confidence-level limits on the parameters of the signal models considered.
}
 
\author{The ATLAS Collaboration}
 
\AtlasRefCode{EXOT-2018-43}

\AtlasJournalRef{Eur. Phys. J. C 81 (2021) 860}
\AtlasDOI{10.1140/epjc/s10052-021-09566-y}

\hypersetup{pdftitle={ATLAS document},pdfauthor={The ATLAS Collaboration}}
 
\begin{document}
 
\maketitle

\section{Introduction}
\label{sec:intro}
% The next lines are included from the .//sections/intro.tex input file
Strong evidence for the existence of a new, non-luminous matter component of the universe, dark matter (DM), arises from astrophysical observations such as
precise measurements of the cosmic microwave background and from gravitational lensing measurements. Through its gravitational interactions, it is suggested that DM constitutes
up to 26\% of the matter--energy content of the universe~\cite{Hinshaw:2012aka,Akrami:2018vks}.
The nature and properties of DM remain largely unknown in the context of the Standard Model (SM) of particle physics.
Under the hypothesis that DM behaves like a weakly interacting massive particle (WIMP)~\cite{Steigman:1984ac}, searches are performed
using multiple,
complementary approaches.  At hadron colliders, searches for WIMP-like DM production crucially rely on one or
more visible particles being produced in association with the sought-after invisible DM candidate.
The experimental signature for DM candidates is missing transverse momentum
(\ptmiss, its modulus denoted by \etmiss) in collision events.
Several models have been proposed in the past decades, with the details of the DM--SM production process
depending on the model assumptions.
 
A class of simplified models for DM searches at the LHC is considered in this paper. It involves a two-Higgs-doublet extended sector
together with an additional pseudoscalar mediator to DM, the 2HDM+$a$ model~\cite{Bauer:2017ota,Abe:2018bpo}.
This class of models represents one of the simplest ultraviolet-complete
and renormalisable frameworks for investigating the broad phenomenology predicted by
spin-0 mediator-based DM models~\cite{Fox:2008kb,Cassel:2009pu,Bai:2010hh,Abdallah:2015ter,Abercrombie:2015wmb,Abe:2018bpo,Cheung:2010zf,Haisch:2013fla,Buckley:2014fba,Buckley:2015ctj,Haisch:2015ioa,Backovic:2015soa,Arina:2016cqj,Haisch:2016gry,Banerjee:2017wxi}.
For the present study, a type-II~\cite{Gunion:2002zf,Branco:2011iw} coupling structure of the Higgs sector to third-generation
fermions is considered. The CP eigenstates are identified with the mass eigenstates, i.e.\ two scalars $h$ and $H$, two
pseudoscalars $A$ and $a$, and charged scalars $H^{\pm}$. Three mixing angles are defined in the model:
$\alpha$ denotes the mixing angle between the two CP-even weak spin-0 eigenstates,
$\tan\beta$ is the ratio of the vacuum expectation values (VEVs) of the two Higgs
doublets and $\theta$ represents the mixing angle of the two CP-odd weak spin-0 eigenstates.
The alignment ($\cos(\beta-\alpha) = 0$) and decoupling limit is assumed, such that the lightest CP-even state
of the Higgs sector, $h$, can be identified with the SM Higgs boson and the electroweak VEV is set to $246~\GeV$.
The pseudoscalar mediator $a$ couples the DM particles, $\chi$, to the SM and mixes with the pseudoscalar
partner of the SM Higgs boson, $A$.
Following the prescriptions in Ref.~\cite{Abe:2018bpo}, the masses of the heavy CP-even Higgs boson $H$
and charged bosons $H^\pm$ are set equal to the mass of the heavy CP-odd partner $A$.
 
\begin{figure}[h!]
\begin{subfigure}[b]{0.24\textwidth}
\centering{\includegraphics[width=\textwidth]{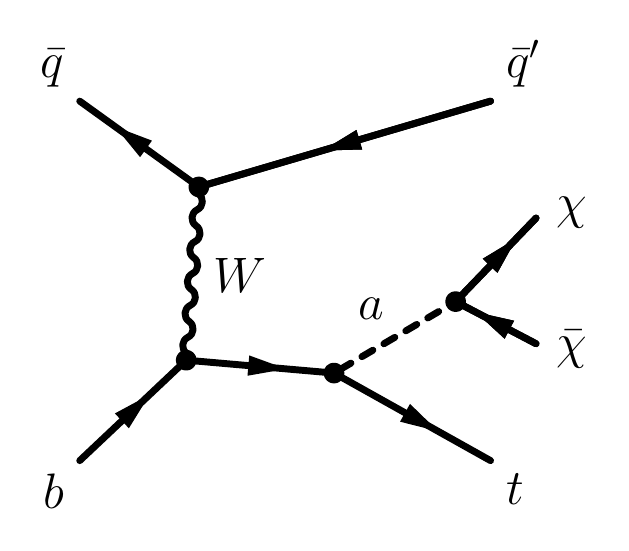}}
\caption{\label{fig:diagram_tj_tchan1}}
\end{subfigure}
\begin{subfigure}[b]{0.24\textwidth}
\centering{\includegraphics[width=\textwidth]{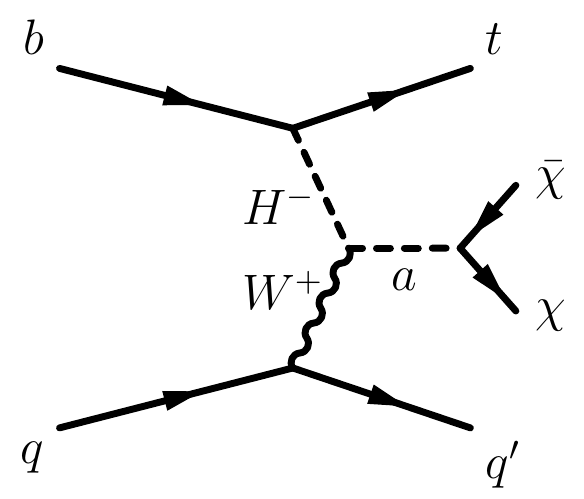}}
\caption{\label{fig:diagram_tj_tchan2}}
\end{subfigure}
\begin{subfigure}[b]{0.24\textwidth}
\centering{\includegraphics[width=\textwidth]{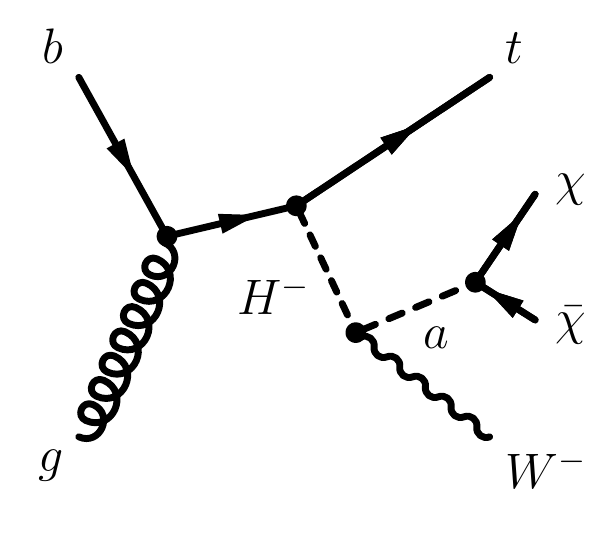}}
\caption{\label{fig:diagram_tw1}}
\end{subfigure}
\begin{subfigure}[b]{0.24\textwidth}
\centering{\includegraphics[width=\textwidth]{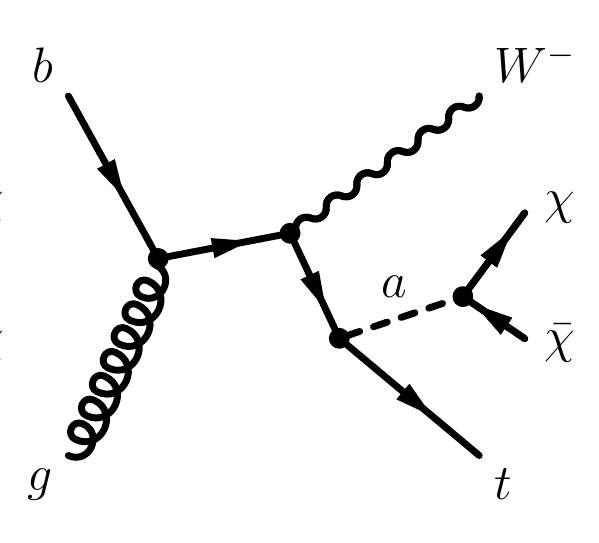}}
\caption{\label{fig:diagram_tw2}}
\end{subfigure}
 
\caption{
\label{fig:THDMpa_feync}
Representative diagrams of the dark-matter particle $\chi$ pair production from the 2HDM+$a$ model
considered in this analysis: \subref{fig:diagram_tj_tchan1}--\subref{fig:diagram_tj_tchan2} through the $t$-channel,
and \subref{fig:diagram_tw1}--\subref{fig:diagram_tw2} through the $tW$ channel.
}
\end{figure}

This set of models offers a rich phenomenology, with a variety of final states that might arise
depending on the production and decay modes of the various bosons composing the Higgs sector,
as investigated in Ref.~\cite{EXOT-2017-32}.
A recent study \cite{Pani:2017qyd} has shown that final-state events characterised by the presence
of \etmiss and a single top quark provide promising sensitivity to 2HDM+$a$ models.
As in SM single top production, three different types of processes contribute at
leading order (LO) in QCD: $t$-channel production, $s$-channel production and
associated production with a $W$ boson ($tW$). In the following, these are collectively referred
to as DM$t$ processes.
The $t$-channel process $pp \rightarrow tj\chi \bar \chi$ receives its dominant contributions from the
two diagrams shown in Figures~\ref{fig:THDMpa_feync}\subref{fig:diagram_tj_tchan1} and~\ref{fig:THDMpa_feync}\subref{fig:diagram_tj_tchan2}.
These two diagrams interfere destructively, ensuring the perturbative unitarity of
the $pp \rightarrow tj\chi \bar \chi$ process. 
The magnitude of the interference decreases with increasing $H^{\pm}$ mass.
In the case of the $tW$ production channel, the two diagrams shown in Figures~\ref{fig:THDMpa_feync}\subref{fig:diagram_tw1} and~\ref{fig:THDMpa_feync}\subref{fig:diagram_tw2}
provide the dominant contributions to the DM$t$ cross section.
As in $t$-channel production, these two diagrams interfere destructively. 
When the decays $H^{\pm}\rightarrow W^{\pm} a$ are kinematically possible,
the charged Higgs bosons are produced on-shell and the cross section of $pp \rightarrow tW\chi \bar \chi$,
assuming  $H^{\pm}$ masses of a few hundred \GeV, increases to produce a sizeable event rate.
Finally, $s$-channel production is relevant in regions of the parameter space
characterised by low $H^{\pm}$ masses ($<$~300~GeV) and it is not directly targeted by the analysis, but its
contribution to the signal is taken into account.
 
This paper presents 
a dedicated search for single top quarks produced in association with DM candidates,
exploiting final-state signatures characterised by the presence of: large \etmiss; jets,
possibly arising from the fragmentation of $b$-hadrons ($b$-jets); and one or two
charged leptons, either electrons or muons ($\ell = e, \mu$). The analysis is conducted
using proton--proton ($pp$) collisions at a centre-of-mass energy $\sqrt{s} = 13$~\TeV\
produced at the LHC and collected by ATLAS between $2015$ and $2018$,
for a dataset corresponding to 139~fb$^{-1}$.  Three analysis channels,
characterised by different lepton or jet multiplicities,
are optimised to target different processes: \tWOneL and \tWTwoL
(single-lepton and dilepton final states, respectively) for the $tW$+DM events and \tjOneL
for $t$-channel DM production.
The results are interpreted in the context of 2HDM+$a$ models, considering various assumptions about the most relevant parameters, $m_a$,  $m_{H^{\pm}}$,
and $\tan\beta$. Furthermore, the mutually exclusive \tWOneL and \tWTwoL analysis channels are statistically combined to maximise the sensitivity
to $tW$+DM processes.
 
Previous searches for 2HDM+$a$ models targeted associated production of DM candidates with Higgs
or $Z$ bosons~\cite{Sirunyan:2020fwm}, as well as DM and a \ttbar pair (referred to as DM\ttbar) (see Ref.~\cite{CMS-EXO-16-050} for CMS
and Ref.~\cite{EXOT-2017-32} and references therein for ATLAS).
This search is targeting the unexplored models within ATLAS where DM produced
in association with single top quarks(for CMS results, see Ref.~\cite{CMS-EXO-18-010}).
The analysis is also sensitive to DM\ttbar processes in regions
of the parameter space where the DM$t$ and DM\ttbar production rates are similar.

% End of text imported from the .//sections/intro.tex input file
 
\section{ATLAS detector}
\label{sec:detector}
% The next lines are included from the .//sections/detector.tex input file
The ATLAS detector~\cite{PERF-2007-01} is a multipurpose particle detector with a forward--backward symmetric cylindrical
geometry and nearly 4$\pi$ coverage in solid angle.\footnote{ATLAS uses a right-handed coordinate system with its origin at the nominal
interaction point in the centre of the detector. The positive $x$-axis is defined by the direction from the interaction point to the centre of the LHC ring, with the positive $y$-axis pointing upwards, while the beam direction defines the $z$-axis. Cylindrical coordinates
$(r,\phi)$ are used in the transverse plane, $\phi$ being the azimuthal angle around the $z$-axis. The pseudorapidity $\eta$ is
defined in terms of the polar angle $\theta$ by $\eta=-\ln\tan(\theta/2)$. Rapidity is defined as $y$ = $0.5 \ln[(E + p_z)/(E-p_z)]$ where $E$
denotes the energy and $p_z$ is the component of the momentum along the beam direction. The angular distance $\DeltaR$ is defined as $\sqrt{(\Delta y)^2 + (\Delta \phi)^2}$.}
The inner tracking detector consists of pixel and microstrip silicon detectors
covering the pseudorapidity region $|\eta|<2.5$, surrounded by a transition radiation tracker
which enhances electron identification in the region $|\eta|<2.0$.
A new inner pixel layer, the insertable B-layer~\cite{ATLAS-TDR-2010-19, PIX-2018-001}, was added at a mean radius of 3.3~cm during the period between Run~1 and Run~2 of the LHC.
The inner detector is surrounded by a thin superconducting solenoid providing an axial 2~T magnetic field and by a fine-granularity lead/liquid-argon
(LAr) electromagnetic calorimeter covering $|\eta|<3.2$. A steel/scintillator-tile calorimeter provides hadronic coverage in the central pseudorapidity
range ($|\eta|<1.7$).
The endcap and forward regions ($1.5<|\eta|<4.9$) of the hadron calorimeter are made of LAr active layers with either copper or tungsten
as the absorber material.
A muon spectrometer with an air-core toroid magnet system surrounds the calorimeters.
Three layers of high-precision tracking chambers provide coverage in the range $|\eta|<2.7$, while dedicated fast chambers allow triggering
in the region $|\eta|<2.4$. The ATLAS trigger system consists of a hardware-based level-1 trigger followed by a software-based high-level trigger~\cite{TRIG-2016-01}.
% End of text imported from the .//sections/detector.tex input file
 
\section{Data and Monte Carlo simulation}
\label{sec:dataandmc}
% The next lines are included from the .//sections/datamc.tex input file
The data analysed in this paper correspond to an integrated luminosity of 139~fb$^{-1}$ of $pp$ collision
data collected between 2015 and 2018 by the ATLAS detector with a centre-of-mass energy of 13~\TeV\ and a 25\,ns proton bunch crossing interval.
The uncertainty in the combined 2015--2018 integrated luminosity is 1.7\%~\cite{ATLAS-CONF-2019-021},
obtained using the LUCID-2 detector~\cite{LUCID2} for the primary luminosity measurements.
All detector subsystems were required to be operational during data taking.
The average number of interactions in the same and nearby bunch crossings (pile-up)
increased from \averagemu$=13.4$ (2015 dataset) to \averagemu$=36.1$ (2018 dataset),
with a highest \averagemu$=37.8$ (2017 dataset) and an average \averagemu$=33.7$.
 
Candidate events were recorded using a combined set of triggers~\cite{TRIG-2016-01} based on the presence of
missing transverse momentum or charged leptons ($\ell = e, \mu$).
The $\etmiss$\ trigger~\cite{ATL-DAQ-PUB-2016-001} is fully efficient for events with reconstructed $\etmiss > 250~\GeV$
and it was used for the single-lepton analysis channels. Furthermore, an OR between $\etmiss$\ and single-lepton triggers
was used for the \tjOneL channel for events with reconstructed $\etmiss < 250~\GeV$.
Triggers based on a single muon (electron) require the presence of a muon (electron) with
transverse momentum \pt (transverse energy \et)  above certain thresholds, and impose data quality
and lepton isolation requirements. The lowest \pt (\et) threshold without trigger prescaling is 24 (26)~\GeV\ for muons (electrons)
and includes a lepton isolation requirement that is not applied for triggers with higher thresholds.
In the two-lepton channel, lower thresholds for electrons and muons must be applied to retain
sensitivity to the signal.
A combined set of two-lepton triggers was used, with the muon (electron) \pt (\et) trigger threshold
depending on the data-taking period. The lepton trigger threshold ranged between
8 and 22~\GeV\ for muons, and between 12 and 24~\GeV\ for electrons. The analysis selections are chosen
to guarantee maximum trigger efficiency, generally above 95\%.
Trigger matching requirements~\cite{TRIG-2016-01} are applied where
the lepton(s) must lie in the vicinity of the corresponding trigger-level object.
 
Dedicated Monte Carlo (MC) simulated samples are used to model SM processes and estimate the expected signal yields.
All samples were produced using the ATLAS simulation infrastructure~\cite{SOFT-2010-01} and $\GEANT4$~\cite{Agostinelli:2002hh},
or a faster simulation based on a parameterisation of the calorimeter response and $\GEANT4$ for the other detector
systems~\cite{SOFT-2010-01}. 
The simulated events are reconstructed with the same algorithms as used for data. They contain a realistic modelling
of pile-up interactions with pile-up profiles matching the ones of each dataset between 2015 and 2018, obtained
by overlaying minimum-bias events simulated using the soft QCD processes of \PYTHIAV{8.186}~\cite{Sjostrand:2007gs}
with the NNPDF2.3 LO set of parton distribution functions (PDFs)~\cite{Ball:2012cx} and the
A3~\cite{ATL-PHYS-PUB-2016-017} set of tuned parameters (tune).
 
Signal MC samples for single top quark production in association with DM include $tW$, $t$-channel and $s$-channel processes.
Samples were produced either varying the $(m_a, m_{H^{\pm}})$ parameters and assuming $\tan\beta$ equal to unity,
or varying the $(\tan\beta, m_{H^{\pm}})$ parameters and setting $m_a=250$~\GeV. Details of other parameter value assumptions are
provided in Section~\ref{sec:result}.
The samples were generated from leading-order (LO) matrix elements using the \MGMCatNLO~\cite{Alwall:2014hca} v2.6.2 generator interfaced
to \PYTHIAV{8.212}~\cite{Sjostrand:2014zea} with the A14 tune~\cite{ATL-PHYS-PUB-2014-021} for the modelling of parton showering (PS),
hadronisation and the description of the underlying event.
Parton luminosities are provided by the five-flavour scheme  NNPDF3.0 NLO~\cite{Ball:2014uwa} PDF set.
Signal cross sections are calculated to LO accuracy in QCD.
Additional simulated samples are used for DM\ttbar\ processes.
They were generated using LO matrix elements, with up to one
extra parton using the \MGMCatNLOV{v2.6.7} generator interfaced to \PYTHIAV{8.244} with the same PDF set and tune as used
for the $tW$, $t$- and $s$-channel processes.  The top quark decay was simulated using \MADSPIN~\cite{Artoisenet:2012st}.
In this case, signal cross sections are calculated to next-to-leading-order (NLO) accuracy using the same version of
\MGMCatNLO as suggested in Ref.~\cite{Backovic:2015soa}.
 
Background samples were simulated using different MC event generators, accurate at NLO or higher order, depending on the process.
All background processes are normalised to the best available theoretical calculation of their respective cross sections.
The event generators, the accuracy of theoretical cross sections, the underlying-event parameter tunes, and the PDF sets
used in simulating the SM background processes most relevant for this analysis are summarised in \Tab{\ref{tab-MCsamples}}.
For all samples, except those generated using \SHERPA~\cite{Gleisberg:2008ta,Gleisberg:2008fv,Cascioli:2011va,Schumann:2007mg,Hoeche:2012yf},
the \evtgen~v1.2.0~\cite{Lange:2001uf} program was used to simulate the properties of the $b$- and $c$-hadron decays.
 
% The next lines are included from the .//tables/tab-mcgenerators.tex input file
\begin{table}[t!]
\caption{List of generators used for the different SM background processes. Diboson includes $WW$, $WZ$ and $ZZ$ production.
Information is given about the underlying-event tunes, the PDF sets and the perturbative QCD highest-order accuracy
(LO, NLO, next-to-next-to-leading order (NNLO), and next-to-next-to-leading-log (NNLL)) used for the normalisation
of the different samples. Diboson cross sections are taken directly from \SHERPA. }
\label{tab-MCsamples}
\begin{center}
\scalebox{0.69}{
\begin{tabular}{lllllll}
\toprule
\multicolumn{2}{c}{\multirow{2}{*}{Process}} & \multirow{2}{*}{Generator} & \multirow{2}{*}{PDF set}  & PS and & \multirow{2}{*}{UE tune}   & Cross-section\\
& & & & frag./hadr. &  & accuracy \\
\midrule
\multicolumn{2}{c}{Top pair ($t\bar{t}$)} & \powbox~v2~\cite{Frixione:2007nw,Nason:2004rx,Frixione:2007vw,Alioli:2010xd} &  NNPDF 3.0 NLO & \PYTHIA 8 & A14 & NNLO+NNLL~\cite{Czakon:2011xx} \\
\multirow{2}{*}{Single-top$\bigg\{$} & $t$-channel  & \powbox~v1~\cite{Frederix:2012dh} & NNPDF 3.0 NLO & \PYTHIA 8 & A14 &NNLO+NNLL~\cite{Kidonakis:2011wy} \\
& $s$- and $Wt$-channel & \powbox~v2~\cite{Re:2010bp} & NNPDF 3.0 NLO & \PYTHIA 8 & A14 & NNLO+NNLL~\cite{Kidonakis:2010ux,Kidonakis:2010tc} \\
\multicolumn{2}{c}{$V$+jets $(V=W/Z)$} & \SHERPA~2.2.1~\cite{Gleisberg:2008ta,Gleisberg:2008fv,Cascioli:2011va,Schumann:2007mg,Hoeche:2012yf} & NNPDF 3.0 NNLO & \SHERPA & Default & NNLO~\cite{Catani:2009sm}\\
\multicolumn{2}{c}{Diboson} & \SHERPA~2.2.1 or 2.2.2 & NNPDF 3.0 NNLO & \SHERPA & Default & NLO\\
\multicolumn{2}{c}{$t\bar{t}+V$, $V=W,Z,h$} & \MGaMC~2.3.3 & NNPDF 3.0 NLO & \PYTHIA 8 & A14 & NLO~\cite{Alwall:2014hca,deFlorian:2016spz}\\
\multicolumn{2}{c}{$tWZ$} & \MGaMC~2.6.7 & NNPDF 3.0 NLO & \PYTHIA 8 & A14 & NLO~\cite{Alwall:2014hca}\\
\bottomrule
\end{tabular}
}
\end{center}
\end{table}
% End of text imported from the .//tables/tab-mcgenerators.tex input file
% End of text imported from the .//sections/datamc.tex input file
 
\section{Event reconstruction and object definitions}
\label{sec:obj}
% The next lines are included from the .//sections/objects.tex input file
Common event-quality criteria and object reconstruction definitions are applied for all analysis channels,
including standard data-quality requirements to select events taken during optimal detector operation.
In addition, in each analysis channel, dedicated selection criteria, which are specific to the objects and
kinematics of interest in those final states, are applied as described in \Sect{\ref{sec:analysis}}.
 
Events are required to have at least one reconstructed interaction vertex with a minimum of two associated tracks each having $\pt>500$~\MeV.
In events with multiple vertices, the one with the highest sum of squared transverse momenta of associated tracks is
chosen as the primary vertex~\cite{ATL-PHYS-PUB-2015-026}.
A set of baseline quality criteria are applied to reject events with non-collision backgrounds or detector noise~\cite{ATLAS-CONF-2015-029}.
 
Two levels of object identification requirements are defined for leptons and jets: baseline and signal.
Baseline leptons and jets are selected with looser identification criteria,
and are used in computing the missing transverse momentum as well as in resolving possible reconstruction ambiguities.
Signal leptons and jets are a subset of the baseline objects, with tighter quality requirements
which are used to define the search regions.
Isolation criteria, defined with a list of tracking-based and calorimeter-based variables,
are used to select signal leptons by discriminating between semileptonic heavy-flavour decays and jets misidentified as leptons.
 
Electron candidates are reconstructed from energy deposits in the electromagnetic calorimeter that are
matched to charged-particle tracks in the inner detector (ID)~\cite{PERF-2017-01}.
Baseline electrons are required to satisfy $\pt>10$~\GeV\ and $|\eta|<2.47$, excluding
the transition region between the barrel and endcap calorimeters ($1.37 < |\eta| < 1.52$).
They are identified using the `loose' likelihood identification operating point as described in Ref.~\cite{PERF-2017-01}.
The number of hits in the innermost pixel layer is used to discriminate between electrons and converted photons.
The longitudinal impact parameter $z_0$ relative to the primary vertex is required to satisfy $|z_0\sin\theta| <0.5$ mm.
Signal electrons are required to also satisfy $\pt>20$~\GeV\ and the `tight' likelihood identification criteria as defined in Ref.~\cite{PERF-2017-01}.
The significance of the transverse impact parameter $d_0$ must satisfy $|d_0/\sigma(d_0)|<5$ for signal electrons.
Signal electrons with $\pt < 200$~\GeV\ are further refined using the `FCLoose' isolation working point,
while those with larger $\pt$ are required to pass the `FCHighPtCaloOnly' isolation working point, as described in Ref.~\cite{PERF-2017-01}.
Corrections for energy contributions due to pile-up are applied.
 
Muon candidates are reconstructed from matching tracks in the ID and muon spectrometer, refined through a global fit which uses the hits from both subdetectors~\cite{PERF-2015-10}.
Baseline muons must have $\pt>10$~\GeV\ and $|\eta|<2.5$, and satisfy the `medium' identification criteria.
Like the electrons, their longitudinal impact parameter $z_0$ relative to the primary vertex is required to satisfy $|z_0\sin\theta| <0.5$~mm.
Signal muons are defined with tighter requirements on their transverse momentum and transverse impact parameter significance:
$\pt>20$~\GeV\ and $|d_0/\sigma(d_0)|<3$.
The `FCLoose' isolation working point is also required for signal muons~\cite{PERF-2015-10}.
 
Jets are reconstructed from topological clusters of energy depositions in the calorimeters
using the anti-$k_t$ algorithm \cite{PERF-2014-07}, with a radius parameter $R=0.4$~\cite{Cacciari:2008gp}.
The average energy contribution from pile-up is subtracted according to the jet area and
the jets are calibrated as described in Ref.~\cite{PERF-2016-04}.
To further reduce the effect of pile-up interactions, the jets with $|\eta|<2.4$ and $\pt<120$~\GeV\ are
required to satisfy the `medium' working point of the jet vertex tagger (JVT),
a tagging algorithm that identifies jets originating from the primary vertex using track information~\cite{PERF-2014-03, Fastjet}.
Baseline jets are selected in the region $|\eta|<4.5$ and have $\pt>20$~\GeV.
The selection of signal jets requires them to be in the region $|\eta|<2.5$ and to have $\pt>30$~\GeV.
 
Jets containing $b$-hadrons are identified as arising from $b$-quarks (`$b$-tagged')
using a multivariate algorithm (MV2c10),
based on the track impact parameters,
the presence of displaced secondary vertices and
the reconstructed flight path of $b$- and $c$-hadrons inside the jet~\cite{FTAG-2018-01}.
These $b$-tagged jets are reconstructed in the region $|\eta|<2.5$ and have $\pt>20$~\GeV.
The $b$-tagging working point provides an efficiency of 77\% for jets containing $b$-hadrons in simulated \ttbar events,
with average rejection of 110 and 4.9 for light-flavour jets and jets containing $c$-hadrons, respectively~\cite{ATL-PHYS-PUB-2017-013}.
 
To resolve the reconstruction ambiguities between electrons, muons and jets, an overlap removal procedure
is applied to baseline objects in a prioritised sequence as follows.
First, if an electron shares the same ID track with another electron, the one with lower \pt is discarded.
Any electron sharing the same ID track with a muon is rejected.
Next, jets that are not $b$-tagged are rejected if they lie within  $\Delta R = 0.2$ of an electron.
Similarly, jets that are not $b$-tagged are rejected if they lie within $\Delta R = 0.2$ of a muon if the
jet has fewer than three associated tracks or the muon is matched to the jet through ghost association~\cite{Cacciari:2008gn}.
Finally, electrons and muons that are close to a remaining jet are discarded
if their distance from the jet is $\Delta R < \min(0.4,0.04+10~\GeV/\pt)$
as a function of the lepton \pt.
 
The missing transverse momentum \ptmiss, with magnitude \met, is calculated as
the negative vectorial sum of the transverse momentum of all baseline reconstructed objects
(electrons, muons, jets and photons~\cite{PERF-2017-02}) and the soft term.
The soft term includes all tracks associated with the primary vertex but not matched to any reconstructed physics object.
Tracks not associated with the primary vertex are not considered in the \met\ calculation,
improving the \met\ resolution by suppressing the effect of pile-up~\cite{PERF-2016-07,ATLAS-CONF-2018-023}.
 
To compensate for differences between data and simulation in trigger,
particle identification and reconstruction efficiency,
correction factors that are usually functions of the relevant kinematic variables
are derived from data and applied to the samples of simulated events.
 
% End of text imported from the .//sections/objects.tex input file
 
\section{Analysis strategy}
\label{sec:analysis}
% The next lines are included from the .//sections/ana_common.tex input file
The search is conducted in three independent analysis channels differing in 
lepton and jet multiplicities to maximise the sensitivity to distinct signal processes.
The \tWOneL\ analysis channel targets $tW$+DM events where one of the $W$ bosons (directly produced or arising from
the top quark decay) decays leptonically (Section~\ref{sec:tW1L}).
The \tWTwoL\ analysis channel targets the same signal processes,
but considers events where both $W$ bosons decay leptonically (Section~\ref{sec:tW2L}).
The two selections are designed to be mutually exclusive.
The results of these two analysis channels are statistically combined to maximise
the sensitivity to the $tW$+DM processes.
Finally, the \tjOneL\ analysis targets $t$-channel production of DM candidates
and requires a single lepton in each event (Section~\ref{sec:tj}).
In all analysis channels, large missing transverse momentum and jets are required.
Event selections and background estimation methods specific to each analysis channel are described in this section,
as are the definitions of the signal, control, and validation regions (SR, CR, and VR, respectively).
 
Dedicated CRs are designed in each analysis channel for the major SM background processes in order
to predict their expected contribution in the SRs.
The CRs and SRs are mutually exclusive, with the CRs enriched in the
major background processes relevant to each analysis channel while minimising the contamination from signal.
The potential signal contamination in the CRs is found to be negligible, at the level of $<$~3\% of the total SM expectation
for all analysis channels.
 
The expected SM backgrounds are first determined independently for each channel,
with a profile likelihood fit~\cite{Baak:2014wma} in a background-only fit.
In this fit, normalisation factors of the backgrounds, for which dedicated CRs are defined, are
adjusted simultaneously to match the data in the associated CRs.
The input to the background-only fit includes the number of events observed in the associated CRs and the
number of events predicted by simulation in each CR for all background processes.
They are both described by Poisson statistics. The systematic uncertainties, described
in \Sect{\ref{sec:sys}}, are included in the fit as nuisance parameters. They are
constrained by Gaussian distributions with widths corresponding to the sizes of the uncertainties
and are treated as correlated, when appropriate, between the various regions.
The product of the various probability density functions forms the likelihood, which the fit maximises
by adjusting the background normalisation and the nuisance parameters.
 
Normalisation and nuisance parameters obtained from the background-only fit
to the control regions are then extrapolated~\cite{Baak:2014wma} to the SRs
to quantify potential excess in data.
The reliability of the MC extrapolation of the SM background estimates outside of the control regions is
verified in dedicated validation regions.
Statistically independent from the corresponding CRs and SRs, these VRs are designed to probe a kinematic region closer to that of the SRs.
The potential signal contamination in the VRs is at the level of $<$~1\% of the total SM expectation
for most validation regions, and between 8\% and 15\% in a few validation regions in the \tWOneL\ analysis channel.
 
In the absence of a significant event excess in the SRs, as determined after the background-only fit,
simultaneous fits of the CRs and SRs are performed to
constrain the parameters of the targeted signal models as well as a generic beyond the standard model (BSM),
referred to as model-dependent and model-independent signal fits as detailed in Section~\ref{sec:result}.
 
% End of text imported from the .//sections/ana_common.tex input file
 
\subsection{Kinematic requirements and event variables}
\label{sec:variables}
% The next lines are included from the .//sections/variables.tex input file
The event selection criteria in each analysis channel
are defined using the physics objects described in Section~\ref{sec:obj} and
the event variables defined in this section.
 
The following variables are defined using
simple combinations of the physics objects in the events.
 
\begin{itemize}
 
\item $\njet$ is the number of jets with $|\eta| < 2.5$ and $\pt > 30$ GeV.
\item $\njetforward$ is the number of jets in the forward region, $2.5<|\eta|<4.5$ and $\pt > 30$ GeV.
\item $\nbjet$ is the number of $b$-jets with $|\eta| < 2.5$ and \pt\ above a given threshold defined in each analysis channel.
 
\item The minimum azimuthal distance $\dphimin$ between the \ptmiss and the $\vec{p}_{\mathrm{T}}$ of each of the four leading jets in the event is useful for rejecting events with mismeasured jet energies leading to \met\ in the event, and is defined as:
 
\begin{equation}
\label{eq:dphimin}
\dphimin = \textrm{min}_{i\leq 4} \Delta\phi
\left(  \ptmiss,
\vec{p}_{\mathrm{T},i}^{\mathrm{jet}}
\right)\,, \notag
\end{equation}
\noindent where $\textrm{min}_{i\leq 4}$ selects the jet that minimises $\Delta\phi$.
 
\item \mll is the invariant mass of the dilepton system in the event.
 
\item
An iterative reclustering approach as defined in Ref.~\cite{Aaboud:2017aeu}
is used to reconstruct the hadronically decaying $W$ bosons.
All the signal jets in the event are first reclustered using the
\akt\ algorithm with a large radius parameter of $R=3.0$.
The radius of each large-radius jet is then iteratively
reduced to an optimal radius, $R(\pt) = 2\times m_{W}/\pt$.
The mass of the reclustered jet, \mWHad, is used in the \tWOneL\ channel.

\item
\mLB\ is the invariant mass of the leading lepton and $b$-jet in the event.

\end{itemize}
 
A set of variables based on transverse mass are defined in order to distinguish between
the signal and SM background processes in the following.
 
\begin{itemize}
 
\item The transverse mass formed by the \ptmiss\ and the leading lepton in the event, \mtw, is used to
reduce the $W$+jets and semileptonic \ttbar~backgrounds. It is defined as:
 
\begin{equation}
\label{eq:mtw}
\mtw  = \sqrt{2 p_{\mathrm{T}}^{\ell} \MET \left(1 - \cos \Delta\phi(\vec{p}_\mathrm{T}^\ell,\ptmiss) \right) }\,. \notag
\end{equation}
 
\item
Similarly, the transverse mass \mtLBMET\ is formed by the \ptmiss\ and the system of the  leading lepton and $b$-jet in  the event
to suppress the $W+b$ background, and is defined as:
\begin{equation}
\mtLBMET  = \sqrt{2 p_{\mathrm{T}}^{\ell_1+b_1} \MET \left(1 - \cos \Delta\phi(\vec{p}_\mathrm{T}^{\ell_1+b_1},\ptmiss) \right) }\,. \notag
\end{equation}
 
\item
Closely related to \mtw, the stransverse mass \mttwo~\cite{Lester:1999tx,Barr:2003rg} is
used to bound the masses of pair-produced particles, such as in $\ttbar$ production,
each of which decays so as to produce a visible particle that can be detected and an invisible particle that contributes to the missing transverse momentum.
In the case of a dilepton final state, it is defined by:
 
\begin{equation}
\mttwo(\vec{p}_{\mathrm{T}}^{\ell_1}, \vec{p}_{\mathrm{T}}^{\ell_2}, \ptmiss) = \min_{\vec{q}_\mathrm{T}}{\left[ \max \left(\mt (\vec{p}_{\mathrm{T}}^{\ell_1},\vec{q}_{\mathrm{T}}),\mt (\vec{p}_{\mathrm{T}}^{\ell_2}, \ptmiss - \vec{q}_{\mathrm{T}}) \right) \right]}\,, \notag
\end{equation}
 
where $\vec{q}_{\mathrm{T}}$ is the transverse momentum vector that minimises the larger of the two transverse masses \mt, and $\vec{p}_{\mathrm{T}}^{\ell_1}$ and $\vec{p}_{\mathrm{T}}^{\ell_2}$ are the leading and subleading transverse momenta of the two leptons in the pair.
For the dileptonic \ttbar\ background events, \mttwo\ has a kinematic endpoint at $m_W$.

\item
The asymmetric stransverse mass \amt\cite{Konar:2009qr,Lester:2014yga}, a variation of \mttwo, is
used in the $\tWOneL$ final state to reduce the number of dileptonic \ttbar\ background events
where one of the leptons is undetected.
For these events, \amt\ has a kinematic endpoint at the top quark mass.
\end{itemize}
 
To improve the selection of single-top events in the \tWTwoL\ channel,
the following quantities based on invariant mass are defined.
 
\begin{itemize}
\item
\minmbl\ is the minimum invariant mass found by combining the leading $b$-jet with each of the leptons,
$\minmbl=\min(m_{b_1\ell_1}, m_{b_1\ell_2})$.
An upper endpoint at approximately 153 \GeV\ or 160$-$170 \GeV\ is expected
for the events with one or two leptonic top quark decays, respectively.
 
\item
 
To further reduce the background with two leptonic top quark decays, such
as \ttbar\ and $\ttbar V$, \mblt, an extended variation of \minmbl, is used in the \tWTwoL\ final state.
It is defined as:
\begin{equation}
\mblt = \min[\max(m_{\ell_{1}j_{1}},m_{\ell_{2}j_{2}}),\max(m_{\ell_{1}j_{2}},m_{\ell_{2}j_{1}})]\,, \notag
\end{equation}
where $m_{\ell_{n}j_{m}}$ is the invariant mass of lepton $\ell_{n}$ and jet $j_{m}$, where $j_{1}$ and $j_{2}$
are the two jets with highest $b$-tag discriminator value.
For the \ttbar\ and $\ttbar V$ backgrounds where both top quarks decay leptonically, \mblt\ has
a kinematic endpoint at approximately 160$-$170 GeV.
 
\end{itemize}
 
Additional variables based on angular separations of the objects are used in the \tjOneL\ analysis to suppress SM background contributions, as
defined below.
 
\begin{itemize}
 
\item
 
\DeltaEtaLB, \DeltaPhiLB, and \DeltaRLB: the pseudorapidity difference, azimuthal angle difference, and angular distance
between the leading lepton and $b$-jet in the event.
 
\item
 
\DeltaPhiLMET: the azimuthal angle difference between the \ptmiss\ and the leading lepton in the event.
 
\end{itemize}
 
Table~\ref{tab:preselection} summarises the trigger and preselection requirements for all analysis channels, in terms
of lepton, jet and $b$-jet multiplicities, as well as transverse momenta and global kinematic variables.
Events with extra baseline leptons are vetoed in addition.
% The next lines are included from the .//tables/tab-preselection.tex input file
\begin{table}[t]
\caption{Summary of the trigger and preselection requirements for the three analysis channels.
Opposite-sign leptons are indicated as (OS).
Events with extra baseline leptons are vetoed in addition.
}
\label{tab:preselection}
\small
\begin{center}
\begin{tabular}{lccc}
\toprule
Variable                      & \tWOneL   &   \tWTwoL      & \tjOneL \\
\midrule
Trigger                       & \etmiss   & dilepton      & \etmiss OR one-lepton \\\midrule
$N_{\ell}^{\mathrm{signal}}$  &  $=1$     &   $=2$ (OS)    &  $=1$  \\
$\pt(\ell_1) \;[\GeV]$        &  $>30$    &   $>25$        & $>30$  \\
$\pt(\ell_2) \;[\GeV]$        &  -        &   $>20$        &  - \\
\midrule
\njet                         & $\geq 3$  &  $\geq 1$      & $\in[1,4]$   \\
$\pt(\mathrm{jet})$\;[\GeV]   &  $>30$    &  $>30$         & $>30$ \\
\nbjet                        &  $\geq1$  &  $\geq 1$      &  $\in[1,2]$   \\
$\pt(b_1)$\;[\GeV]            &  $>50$    &  $>50$         & $>50$ \\
\midrule
$\met \;[\GeV]$               & $>250$    &  $>200$        & $>200$ \\
$\mtw \;[\GeV]$               & $>30$     &   -            & $>60$  \\
$m_{\ell\ell}\;[\GeV]$        &   -       & $\geq40$, $\notin [71,111]$ ($ee/\mu\mu$)       & - \\
$\dphimin$ [rad]            & $>0.5$    &   -           & $>0.5$  \\
\bottomrule
\end{tabular}
\end{center}
\end{table}
% End of text imported from the .//tables/tab-preselection.tex input file
 
% End of text imported from the .//sections/variables.tex input file
 
% The next lines are included from the .//sections/ana_onelep.tex input file
\subsection{Single-lepton \tWOneL analysis channel}
\label{sec:tW1L}
 
Events with exactly one electron or muon are first selected for the SR if they also contain at least three jets,
exactly one of which must be $b$-tagged, and satisfy the preselection requirements described in Table~\ref{tab:preselection}.
The dominant SM background contributions in the channel are
\ttbar, $W$+jets, and single top ($Wt$ channel) production.
Discriminating variables,  \etmiss, \mtw, \mWHad\ and the asymmetric stransverse mass \amt\ as described in \Sect{\ref{sec:variables}},
are used to further separate the signal from backgrounds.
A `genetic algorithm'~\cite{Sastry2005} is used to optimise a baseline signal region defined as in Table~\ref{tab:regions_tW1L}.
To increase the sensitivity to different signal model parameters,
a binned \etmiss\ distribution is used as the final input for the statistical analysis.
The binning is chosen to be $[250, 300]$~\GeV, $[300, 400]$~\GeV, $[400, 500]$~\GeV, $[500, 600]$~\GeV\ and
$\geq 600$~\GeV, referred to as Bins [0$-$4].
 
The acceptance times detector efficiency for the $tW$+DM signal processes
after applying all selection criteria is between 0.3\% and 5.1\% in the parameter space of $\tan\beta=1$,
$m_a\in [100, 450]$~\GeV\ and $m_{H^{\pm}}\in [400, 1500]$~\GeV,
and between 0.2\% and 4.8\% in the parameter space of $m_a=250$~\GeV,
$\tan\beta\in [0.5, 30]$ and $m_{H^{\pm}}\in [400, 1500]$~\GeV.
 
\begin{table}[h!]
\caption{Summary of signal, control and validation region definitions used in the \tWOneL\ analysis channel.
The `--' entries represent an inclusive selection with no requirements.
The $W+$jets control and validation regions are each split into two regions with opposite lepton charges as described in the text.
}
\label{tab:regions_tW1L}
\small
\begin{center}
\begin{tabular}{p{2.5cm}|c|cc|cccc}
\toprule
Variable   &  SR   & CR($\mathrm{\ttbar}$) & CR($\mathrm{W}$) & VR1($\mathrm{\ttbar}$) & VR2($\mathrm{\ttbar}$) & VR1($\mathrm{W}$) & VR2($\mathrm{W}$) \\
\midrule
\nbjet            &   $=1$ &  $\geq 2$             &  $=1$            &  $=1$         &   $=1$          &   $=1$        &  $=1$    \\
$\pt(b_2)$ [\GeV] &  $<50$ &  $>50$                &  $<50$           &  $<50$        &   $<50$         &  $<50$        & $<50$      \\
\midrule
\mWHad [\GeV]     & $>60$  &  --                   &  $<60$           &  --        &  $<60$          &  $>60$        &  $<60$    \\
\mtw   [\GeV]     & $>200$ &  $>200$               & $\in[40,100]$    & $>200$        &  $>200$         & $\in[40,100]$ &  $>100$    \\
\amt   [\GeV]     & $>220$ &  $<220$               & $>220$           & $<220$        &  $>220$         &  $>220$       &  $>220$    \\
\bottomrule
\end{tabular}
\end{center}
\end{table}
 
Dominant background contributions from the $t\bar{t}$ and $W$+jets processes
are estimated using MC simulation and the dedicated CRs.
The contribution from multijet production, where the lepton is a misidentified jet or originates
from a heavy-flavour hadron decay or photon conversion, is found to be negligible.
The remaining sources of background (single-top, $Z$+jets, diboson, \ttv, and \tWZ\ production,
as well as rarer processes such as triboson, \tttt, and \ttWW), are estimated from simulation.
 
Dedicated control regions \CRtttWOneL\ and \CRWtWOneL, defined in \Tab{\ref{tab:regions_tW1L}},
are designed for the $t\bar{t}$ and $W$+jets background estimations.
Compared to the SR, the acceptance for $t\bar{t}$ events is increased in \CRtttWOneL\ by requiring at least two $b$-jets, inverting
the selection on \amt\ and removing the requirement on \mWHad.
To increase the acceptance of the $W+$jets events and hence the sample size,
\CRWtWOneL\ is first selected by requiring $40<\mtw<100~\GeV$ and $\mWHad<60~$\GeV.
To exploit the lepton charge asymmetry of the $W$+jets events relative to the remaining backgrounds,
it is subsequently split into two regions with opposite lepton charges, \CRWtWOneLp\ and \CRWtWOneLm.
Normalisation factors, $\mu^{}_{\ttbar}$\ and $\mu^{}_{W+\mathrm{jets}}$, defined as the
ratio of the number of observed events to the SM prediction,
are found to be $0.96\pm 0.08$ and $1.01\pm0.05$ after the background-only fit for the $t\bar{t}$ and $W+$jets processes, respectively.

\begin{figure}
\begin{center}
\begin{subfigure}[b]{0.32\textwidth}
\centering\includegraphics[width=\textwidth]{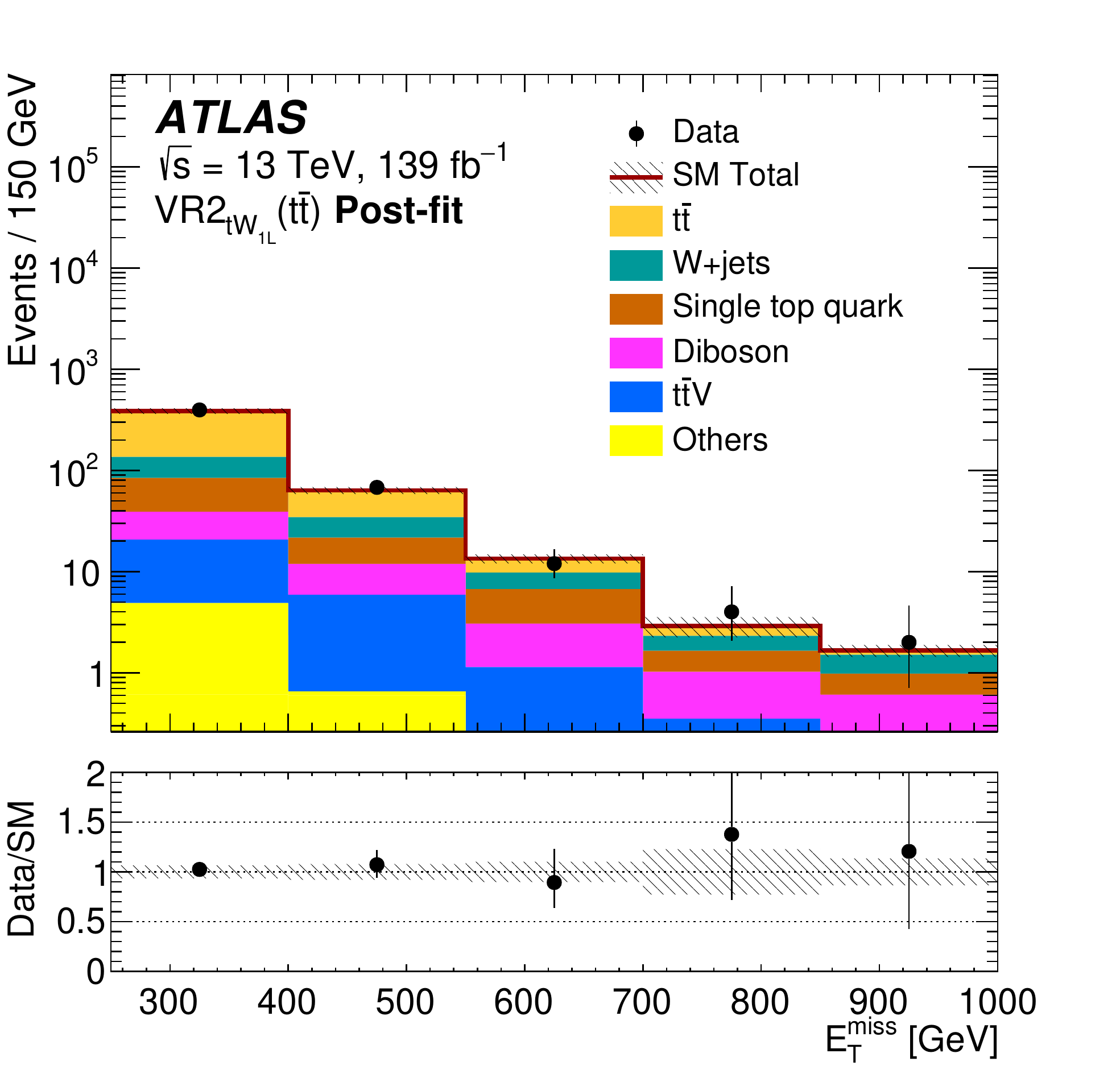}
\caption{\label{fig:tW1L_VRtt_VR2_met}}
\end{subfigure}
\begin{subfigure}[b]{0.32\textwidth}
\centering\includegraphics[width=\textwidth]{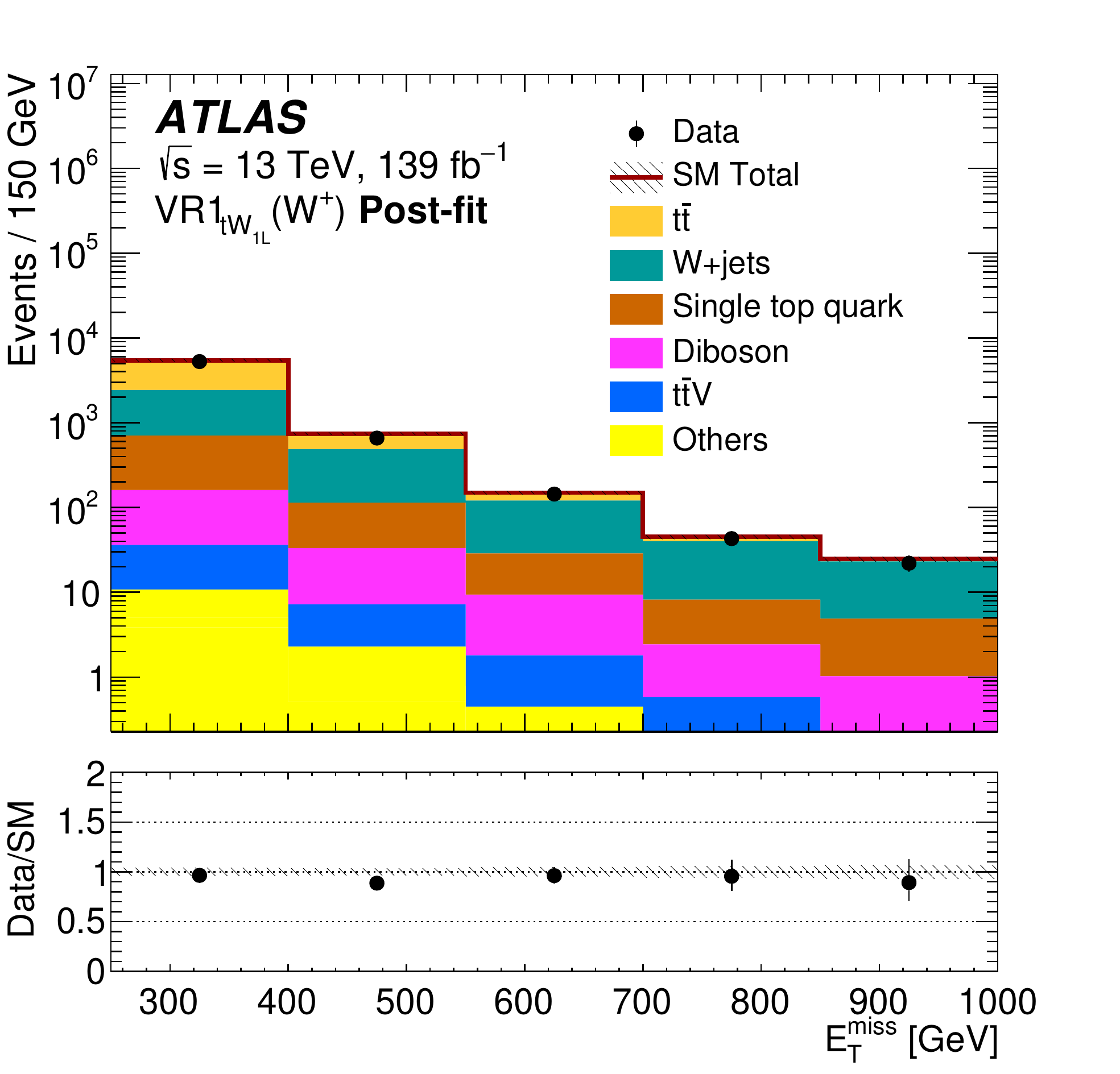}
\caption{\label{fig:tW1L_VRW_VR1p_met}}
\end{subfigure}
\begin{subfigure}[b]{0.32\textwidth}
\centering\includegraphics[width=\textwidth]{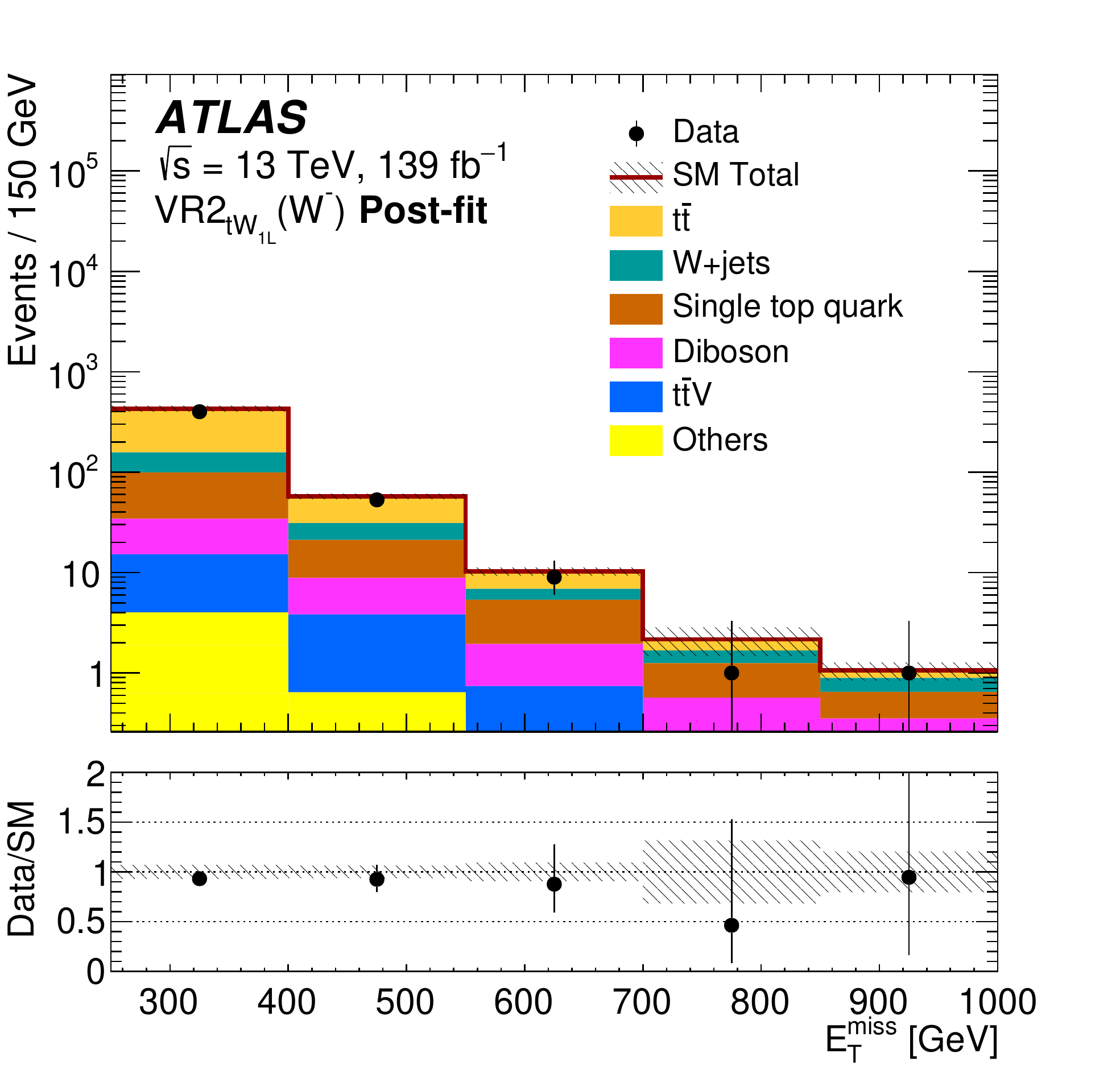}
\caption{\label{fig:tW1L_VRW_VR2m_met}}
\end{subfigure}
\end{center}
\caption{The \etmiss\ distributions after the background-only fit (post-fit) are shown in three representative validation regions:
\subref{fig:tW1L_VRtt_VR2_met}\VRBtttWOneL, \subref{fig:tW1L_VRW_VR1p_met}\VRAWtWOneLp\ and
\subref{fig:tW1L_VRW_VR2m_met}\VRBWtWOneLm.
The uncertainty bands plotted include all statistical and systematic uncertainties.
The `Others' category includes contributions from $Z$+jets and \tWZ\ production, and
rare processes such as triboson, \tttt, \ttWW, and Higgs boson production processes.
The overflow events, where present, are included in the last bin.
The lower panels show the ratio of data to the background prediction.
The hatched error bands indicate the combined experimental and MC statistical uncertainties on these background predictions.
}
\label{fig:tW1L:VRs}
\end{figure}

\begin{figure}[!ht]
\centering
\includegraphics[width=0.90\textwidth]{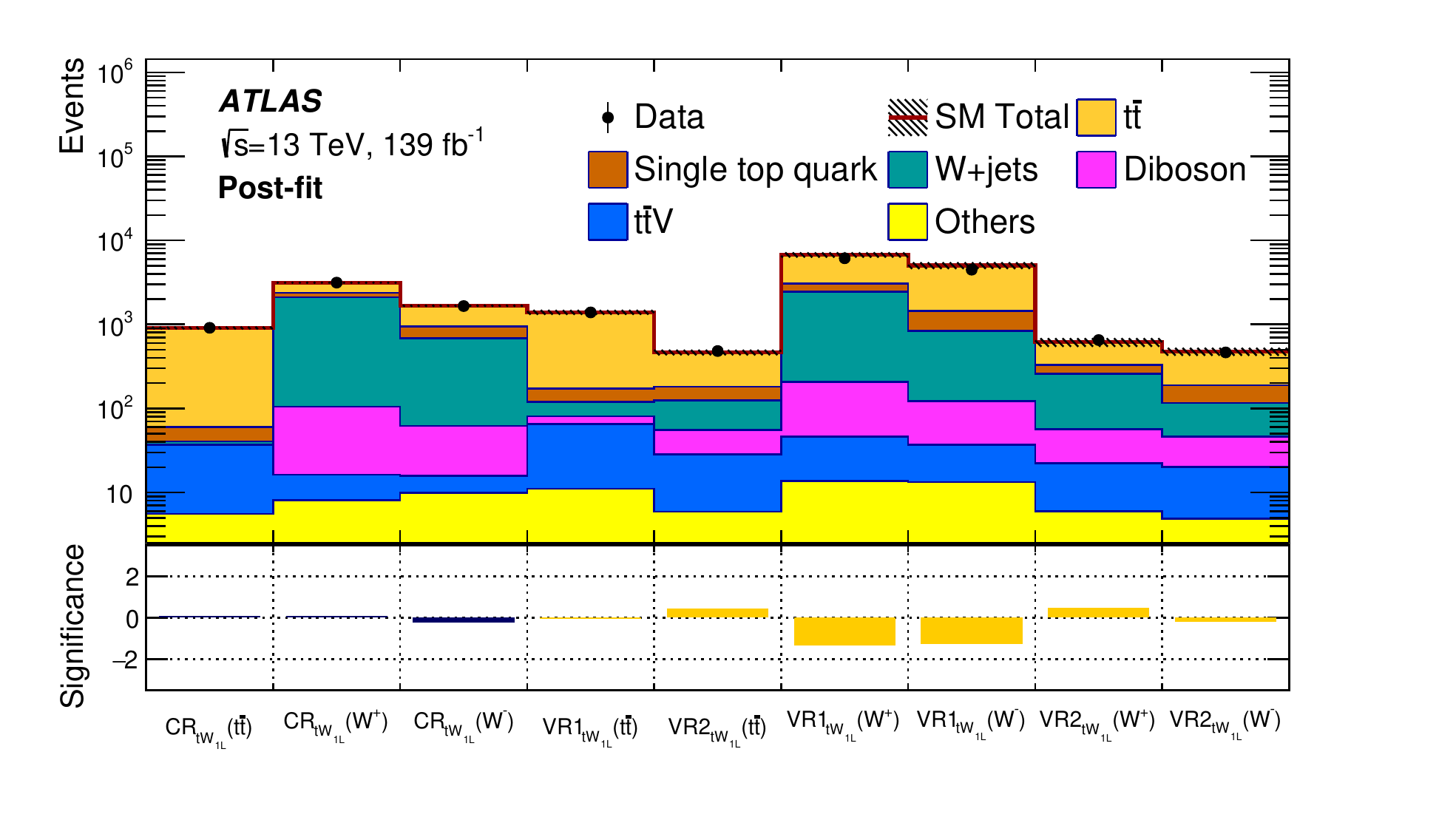}
\caption{ Comparison of the predicted backgrounds with the observed numbers of events in the
CRs and VRs associated with the \tWOneL\ channel.
The normalisation of the backgrounds is obtained from the background-only fit to the CRs.
The `Others' category includes contributions from $Z$+jets and \tWZ\ production, and
rare processes such as triboson, \tttt, \ttWW, and Higgs boson production processes.
The upper panel shows the observed number of events and the predicted background yield.
All uncertainties are included in the uncertainty band.
The lower panel shows the significances in each region.
}
\label{fig:tW1L:pulls-CRVR}
\end{figure}

To validate the \ttbar\ background predictions and the reliability of
MC extrapolation in \mWHad\ and \amt, two validation regions, \VRAtttWOneL\ and \VRBtttWOneL,
are defined by reversing the SR selection requirements on \amt\ and \mWHad\ respectively, as shown in Table~\ref{tab:regions_tW1L}.
To increase the sample size, the SR selection requirement on the \mWHad\ is removed in the \VRAtttWOneL\ region.
Similarly, for the $W+$jets background processes, two validation regions, \VRAWtWOneL\ and \VRBWtWOneL, are
defined by varying the SR selection requirements on \mtw\ and \mWHad\ shown in Table~\ref{tab:regions_tW1L}, respectively.
Each of the $W+$jets validation regions is split into two regions with opposite lepton charge.
\Fig{\ref{fig:tW1L:VRs}} shows the post-fit \etmiss\ distributions in the representative validation regions.
Good agreement is observed between data and SM expectation in all validation regions.
The observed yield, post-fit background estimates and significance~\cite{SignificanceCalc}
in each CR and VR are shown in \Fig{\ref{fig:tW1L:pulls-CRVR}} after the background-only fit.
Since the $W+$jets CR is split into two regions with opposite lepton charges sharing the same
normalisation factor, the significances in the CRs are shown explicitly.
The data event yields are found to be consistent with background expectations.
 
% End of text imported from the .//sections/ana_onelep.tex input file
% The next lines are included from the .//sections/ana_dilep.tex input file
\subsection{Dilepton \tWTwoL analysis channel}
\label{sec:tW2L}

\begin{figure}
\begin{center}
\begin{subfigure}[b]{0.32\textwidth}
\centering\includegraphics[width=\textwidth]{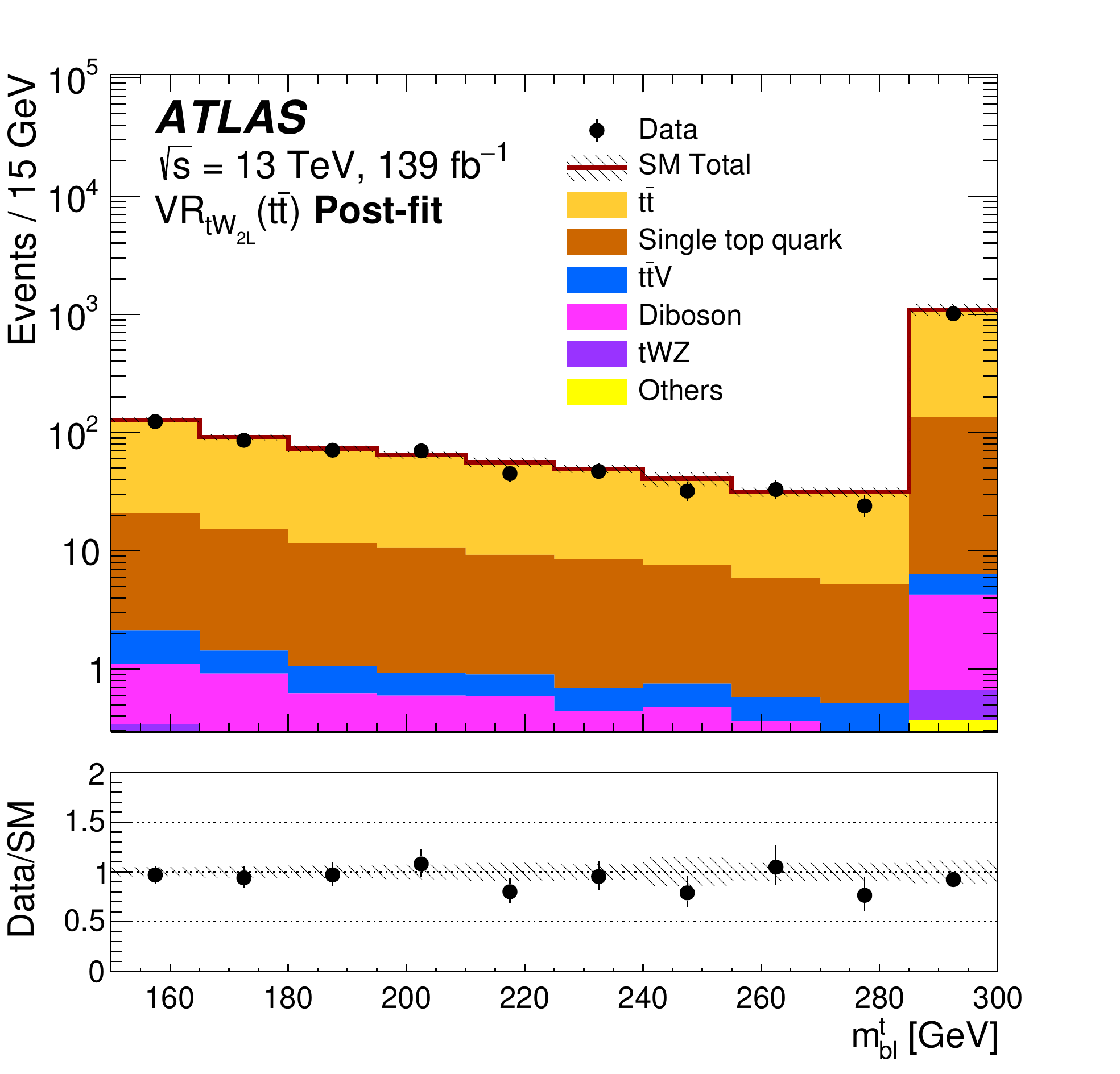}
\caption{\label{fig:tW1L_VRtt_mblt}}
\end{subfigure}
\begin{subfigure}[b]{0.32\textwidth}
\centering\includegraphics[width=\textwidth]{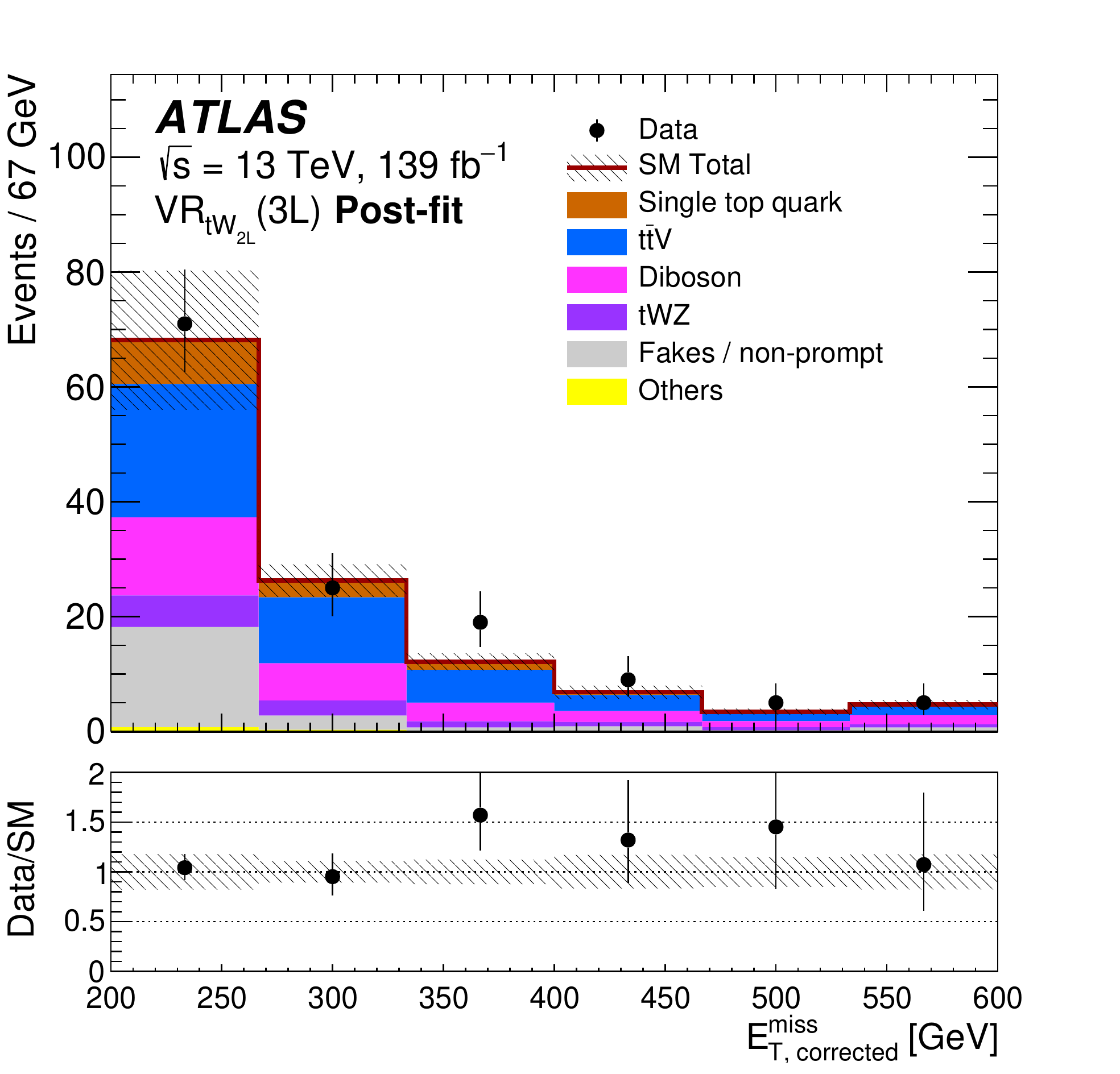}
\caption{\label{fig:tW1L_VR3L_met}}
\end{subfigure}
\begin{subfigure}[b]{0.32\textwidth}
\centering\includegraphics[width=\textwidth]{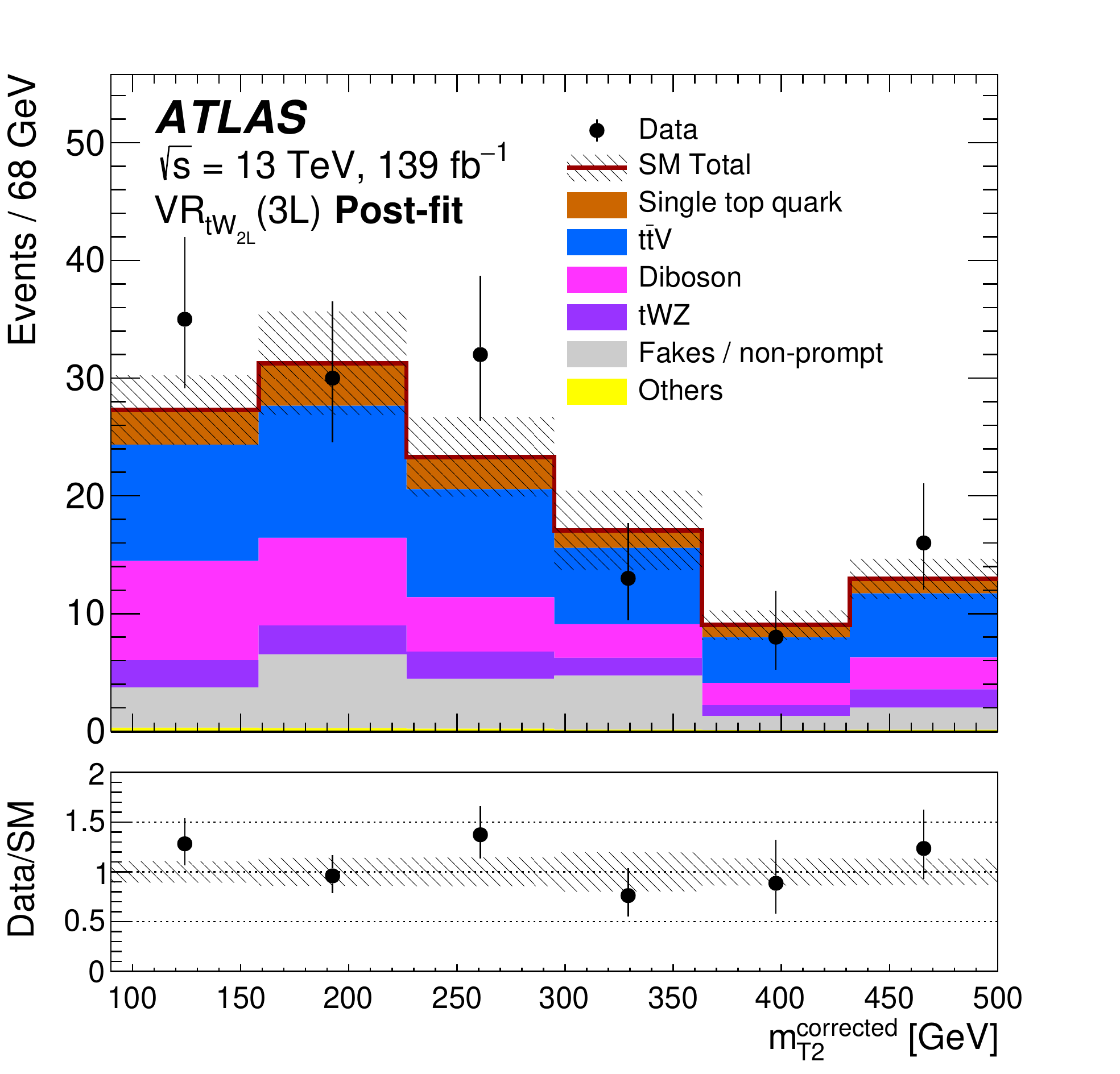}
\caption{\label{fig:tW1L_VR3L_mt2}}
\end{subfigure}
\end{center}
\caption{The kinematic distributions in the \ttbar\ and $3\ell$ validation regions of
the \tWTwoL\ analysis channel after the background-only fit:
\subref{fig:tW1L_VRtt_mblt} \mblt\ in \VRtttWTwoL,
\subref{fig:tW1L_VR3L_met} \etmiss\ in \VRThreeLTwoL and
\subref{fig:tW1L_VR3L_mt2} \mttwo\ in \VRThreeLTwoL.
The uncertainty bands plotted include all statistical and systematic uncertainties.
The `Others' category includes contributions from rare processes such as triboson,
\tttt, \ttWW, and Higgs boson production processes.
The overflow events, where present, are included in the last bin.
Since the \mblt\ is defined for events with at least two jets, the
events with exactly one jet are included in the overflow bin.
The lower panels show the ratio of data to the background prediction.
The hatched error bands indicate the combined experimental and MC statistical uncertainties on these background predictions.
}
\label{fig:tW2L_VRs}
\end{figure}

\begin{figure}[!ht]
\centering
\includegraphics[width=0.90\textwidth]{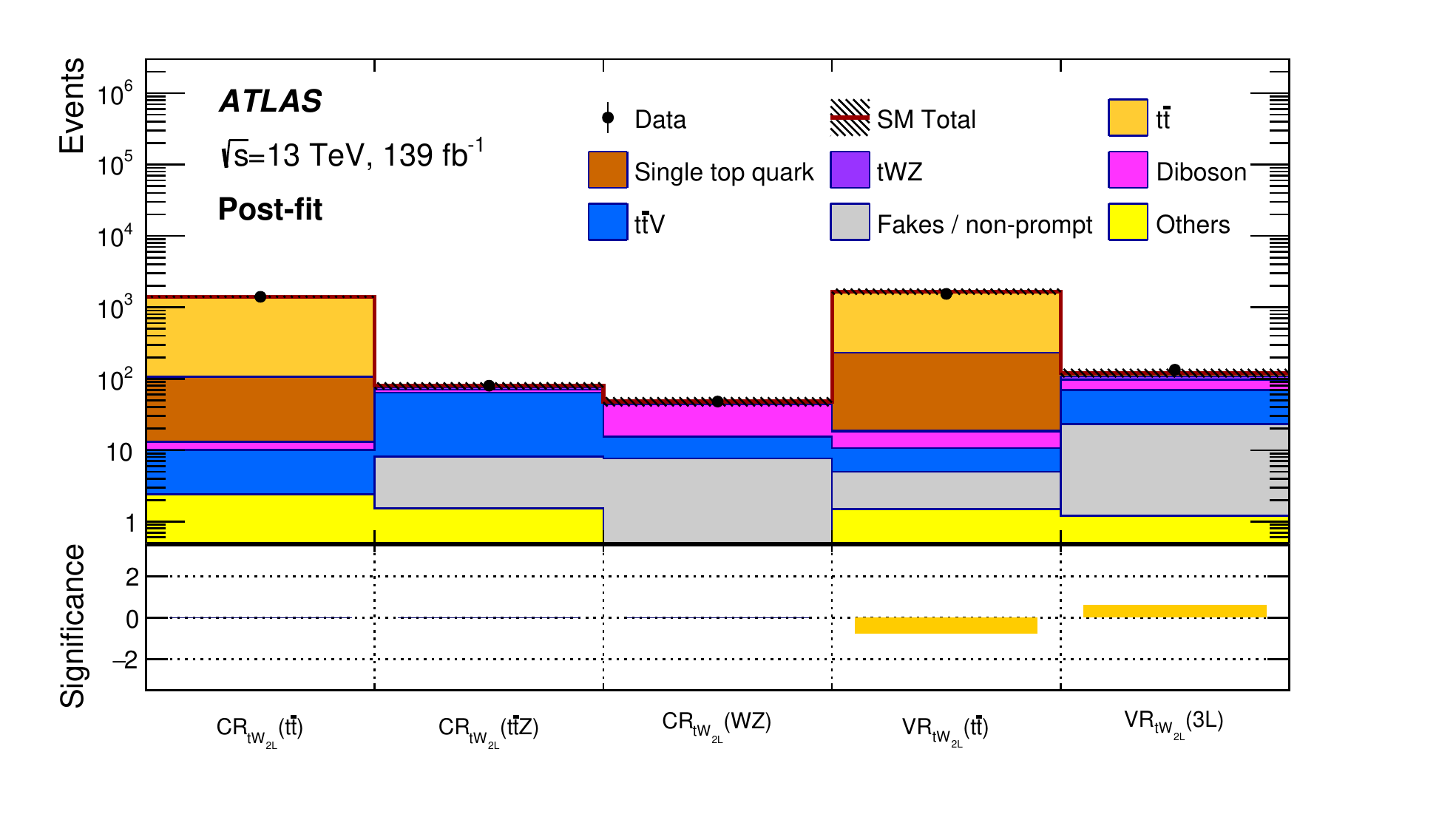}
\caption{
Comparison of the predicted backgrounds with the observed numbers of events in the
CRs and VRs associated with the \tWTwoL\ channel.
The normalisation of the backgrounds is obtained from the background-only fit to the CRs.
The `Others' category includes contributions from rare processes such as triboson,
\tttt, \ttWW, and Higgs boson production processes.
The upper panel shows the observed number of events and the predicted background yield.
All uncertainties are included in the uncertainty band.
The lower panel shows the significances estimated for each region.
}
\label{fig:tW2L:pulls-CRVR}
\end{figure}

Events with exactly two oppositely charged leptons (electron or muon)
are first selected for the SR if they also contain at least one signal jet,
at least one of which must be $b$-tagged with $\pt>50$~\GeV,
and satisfy the preselection requirements described in Table~\ref{tab:preselection}.
The dominant SM background contributions in the channel after these selections are
from the \ttbar, $\ttz$, and \tWZ\ processes, followed by that of diboson events.
The contribution from misidentified or non-prompt lepton backgrounds
(referred to as `Fakes /non-prompt' in Figures~\ref{fig:tW2L_VRs} and~\ref{fig:tW2L:pulls-CRVR})
is found to be negligible in the signal region.
 
Discriminating variables, \minmbl, \mblt, \mttwo and \dphimin\ as defined in \Sect{\ref{sec:variables}},
are used to define the final signal region as shown in Table~\ref{tab:regions_tW2L}.
 
The acceptance times detector efficiency after applying all selection criteria for the $tW$+DM
signal processes is between 0.07\% and 0.7\% in the parameter space of $\tan\beta=1$,
$m_a\in [100, 450]$~\GeV\ and $m_{H^{\pm}}\in [400, 1500]$~\GeV,
and between 0.05\% and 0.6\% in the parameter space of $m_a=250$~\GeV,
$\tan\beta\in [0.5, 30]$ and $m_{H^{\pm}}\in [400, 1500]$~\GeV.
 
\begin{table}[h!]
\caption{Summary of signal, control and validation region definitions used in the \tWTwoL\ analysis channel.
The `--' entries represent an inclusive selection with no requirements.
In the final states with three leptons, the corrected \met, \minmbl\ and \mttwo\  variables are used instead
as described in the main text.
The selection requirement on the corrected \minmbl\ in the VR$(3\ell)$ region varies
according to the jet and $b$-jet multiplicity as described in the main text.
Events with additional baseline leptons are vetoed.
}
\label{tab:regions_tW2L}
\small
\begin{center}
\begin{tabular}{p{2.5cm}|c|ccc|cc}
\toprule
Variable              &   SR             &  CR($\mathrm{\ttbar}$) & CR($\mathrm{\ttbar Z}$)  & CR($\mathrm{WZ}$)  & VR($\mathrm{\ttbar}$) & VR($\mathrm{3\ell}$) \\
\midrule
$N_{\ell}^\mathrm{signal}$ 	 &  $=2$            &   $=2$           & $=3$                &  $=3$             & $=2$             &  $=3$  \\
&  (OS)            &   (OS)           & ($\geq1$ SFOS)      & ($\geq1$ SFOS)    & (OS)             & ($\geq1$ SFOS) \\
$\pt(\ell_3)$  [\GeV]        &  -               &   -              & $>20$               &  $>20$            & -                & $>20$   \\
$m_{ee/\mu\mu}$ [\GeV]       & $\notin[71,111]$ & $\notin[71,111]$ & $\in[71,111]$       & $\in[71,111]$     & $\notin[71,111]$ & $\in[71,111]$ \\
\midrule
\njet                        & $\geq1$          &   $\geq1$         & $\geq3$             & $\in[1,3]$        & $\geq1$  		    & $\geq1$ \\
\nbjet                       & $\geq1$          &   $\geq1$         & $\geq1$             &   $=1$            & $\geq1$     	 	  & $\geq1$   \\
&                  &               & ($\geq2$ if $\njet=3$) &   			           &             	 	  &   \\
\midrule
\minmbl        [\GeV]        & $<170$           &  $<170$          &  $<170$             &  $>170$           & $<170$			      & varies   \\
\mblt          [\GeV]        & $>150$           &  $<150$          &  --                 &  --               & $>150$           & --    \\
\mttwo         [\GeV]        & $>130$           &  $\in[40,80]$    &  $>90$              &  $>90$            & $\in[40,80]$     & $>90$   \\
\dphimin       [rad]         & $>1.1$           &  $>1.1$          &  --              & --               & $>1.1$           & --   \\
\bottomrule
\end{tabular}
\end{center}
\end{table}
 
The contributions from the $t\bar{t}$, $\ttbar V$ (with $V=W$ or $Z$ boson) and diboson background processes
are estimated from MC simulation and dedicated CRs.
The remaining sources of background, including the irreducible \tWZ\ process, which is dominated
by the $Z\to \nu\nu$ component, single top quark production,  $\ttbar h$ production
and other rarer processes such as \tttt\ and \ttWW, are estimated from simulation.
 
The acceptance for $t\bar{t}$ events is increased in \CRtttWTwoL\ by requiring a low value of
\mttwo\ and inverting the SR selection criteria on \mblt.
 
The $\ttbar V$ contribution is dominated by the $\ttz$ component (about 80\% of $\ttbar V$ in the SR),
especially where $Z\to \nu\nu$.
A dedicated control region, \CRttZtWTwoL, is defined by first selecting three leptons,
where at least one same-flavour--opposite-charge (SFOS) pair is required to be consistent with coming from a $Z$ boson decay
with an invariant mass within a window of $[71, 111]$~\GeV.
If more than one such pair is present in the event, the pair which has an invariant mass closest to the $Z$ boson mass is chosen.
The purity of $\ttz$ events is further increased by requiring at least three jets.
To reduce the diboson background in this region, events with exactly one $b$-jet and three jets are rejected.
 
Due to the presence of three leptons in this region,
the background contribution from misidentified or non-prompt leptons
becomes non-negligible and is estimated using a data-driven matrix method (MM) as described in Refs.~\cite{TOPQ-2010-01,TOPQ-2011-01}.
Two types of lepton identification criteria, `tight' and `loose' are defined in
the evaluation, corresponding to the baseline and signal lepton selections described in Section~\ref{sec:obj}.
The number of events containing misidentified or non-prompt leptons in the $\ttz$ CR
is estimated from the number of observed events with tight or loose leptons
using as input the probability for loose prompt, misidentified or non-prompt leptons
to satisfy the tight criteria.
The probability for prompt loose leptons to pass the tight selection is determined from $\ttz$ MC simulation.
The equivalent probability for loose misidentified or non-prompt leptons
to pass the tight selection is measured in a \ttbar-enriched region with two same-sign leptons (electrons or muons)
and a least one $b$-tagged jet, which is dominated by events
with at least one misidentified or non-prompt lepton.
 
In the \CRttZtWTwoL\ region, to mimic the event topology of the $\ttz$ background in the signal region,
a corrected \ptmiss\ is obtained by vectorially adding the transverse momenta of the SFOS pair,
and it is subsequently used to calculate a transverse mass (\mtw) with
the third lepton, referred to as corrected \mttwo.
The two leptons from the SFOS pair are excluded in the calculation of \minmbl,
which effectively becomes the invariant mass  of the third lepton and the leading $b$-jet.
To improve the estimation of the dominant background from the $WZ$ process in the \CRttZtWTwoL,
a dedicated $WZ$ CR, \CRWZtWTwoL, is defined by inverting the \CRttZtWTwoL\ selection requirements
on the jet multiplicity and the corrected \minmbl.
This CR is also used to aid in the estimation of all diboson processes in the SR.
Normalisation factors $\mu^{}_{\ttbar}$, $\mu^{}_{\ttbar V}$ and $\mu^{}_\mathrm{Diboson}$
are found to be $1.00\pm 0.03$, $0.76\pm 0.26$ and $0.80\pm0.16$ after the background-only fit
for the \ttbar, $\ttbar V$ and diboson processes, respectively.

A validation region, \VRtttWTwoL, is defined in order to validate the \ttbar\ background predictions
by applying all the signal selection criteria, apart from requiring lower values of \mttwo, as shown in Table~\ref{tab:regions_tW2L}.
For the background predictions of the $\ttbar V$ and diboson processes,
a $3\ell$ validation region, \VRThreeLTwoL, is defined with selection requirements similar to those of
the \CRttZtWTwoL\ and \CRWZtWTwoL.
To ensure that the \VRThreeLTwoL\ is orthogonal to those two CRs,
the selection on the corrected \minmbl\ variable is varied according to the jet and
$b$-jet multiplicities.
For the events with exactly one $b$-jet, the corrected \minmbl\ is required to be
larger than 170~\GeV\ if $\njet>3$, or smaller than 170~\GeV\ if $\njet\leq3$.
For the events with more than one $b$-jet and $\njet>2$, the corrected \minmbl\ is
required to be larger than 170~\GeV.
To increase the sample size in this region, the \pt\ threshold for the
$b$-tagged jets is reduced to 40~\GeV.

\Fig{\ref{fig:tW2L_VRs}} shows the post-fit kinematic distributions in the validation regions.
Good agreement is observed between data and the SM expectation in all validation regions.
The observed yield, post-fit background estimates and significance~\cite{SignificanceCalc}
in each CR and VR are shown in \Fig{\ref{fig:tW2L:pulls-CRVR}} after the background-only fit.
The data event yields are found to be consistent with background expectations.

% End of text imported from the .//sections/ana_dilep.tex input file
% The next lines are included from the .//sections/ana_tj.tex input file
\subsection{Single-lepton \tjOneL analysis channel}
\label{sec:tj}

Events with exactly one electron or muon are first selected for the SR if they also contain 1--4 jets with $\pt>30$~\GeV,
one or two of which must be $b$-tagged, and satisfy the preselection requirements described in Table~\ref{tab:preselection}.
The fourth jet in the event, if present, is required to have $\pt<50$~\GeV.
The second $b$-tagged jet is required to have $\pt>30$~\GeV.
The dominant SM background contributions in this channel are from
$t\bar{t}$, $W$+jets, and single top ($Wt$ channel) production.
Discriminating variables, \etmiss, \mtw, \njetforward\ and \DeltaPhiLB\ as described in \Sect{\ref{sec:variables}},
are used to define the signal region as shown in Table~\ref{tab:regions_tj1L}.
 
To further improve the sensitivity, a boosted decision tree (BDT), provided by the Toolkit for Multivariate Analysis (TMVA)~\cite{TMVA},
is trained to distinguish between signal and background processes, using events passing the preselection defined in Table~\ref{tab:preselection}.
BDT training settings found to be optimal for this analysis include number of trees
set to 1500 with a maximum depth of 5 and gradient boosting. Cross-validation is performed to ensure there is no over-training.
The following nine kinematic variables defined in \Sect{\ref{sec:variables}} are used as input:
\begin{itemize}
\item $\pt$ and $\eta$ of the highest-$\pt$ jet: $\pt(j_1)$ and $\eta(j_1)$.
\item The transverse masses: $\mtw$ and $\mtLBMET$.
\item $\eta^{\ell b}$ of the leading lepton and $b$-jet system.
\item The invariant mass of, and angular distances between, the highest-$\pt$ lepton and $b$-jet: $\mLB$, $\DeltaPhiLB$,
and $\DeltaRLB$.
\item The azimuthal separation between the highest-$\pt$ lepton and missing transverse momentum, $\DeltaPhiLMET$.
\end{itemize}
To explore the kinematic features in different regions of the signal parameter space,
samples with different signal model parameters are used as an ensemble in the training.
A binned distribution of the BDT output score above 0.6 is then used
to extract the final results in the signal regions.
The binning of the distribution is optimised as [0.6, 0.75], [0.75, 0.85], [0.85, 0.9] and [0.9, 1.0],
referred to as Bins [0--3].
 
The acceptance times detector efficiency after applying all selection criteria for $t$-channel
production in the signal models is between 0.37\% (0.36\%) and 0.73\% (0.67\%) in the parameter
space of $m_a=250$~\GeV, $\tan\beta= 0.3~(0.5)$ and $m_{H^{\pm}}\in [500, 1750]$~\GeV.
 
\begin{table}[h!]
\caption{Summary of signal, control and validation region definitions used in the \tjOneL\ analysis channel.
The `--' entries represent an inclusive selection with no requirements.
}
\label{tab:regions_tj1L}
\small
\begin{center}
\begin{tabular}{p{2.5cm}|c|cc|cc}
\toprule
Variable        &   SR          &  CR($\mathrm{\ttbar}$) &   CR($\mathrm{W}$)      & VR($\mathrm{\ttbar}$) &  VR($\mathrm{W}$) \\
\midrule
\njet                   &  $\in[1,4]$   & $\in[1,4]$    &   $\in[1,2]$   & $\in[1,4]$   &  $\in[1,2]$ \\
\nbjet                  &  $\in[1,2]$   &  $=2$         &    $=1$        & $\in[1,2]$   &   $=1$ \\
\njetforward            &  $>0$         &  --           &     --         &   --         &   -- \\
\midrule
\met            [\GeV]  & $>225$        &  $>225$       & $>225$         &  $>225$      &   $>225$ \\
\mtw            [\GeV]  & $>100$        &  $>100$       & $\in[60,100]$  &  $>100$      & $\in[60,100]$ \\
\midrule
$\DeltaPhiLB$   [rad]   &  $<1.2$       &  $>1.5$       &  $>1.5$        &  $<1.5$      &   $<1.5$ \\
BDT score               &  $>0.6$       &  incl.        &  incl.         &  $<0.5$      &   incl. \\
\bottomrule
\end{tabular}
\end{center}
\end{table}
 
Similarly to the \tWOneL\ analysis channel, dominant backgrounds
from the $\ttbar$ and $W$+jets processes
are estimated using MC simulation and dedicated CRs.
The contribution from multijet production is found to be negligible.
The remaining sources of background (single-top, $Z$+jets, diboson, $t\bar{t}V$, \tth, \tWZ\ production and rarer processes such as
triboson, \tttt, and \ttWW) are estimated from simulation.
 
Dedicated control regions \CRtttjOneL and \CRWtjOneL\ are designed
to estimate the $t\bar{t}$ and $W$+jets background processes, respectively, as shown in \Tab{\ref{tab:regions_tj1L}}.
Compared to the SR, the acceptance for $t\bar{t}$ events is increased in \CRtttjOneL\ by requiring exactly
two $b$-jets and large \DeltaPhiLB\ values.
The contribution from $W$+jets events in the \CRWtjOneL\ is enhanced
by selecting events with one or two jets, exactly one $b$-jet, and low \mtw\ and large \DeltaPhiLB\ values. No splitting based on the $W$ boson charge is applied.
The normalisation factors $\mu^{}_{\ttbar}$\ and $\mu^{}_{W+\mathrm{jets}}$
are found to be $1.00\pm 0.27$ and $1.10\pm0.13$ for the $t\bar{t}$ and $W+$jets processes, respectively.
 
\begin{figure}
\begin{center}
\begin{subfigure}[b]{0.32\textwidth}
\centering\includegraphics[width=\textwidth]{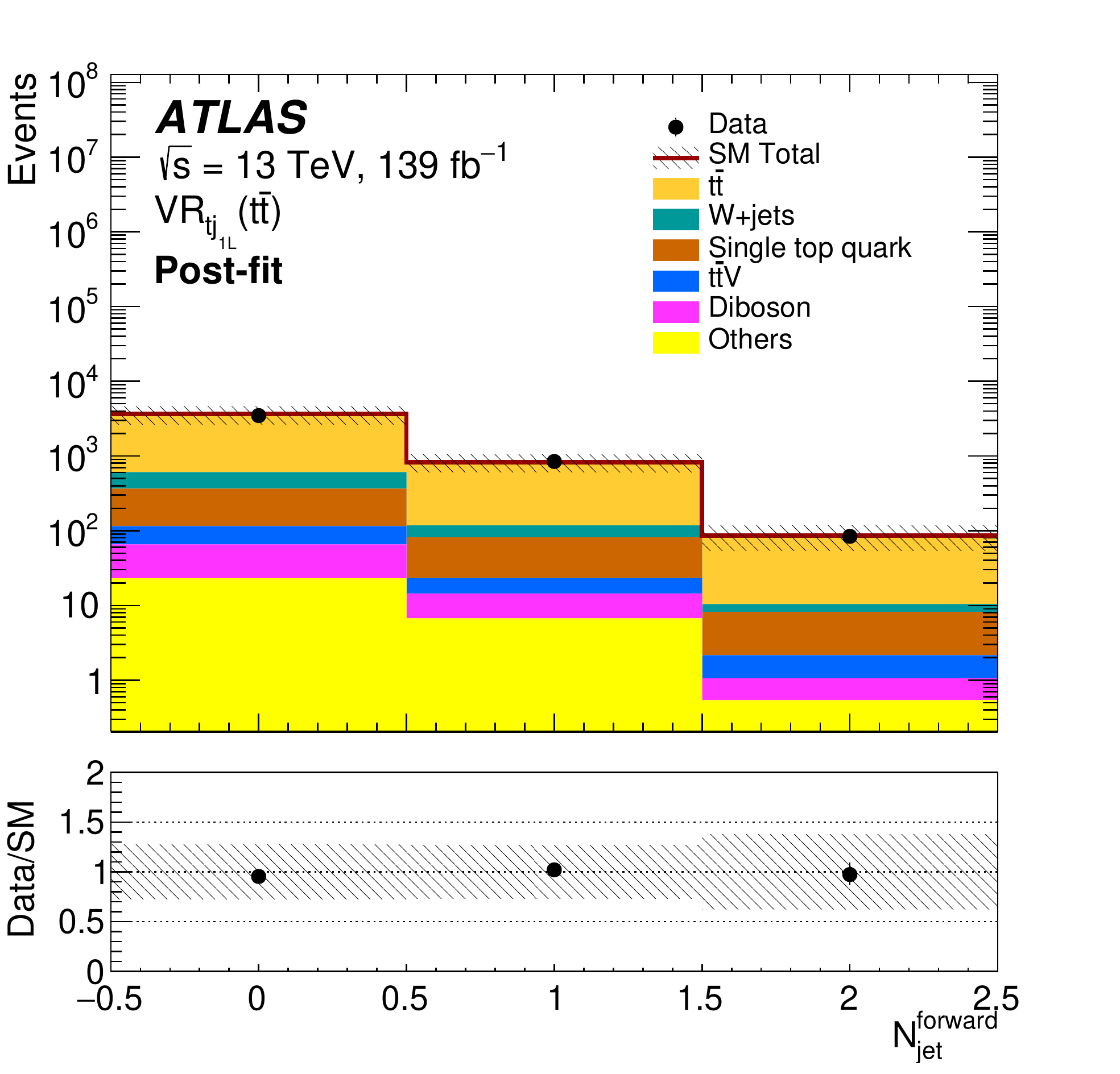}
\caption{\label{fig:tj1L_ttbarVR_nGoodForward}}
\end{subfigure}
\begin{subfigure}[b]{0.32\textwidth}
\centering\includegraphics[width=\textwidth]{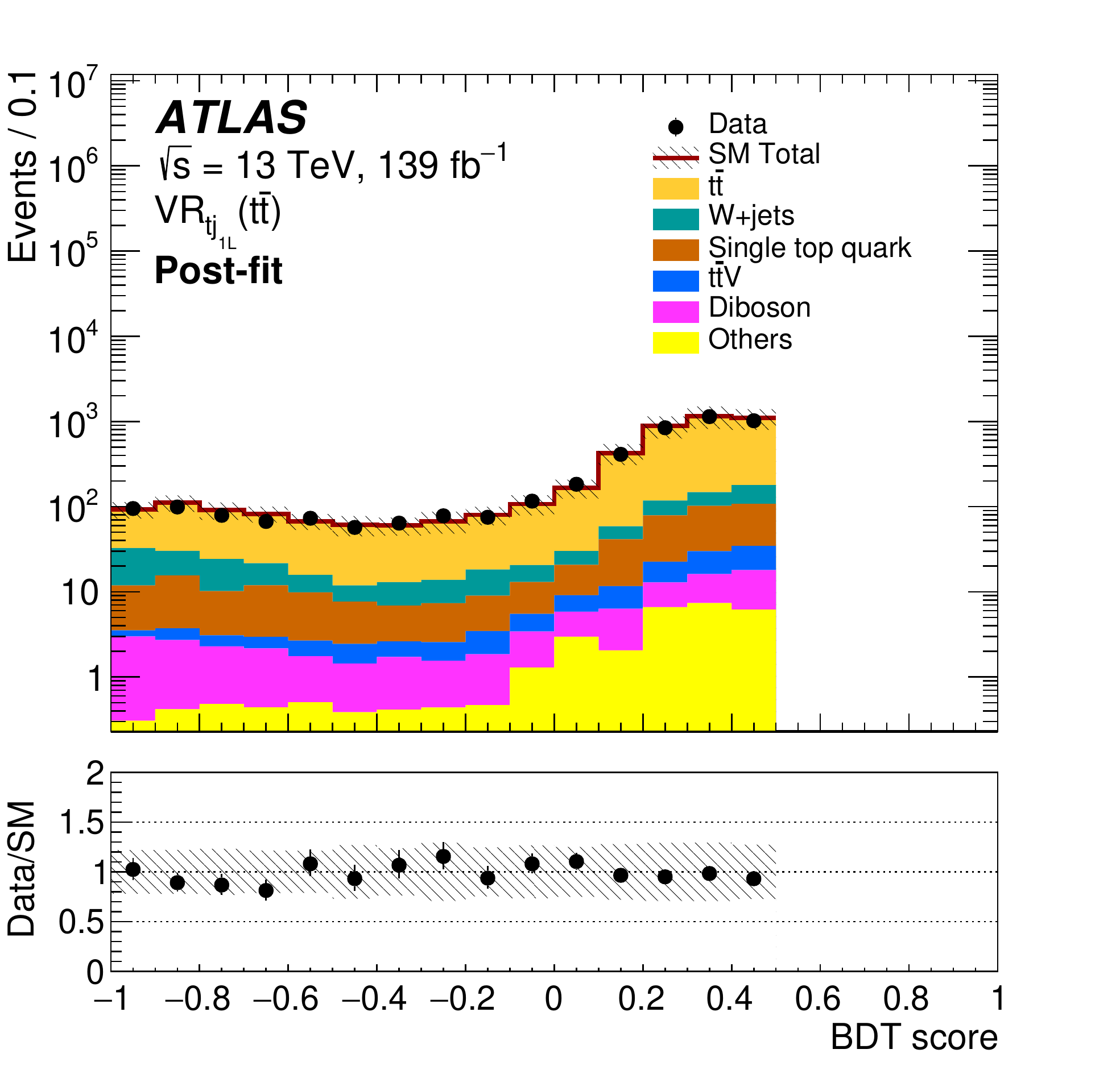}
\caption{\label{fig:tj1L_ttbarVR_BDT}}
\end{subfigure}
\begin{subfigure}[b]{0.32\textwidth}
\centering\includegraphics[width=\textwidth]{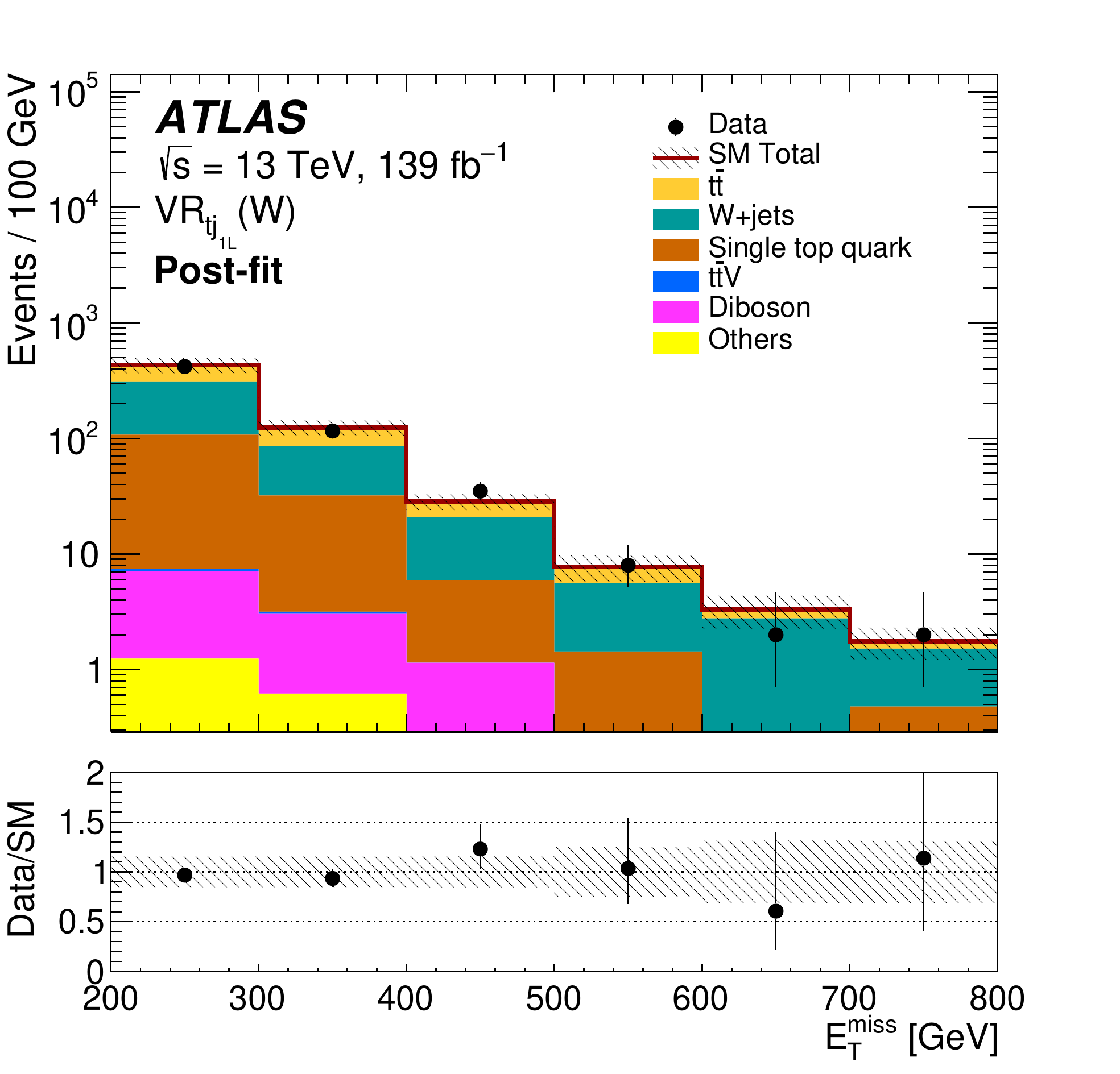}
\caption{\label{fig:tj1L_wjetsVR_MET}}
\end{subfigure}
\end{center}
\caption{Distributions in the \ttbar\ and $W+$jets validation regions in the \tjOneL\ analysis channel:
\subref{fig:tj1L_ttbarVR_nGoodForward} \njetforward\ in \VRtttjOneL,
\subref{fig:tj1L_ttbarVR_BDT} BDT score in \VRtttjOneL
and \subref{fig:tj1L_wjetsVR_MET} \met\ in \VRWtjOneL.
The normalisation of the backgrounds is obtained from the background-only fit to the CRs.
The uncertainty bands plotted include all statistical and systematic uncertainties.
The `Others' category includes contributions from $Z$+jets and \tWZ\ production, and
rare processes such as triboson, \tttt, \ttWW, and Higgs boson production processes.
The overflow events, where present, are included in the last bin.
The lower panels show the ratio of data to the background prediction.
The hatched error bands indicate the combined experimental and MC statistical uncertainties on these background predictions.
}
\label{fig:tj1L:VRs}
\end{figure}

\begin{figure}[!ht]
\centering
\includegraphics[width=0.90\textwidth]{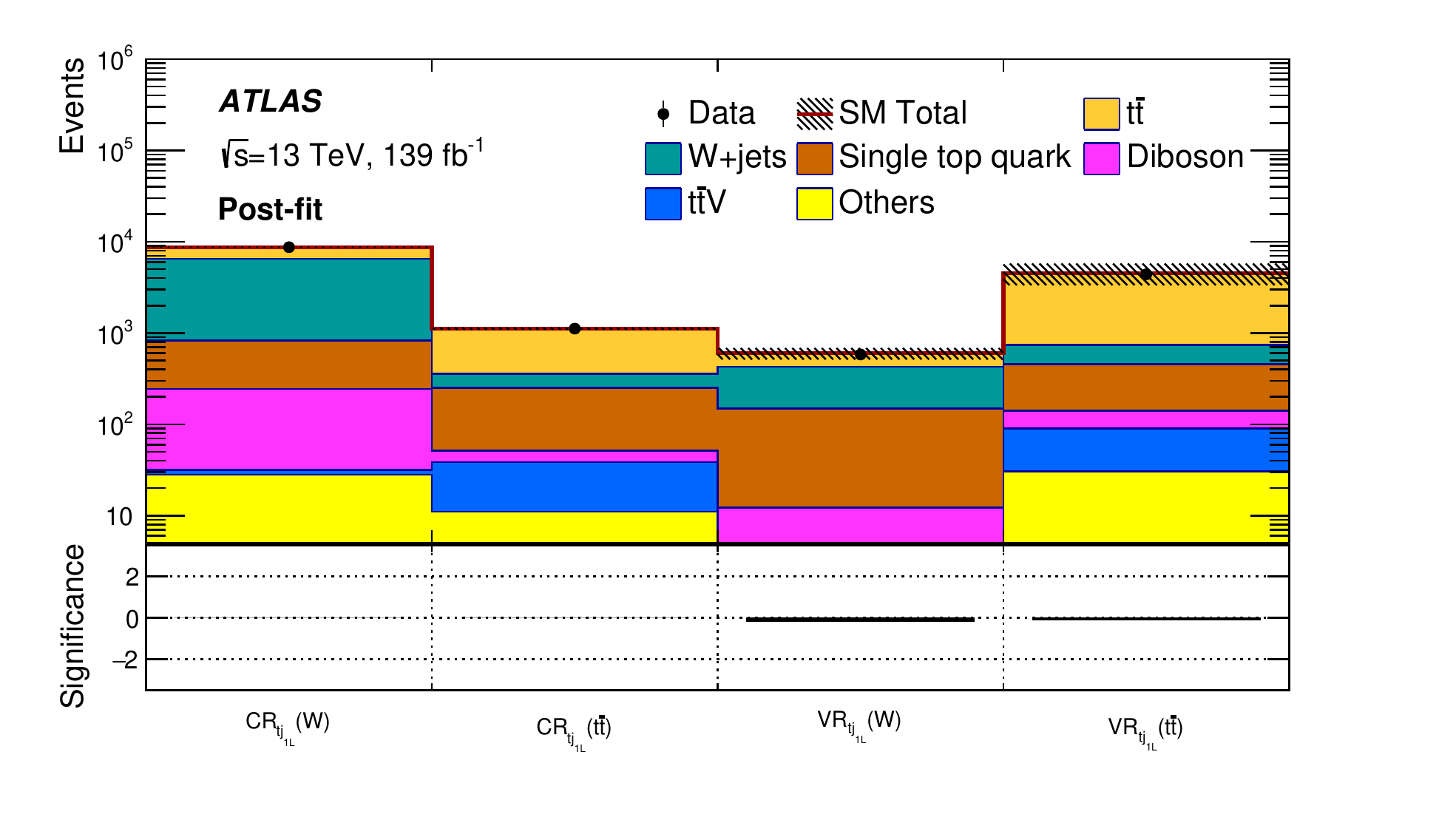}
\caption{
Comparison of the predicted backgrounds with the observed numbers of events in the
CRs and VRs associated with the \tjOneL\ channel.
The normalisation of the backgrounds is obtained from the background-only fit to the CRs.
The upper panel shows the observed number of events and the predicted background yield.
The `Others' category includes contributions from $Z$+jets and \tWZ\ production, and
rare processes such as triboson, \tttt, \ttWW, and Higgs boson production processes.
All uncertainties are included in the uncertainty band.
The lower panel shows the significance for each region.
}
\label{fig:tj1L:pulls-VR}
\end{figure}

To validate the \ttbar\ background predictions, a validation region \VRtttjOneL\ is
defined by requiring a BDT score that is lower than in the SR definition,
as shown in Table~\ref{tab:regions_tj1L}.
For the $W$+jets background, a validation region \VRWtjOneL\ is defined by
requiring a lower \mtw\ value than in the SR definition, as
shown in Table~\ref{tab:regions_tj1L}.
To ensure orthogonality to the corresponding CRs, events
in these two VRs are required to have low $\DeltaPhiLB$.
\Fig{\ref{fig:tj1L:VRs}} shows the post-fit distribution of representative
kinematic variables and the BDT score for these two validation regions.
Good agreement is observed between data and expectation in all validation regions.
The observed yield, post-fit background estimates and significance~\cite{SignificanceCalc}
in each CR and VR are shown in \Fig{\ref{fig:tj1L:pulls-VR}} after the background-only fit.
The data event yields are found to be consistent with background expectations.

% End of text imported from the .//sections/ana_tj.tex input file
 
\clearpage
\section{Systematic uncertainties}
\label{sec:sys}
\begin{figure}[!b]
\begin{center}
\includegraphics[width=0.85\textwidth]{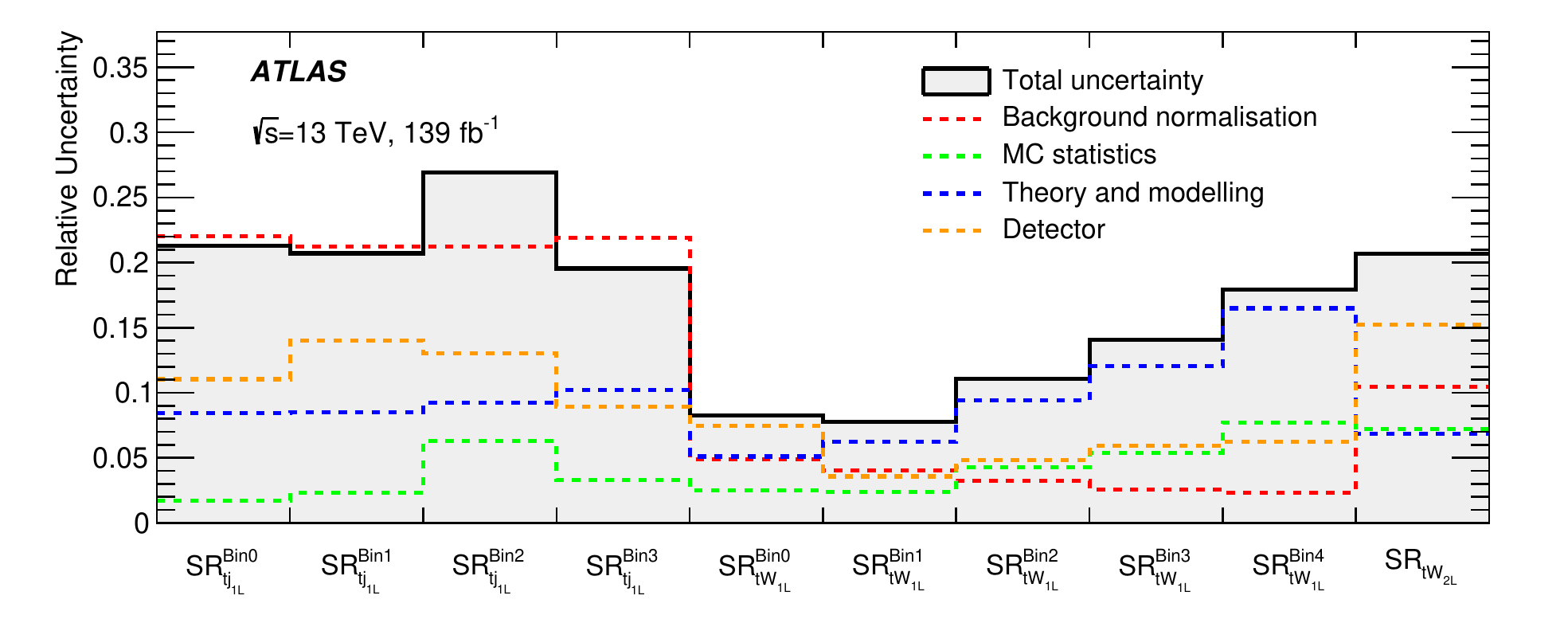}
\end{center}
\caption{Relative uncertainties in the total background yield in each SR for the three analysis channels, including the contribution
from the different sources of uncertainty. The `Detector' category contains all detector-related systematic uncertainties
and is dominated by the jet energy scale and resolution. The `Background normalisation' represents the uncertainty in the
fitted normalisation factors, including the available event counts in the CRs.
Individual uncertainties can be correlated, and do not necessarily add up in quadrature to the total background uncertainty.}
\label{fig:SRsyst}
\end{figure}

% The next lines are included from the .//sections/systematics.tex input file
Several sources of experimental and theoretical systematic uncertainty in the signal and background estimates are considered.
Their impact is reduced through the normalisation of the dominant backgrounds in the control regions defined with kinematic
selections resembling those of the corresponding signal region. Uncertainties are included as nuisance parameters,
common across all regions, with Gaussian constraints in the
likelihood fits, taking into account correlations between different regions.
Uncertainties due to the numbers of events in the CRs are also included in the fit for each region.
The magnitude of the contributions arising from uncertainties on the background normalisation factors $\mu$
and on the detector, theoretical modelling and statistics of the MC samples
are summarised in Figure~\ref{fig:SRsyst} as a relative uncertainty in the total background yield
for each SR in the three analysis channels.
 
Dominant detector-related systematic uncertainties arise from the jet energy scale and resolution,
and from the $b$-tagging efficiency and mis-tagging rates.
 
The uncertainties in the jet energy scale and resolution are based on their respective
measurements in data~\cite{PERF-2016-04} and are derived as a function of the \pT\ and $\eta$ of the jet, as
well as of the pile-up conditions and the jet flavour composition (light-quark, $b$-quark, or gluon-initiated jets)
of the selected jet sample.
Their contributions to the SRs are the dominant experimental uncertainty components and
are almost equivalent in all analysis channels.
The systematic uncertainty in the $b$-tagging efficiency is the second largest experimental uncertainty.
It ranges from 4.5\% for \bjets with $\pt \in [35, 40]$~\GeV\ up to 7.5\% for \bjets with high \pt ($> 100$~\GeV).
The $b$-tagging uncertainty is estimated by varying  the $\eta$-, \pt- and flavour-dependent scale factors applied
to each jet in the simulation within a range that reflects the systematic uncertainty in the measured tagging
efficiency and mis-tag rates in data ~\cite{Aad:2019aic}.
The uncertainties associated with trigger requirements, pile-up modelling, and lepton reconstruction and energy measurements have a small or negligible impact
on the final results; however, the lepton, photon and jet-related uncertainties are propagated to
the calculation of the \etmiss, and additional uncertainties due to the energy scale and resolution
of the soft term are included in the \etmiss. Finally, uncertainties in estimates of the non-prompt
or misidentified leptons background are found to be below 1\% in the \tWTwoL analysis channel and negligible for
single-lepton selections.
 
Uncertainties in the modelling of the SM background processes in MC simulation and their theoretical
cross-section uncertainties are also taken into account. Furthermore, for these processes the 1.7\% uncertainty in the combined 2015--2018 integrated luminosity is included.
 
Modelling uncertainties in the \ttbar and single-top backgrounds are dominant in all SRs for the \tWOneL and \tjOneL analysis channels,
and the second leading source of uncertainty for the \tWTwoL SR. They are computed as the difference between the predictions
from nominal samples and those from additional samples differing in hard-scattering generator
and parameter settings, or by using internal weights assigned to the events
depending on the choice of renormalisation and factorisation scales ($\mu_\mathrm{R}$ and $\mu_\mathrm{F}$, respectively, varied independently
by factors of 2 and 0.5), initial- and final-state radiation parameters, and PDF sets.
The impact of the PS and hadronisation model is evaluated by comparing the nominal generator
with a \powbox\ sample interfaced to \HERWIGV{7}~\cite{Bahr:2008pv,Bellm:2015jjp}, using the H7UE set of tuned
parameters~\cite{Bellm:2015jjp}. To assess the uncertainty due to the choice of hard-scattering generator
and matching scheme, an alternative generator set-up using \MGaMC\ interfaced to \pythia~8 is employed.
For single-top $Wt$ production, the impact of interference between single-resonant and
double-resonant top quark production and on the implementation of the $W$ lineshape in the
generator is estimated in all analysis channels by comparing the nominal sample generated using
the diagram removal method with alternative samples, including those generated using the diagram
subtraction method~\cite{Frixione_2008}.
For the \tWTwoL selection, this results in a 100\% uncertainty in the subdominant $Wt$ contribution.
 
For the \ttbar+$W/Z$ background, uncertainties due to parton shower and hadronisation modelling
are evaluated by comparing the predictions from \MGaMC\ interfaced to \pythia8 and \HERWIGV{7},
while the uncertainties related to the choice of renormalisation and factorisation scales are
assessed by varying the corresponding event generator parameters up and down by a factor of two
around their nominal values.
Their contribution is dominant in the \tWTwoL analysis channel
and subdominant or small in all other SRs.
A similar approach is used to assess the uncertainties in the \tWZ\ process,
with an additional 20\% uncertainty assigned to account for uncertainties in the effects of
interference between the \ttbar+$W/Z$ and \tWZ\ processes. The 20\% is assigned on the basis of preliminary
comparisons of alternative approaches developed to evaluate interference effects in the
$\ttbar$--$tW$~\cite{Demartin:2016axk} and \ttz--\tWZ\ processes~\cite{Bylund:2016qau}.
 
Finally, modelling and normalisation uncertainties in minor backgrounds are also considered. For diboson and $W/Z$+jets events, they are
estimated by varying the renormalisation, factorisation and resummation scales up and down by a factor of two
around the values used to generate the nominal samples.
For \ttWW, $tZ$, $\ttbar H$, $Wh$, $Zh$, \tttt, and triboson production processes, experimental and theoretical
uncertainties in the event yields are also evaluated and found to be negligible.
 
For the DM signal processes, both the experimental and theoretical uncertainties in the expected signal yields are considered, including the aforementioned luminosity uncertainty.
Experimental uncertainties are found to be 3--35\% (2.5--11\%) across the $m_a$--$m_{H^{\pm}}$ and $m_a$--$\tan\beta$
planes for the \tWOneL (\tWTwoL) analysis channel, and in the range 3--14\% as a function of $m_{H^{\pm}}$ for the \tjOneL
selection, independently of $\tan\beta$. In all SRs, the dominant uncertainty in the
signal yields is found to be from the jet energy scale and resolution, followed by uncertainties in $b$-tagging rates.
Larger uncertainties for the \tWOneL selections are found for the highest \met-binned region, where MC statistical fluctuations are also relevant.
In the modelling of the signal samples, uncertainties due to the variations of the renormalisation and factorisation scales are dominant.
They are evaluated using a variation scheme wherein $\mu_\mathrm{R}$ and $\mu_\mathrm{F}$ are scaled simultaneously by either a factor of 2 or 0.5.
For the PS and hadronisation uncertainties, alternative samples with varied A14 tune parameter values are used.
The effect of each systematic variation on the acceptance and efficiency is evaluated for each analysis channel SR
by comparing the variation samples with the corresponding nominal sample.
The impact on the total yields for $tW$+DM, $\ttbar$+DM and $t$-channel production processes is also evaluated for each
signal scenario and found to be between 5\% and 15\%.
For the \tWOneL and \tWTwoL analysis channels, the uncertainties vary between 5\% and 30\% across the $m_a$--$m_{H^{\pm}}$ and $m_a$--$\tan\beta$  planes,
with the largest values obtained for samples characterised by low values of the ${H^{\pm}}$ mass and
independently of $\tan\beta$. For the \tjOneL analysis channel, uncertainties are found to be between 15\% and 5\% as a
function of increasing $m_{H^{\pm}}$ for all $\tan\beta$ values considered.
 
% End of text imported from the .//sections/systematics.tex input file

\section{Results}
\label{sec:result}
\begin{figure}[!b]
\begin{center}
\includegraphics[width=0.85\textwidth]{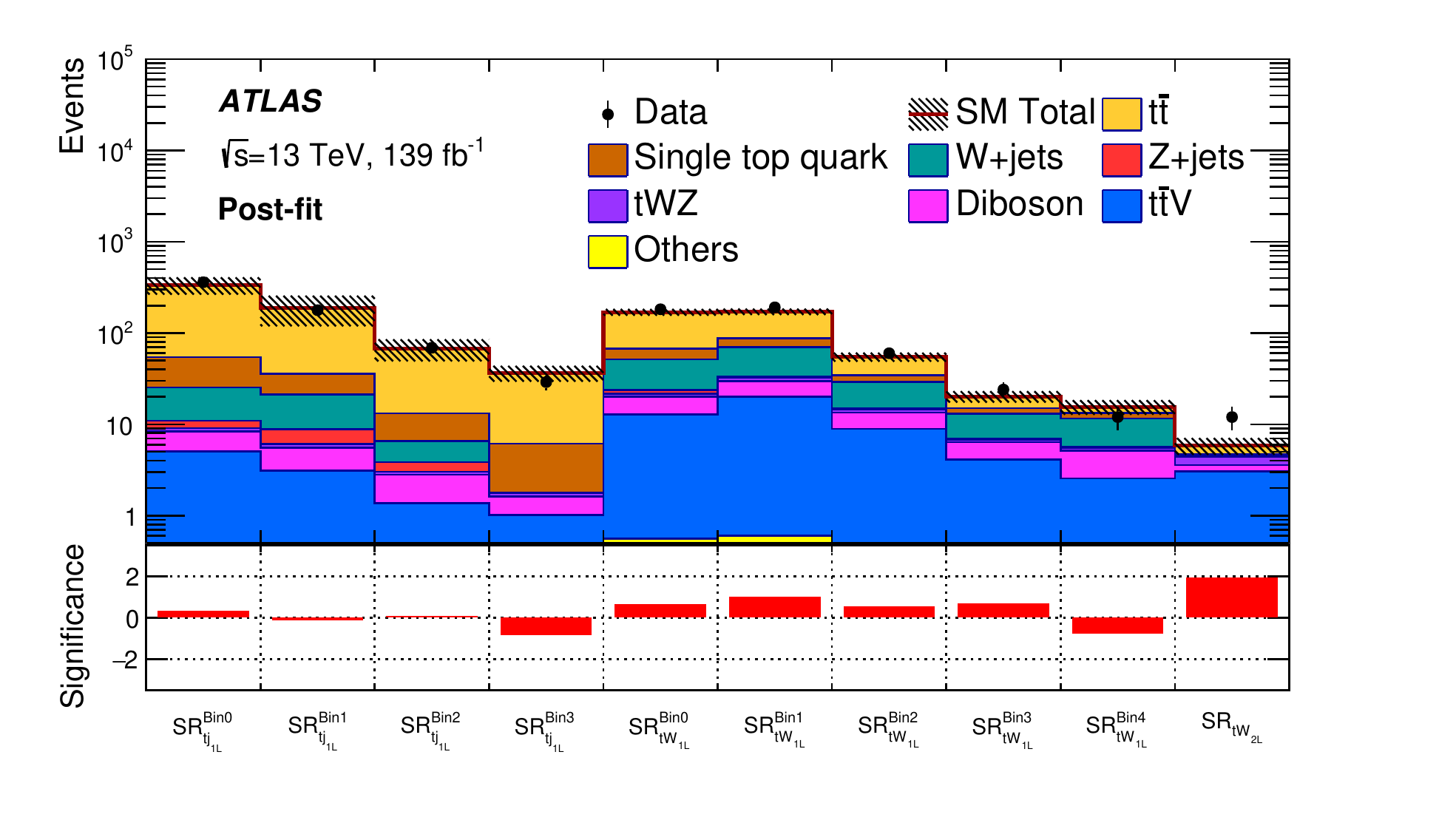}
\end{center}
\caption{Results of the background-only fit extrapolated to all SRs. The normalisation of the backgrounds is obtained from the fit to the CRs.
The upper panel shows the observed number of events and the predicted background yields.
The `Others' category includes contributions from rare processes such as triboson, \tttt, \ttWW, and Higgs boson production processes.
All uncertainties defined in Section~\ref{sec:sys} are included in the uncertainty band.
The lower panel shows the significance in each SR. The significance calculation is performed as described in Ref.~\cite{SignificanceCalc}.}
\label{fig:SRpulls}
\end{figure}

% The next lines are included from the .//sections/results.tex input file
The event yields for all SRs in the three analysis channels are reported in Tables~\ref{tab:fit_results1} and~\ref{tab:fit_results2}
and are summarised in Figure~\ref{fig:SRpulls}, where the significance for each of the SRs is also presented.
The SM background expectations resulting from background-only fits are shown along with their statistical
plus systematic uncertainties.
No significant deviations from the expected yields are observed in any of the signal regions considered.
The largest background contribution in the \tWOneL\ and \tjOneL\ analysis channel SRs arises from \ttbar\ production,
whilst the contribution from \ttv\ is largest in the \tWTwoL\ SRs, with subdominant contributions from
the \ttbar, single-top (including \tWZ) and diboson processes. Other non-negligible
background sources are $W$+jets and $Z$+jets production.
 
% The next lines are included from the .//tables/tWyields.tex input file
\begin{table}[h!]
\caption{Background-only fit results for the \tWOneL and \tWTwoL signal regions.
The backgrounds which contribute only a small amount (rare processes such as triboson, \ttbar\ttbar, \ttbar$WW$ and Higgs boson production processes, and non-prompt or misidentified leptons background) are grouped and labelled as `Others'.
The quoted uncertainties of the fitted SM background include both the statistical and systematic uncertainties. }
\label{tab:fit_results1}
\begin{center}
\setlength{\tabcolsep}{0.0pc}
{\small
\begin{tabular*}{\textwidth}{@{\extracolsep{\fill}}lcccccc}
\toprule
& SR$^{\mathrm{Bin 0}}_{\tWOneL}$  & SR$^{\mathrm{Bin 1}}_{\tWOneL}$  & SR$^{\mathrm{Bin 2}}_{\tWOneL}$ & SR$^{\mathrm{Bin 3}}_{\tWOneL}$   & SR$^{\mathrm{Bin 4}}_{\tWOneL}$ & SR$^{}_{\tWTwoL}$      \\[-0.05cm]
\midrule
Observed events          & $182$              & $191$              & $60$              & $24$              & $12$         & $12$           \\
\midrule
Fitted SM bkg events         & $169 \pm 14$~~          & $171 \pm 13$~~          & $55 \pm 6$~~          & $20.1 \pm 2.8$~~          & $15.6 \pm 2.8$~~      & $5.9 \pm 1.2$        \\
\midrule
\ttbar         & $101 \pm 12$~~          & $84 \pm 12$          & $20 \pm 5$~~          & $5.1 \pm 1.7$          & $2.3 \pm 1.5$           & $1.2 \pm 0.9$   \\
Single top         & $16.3 \pm 5.2$~~          & $17.3 \pm 5.2$~~          & $5.4 \pm 3.2$          & $2.0 \pm 1.8$          & $1.7_{-1.7}^{+2.0}$\,~ & $0.26_{-0.26}^{+0.27}$~~              \\
$W$+jets         & $28 \pm 4$~~          & $37.0 \pm 4.3$~~          & $14.2 \pm 2.4$~~          & $6 \pm 1$          & $5.9 \pm 1.1$   & --           \\
$Z$+jets         & $2.0 \pm 0.9$          & $1.1 \pm 0.7$          & $0.3 \pm 0.1$          & $0.15 \pm 0.04$          & $0.15 \pm 0.02$     & --         \\
Diboson         & $7.2 \pm 1.7$          & $9.6 \pm 2.0$          & $4.6 \pm 1.0$          & $2.2 \pm 0.5$          & $2.7 \pm 0.6$   & $0.5 \pm 0.2$           \\
\ttbar $V$          & $12.3 \pm 1.4$~~          & $19.5 \pm 3.5$~~          & $8.7 \pm 1.2$          & $4.0 \pm 0.7$          & $2.5 \pm 0.5$    & $2.9 \pm 0.7$           \\
$tWZ$         & $1.7 \pm 0.2$          & $2.4 \pm 0.5$          & $1.17 \pm 0.15$          & $0.42 \pm 0.09$          & $0.39 \pm 0.09$   & $0.8 \pm 0.1$           \\
Others         & $0.6 \pm 0.1$          & $0.6 \pm 0.1$          & $0.17 \pm 0.02$          & $0.06 \pm 0.02$          & $0.03 \pm 0.01$   & $0.16 \pm 0.08$           \\
\bottomrule
\end{tabular*}
}
\end{center}
\end{table}
% End of text imported from the .//tables/tWyields.tex input file
% The next lines are included from the .//tables/tjyields.tex input file
\begin{table}[h!]
\caption{
Background-only fit results for the \tjOneL signal regions.
The backgrounds which contribute only a small amount ($Z$+jets, rare processes such as $tWZ$, triboson, \ttbar\ttbar, \ttbar$WW$ and Higgs boson production processes) are grouped and labelled as `Others'. The quoted uncertainties of the fitted SM background include both the statistical and systematic uncertainties. }
\label{tab:fit_results2}
\begin{center}
\setlength{\tabcolsep}{0.0pc}
{\small
\begin{tabular*}{\textwidth}{@{\extracolsep{\fill}}lcccc}
\toprule
& SR$^{\mathrm{Bin 0}}_{\tjOneL}$             & SR$^{\mathrm{Bin 1}}_{\tjOneL}$ & SR$^{\mathrm{Bin 2}}_{\tjOneL}$  & SR$^{\mathrm{Bin 3}}_{\tjOneL}$              \\[-0.05cm]
\midrule
Observed events          & $360$              & $178$              & $69$              & $29$                    \\
\midrule
Fitted SM bkg events         & $335 \pm 74$~~          & $187 \pm 40$~~          & $67 \pm 18$          & $37 \pm 7$~~              \\
\midrule
$t\bar{t}$         & $280 \pm 75$~~          & $151 \pm 42$~~          & $54 \pm 16$          & $30 \pm 8$~~              \\
$W$+jets        & $14.4 \pm 5.1$~~          & $12.4 \pm 8.7$~~          & $2.7_{-2.7}^{+7.5}$~          & $<$0.1              \\
Single top         & $27 \pm 14$          & $13.2 \pm 7.5$~~          & $5.7 \pm 3.7$          & $3.8 \pm 2.5$              \\
$t\bar{t}V$         & $5.0 \pm 1.6$          & $3.1 \pm 1.0$          & $1.4 \pm 0.6$          & $1.3 \pm 0.5$              \\
Diboson         & $3.3 \pm 1.0$          & $2.5 \pm 0.8$          & $1.4 \pm 0.5$          & $0.8 \pm 0.3$              \\
Others         & $4.9 \pm 2.4$          & $4.8 \pm 2.0$          & $2.0 \pm 0.7$          & $1.0 \pm 0.4$              \\
 
\bottomrule
\end{tabular*}
}
\end{center}
\end{table}
% End of text imported from the .//tables/tjyields.tex input file
 
\begin{figure}[hp!]
\begin{center}
\begin{subfigure}[b]{0.31\textwidth}
\centering\includegraphics[width=\textwidth]{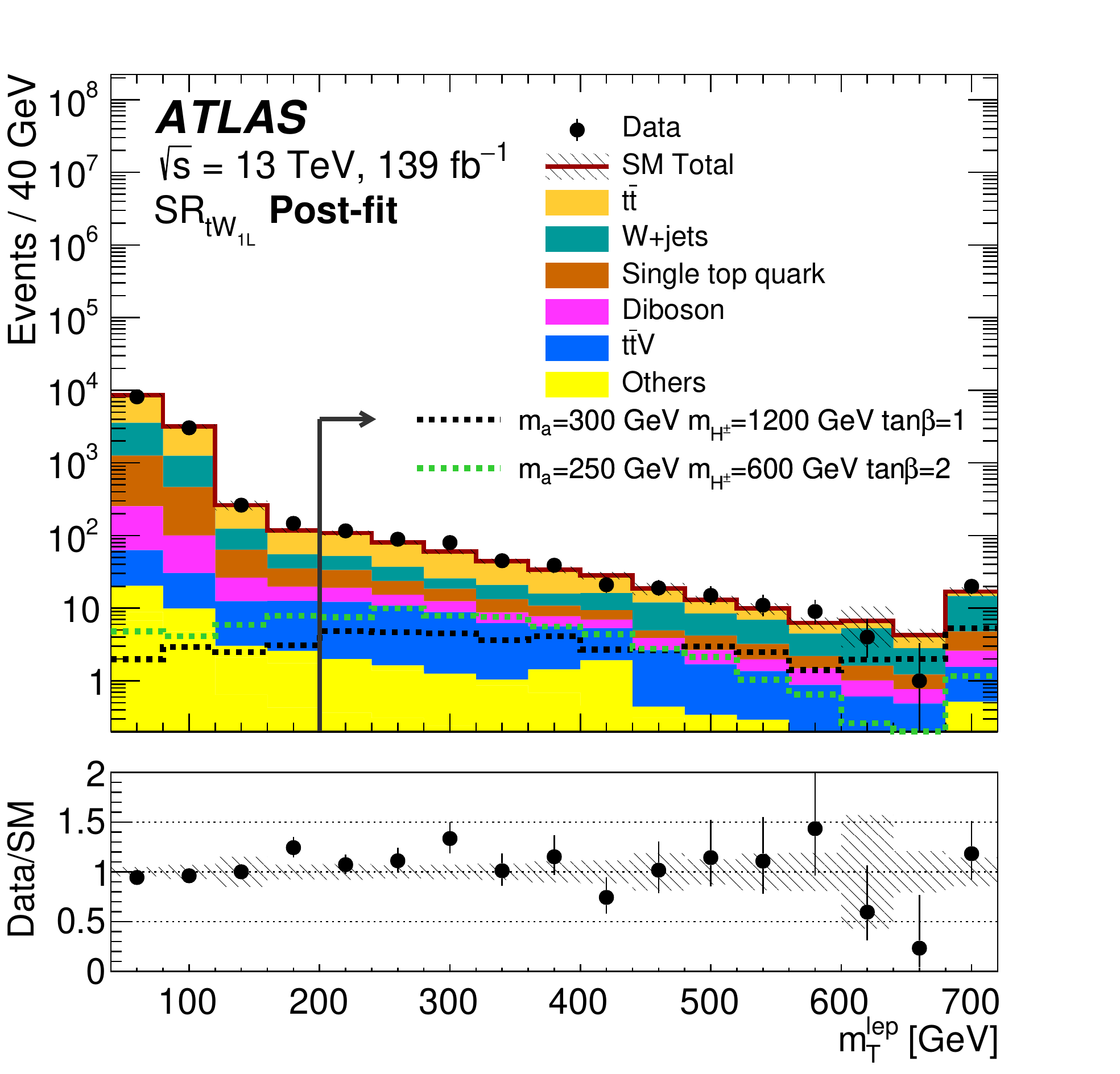}
\caption{\label{fig:tW1L_Nm1_mt}}
\end{subfigure}
\begin{subfigure}[b]{0.31\textwidth}
\centering\includegraphics[width=\textwidth]{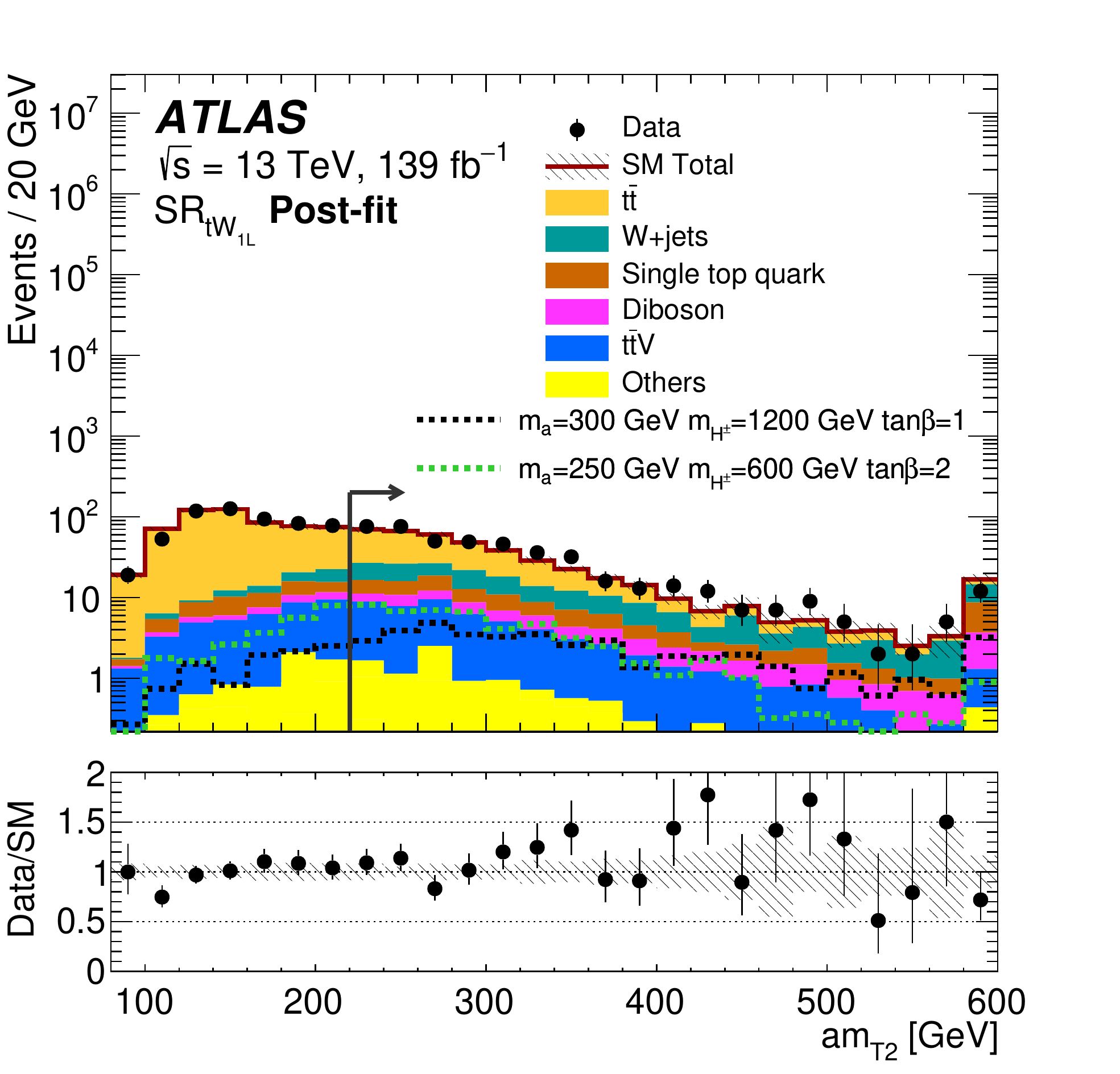}
\caption{\label{fig:tW1L_Nm1_amt2}}
\end{subfigure}
\begin{subfigure}[b]{0.31\textwidth}
\centering\includegraphics[width=\textwidth]{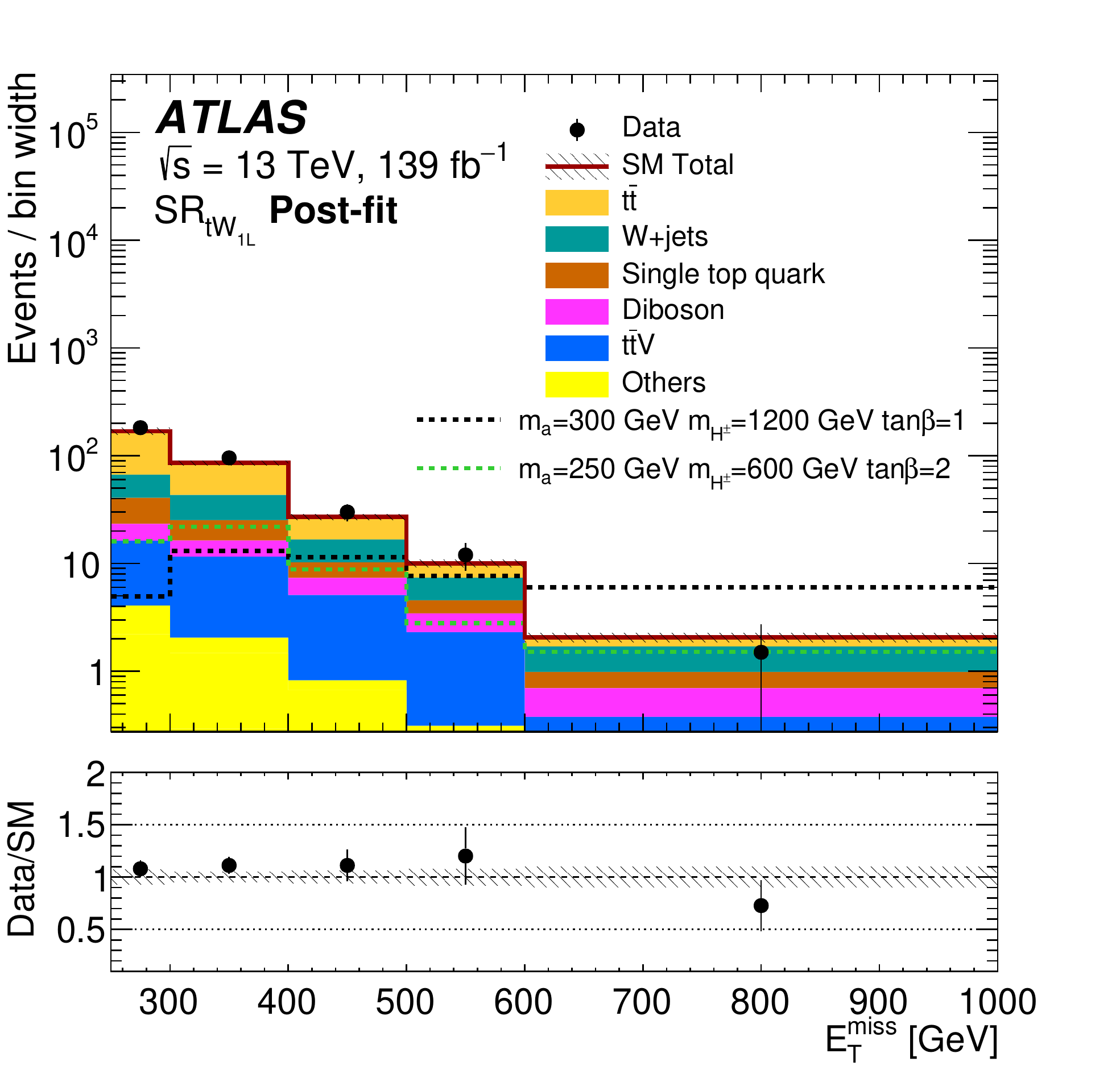}
\caption{\label{fig:tW1L_Nm1_met}}
\end{subfigure}
\end{center}
\caption{
Representative distributions of \subref{fig:tW1L_Nm1_mt}~\mtw, \subref{fig:tW1L_Nm1_amt2}~\amt\ and \subref{fig:tW1L_Nm1_met}~\met
in the \tWOneL\ channel. Observed data are compared with
the SM background predictions extrapolated from the background-only fit.
All SR selections except the one on the quantity shown are applied.  The SR requirement is indicated by the arrow.
The `Others' category includes contributions from $Z$+jets and \tWZ\ production, and
rare processes such as triboson, \tttt, \ttWW, and Higgs boson production processes.
The expected distributions for representative scenarios with different $m_a$,  $m_{H^{\pm}}$, and $\tan\beta$ are
shown for illustrative purposes.
The overflow events, where present, are included in the last bin.
The lower panels show the ratio of data to the background prediction.
The hatched error bands indicate the combined experimental and MC statistical uncertainties on these background predictions.
}
\label{fig:tW1L_SROpt_Nminus1}
\end{figure}

\begin{figure}[hp!]
\begin{center}
\begin{subfigure}[b]{0.31\textwidth}
\centering\includegraphics[width=\textwidth]{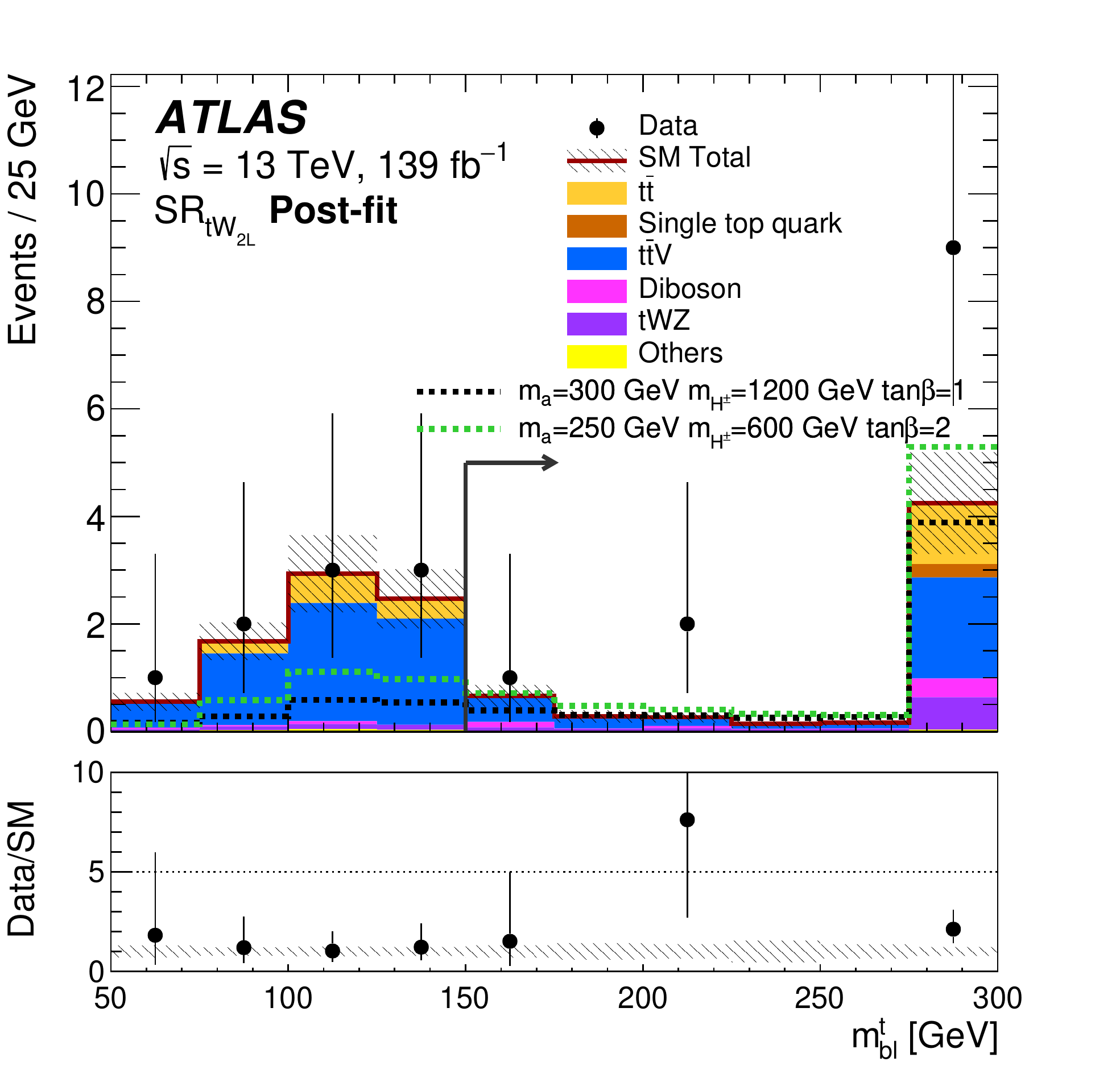}
\caption{\label{fig:tW2L_Nm1_minmax}}
\end{subfigure}
\begin{subfigure}[b]{0.31\textwidth}
\centering\includegraphics[width=\textwidth]{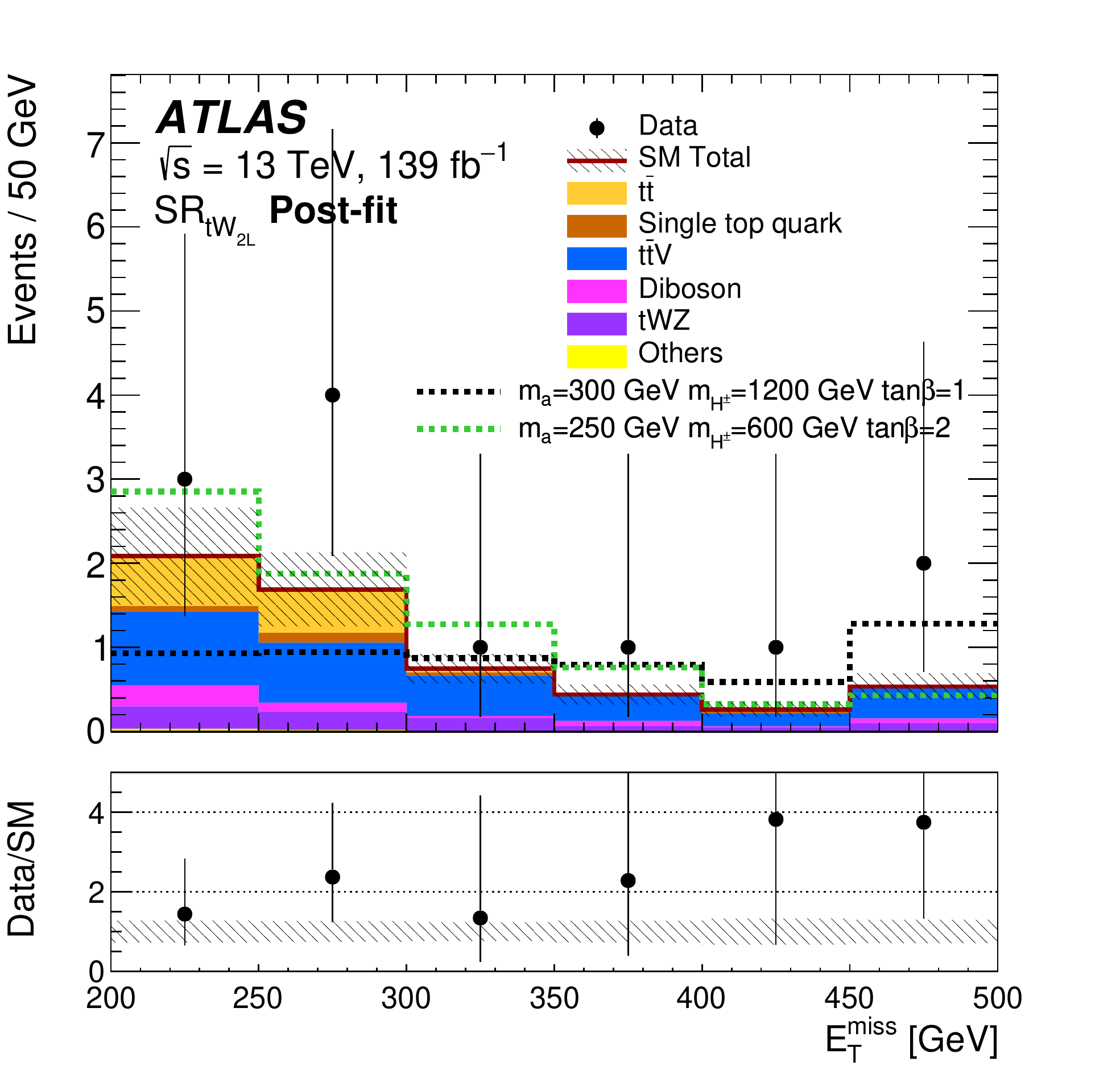}
\caption{\label{fig:tW2L_Nm1_met}}
\end{subfigure}
\begin{subfigure}[b]{0.31\textwidth}
\centering\includegraphics[width=\textwidth]{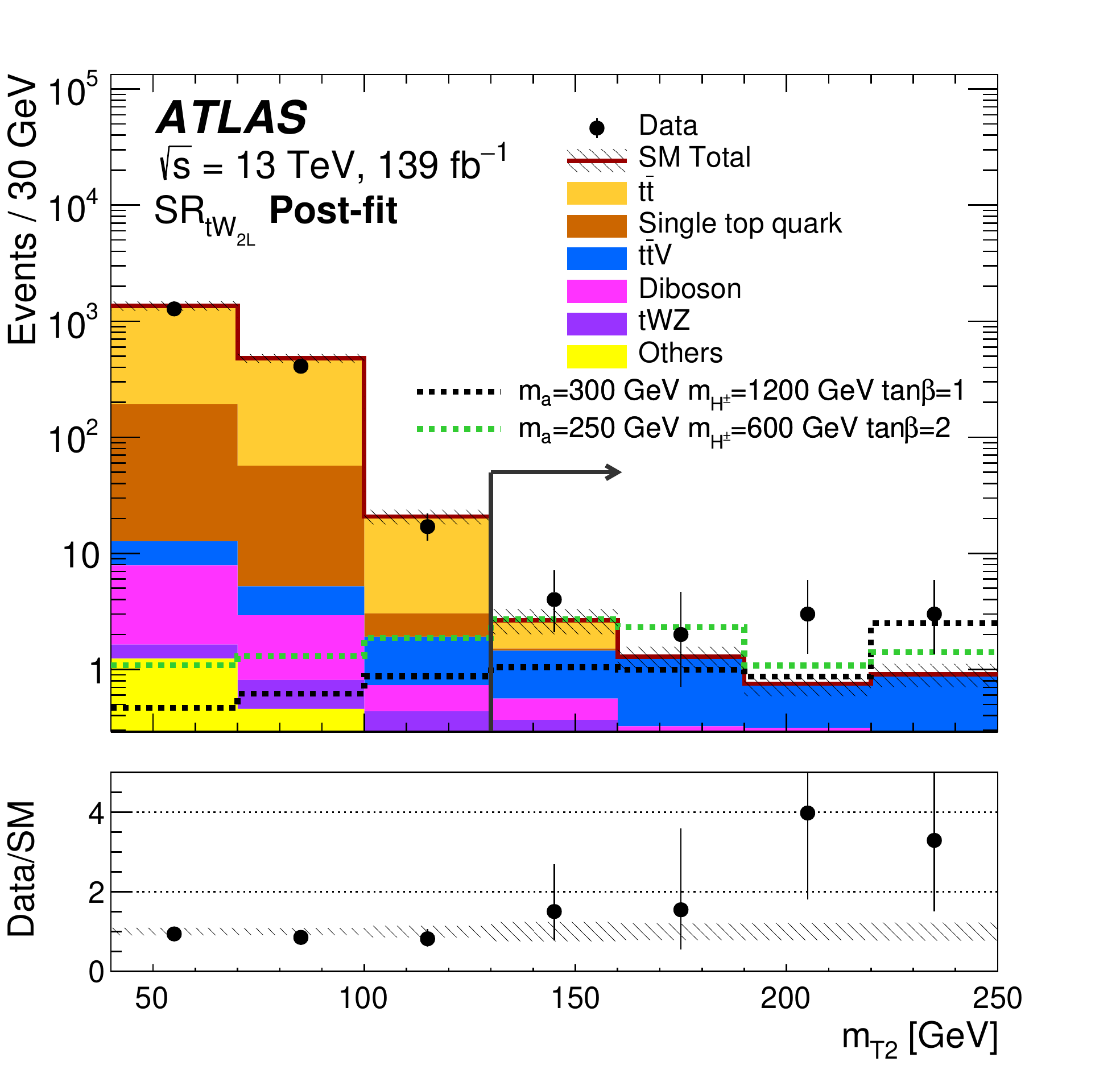}
\caption{\label{fig:tW2L_Nm1_mt2}}
\end{subfigure}
\end{center}
\caption{Representative distributions of \subref{fig:tW2L_Nm1_minmax}~\mblt, \subref{fig:tW2L_Nm1_met}~\met\ and \subref{fig:tW2L_Nm1_mt2}~\mttwo\
in the \tWTwoL\ channel.
Observed data are compared with the SM background predictions extrapolated from the background-only fit.
All SR selections except the one on the quantity shown are applied. The SR requirement is indicated by the arrow.
As the \mblt\ is defined for events with at least two jets, the events with exactly one jet are included in the overflow bin.
The `Others' category includes contributions from rare processes such as triboson, \tttt, \ttWW, and Higgs boson production processes.
The expected distributions for representative scenarios with different $m_a$,  $m_{H^{\pm}}$, and $\tan\beta$ are
shown for illustrative purposes.
The overflow events, where present, are included in the last bin.
The lower panels show the ratio of data to the background prediction.
The hatched error bands indicate the combined experimental and MC statistical uncertainties on these background predictions.
}
\label{fig:tW2L_SROpt_Nminus1}
\end{figure}

\begin{figure}[hp!]
\begin{center}
\begin{subfigure}[b]{0.31\textwidth}
\centering\includegraphics[width=\textwidth]{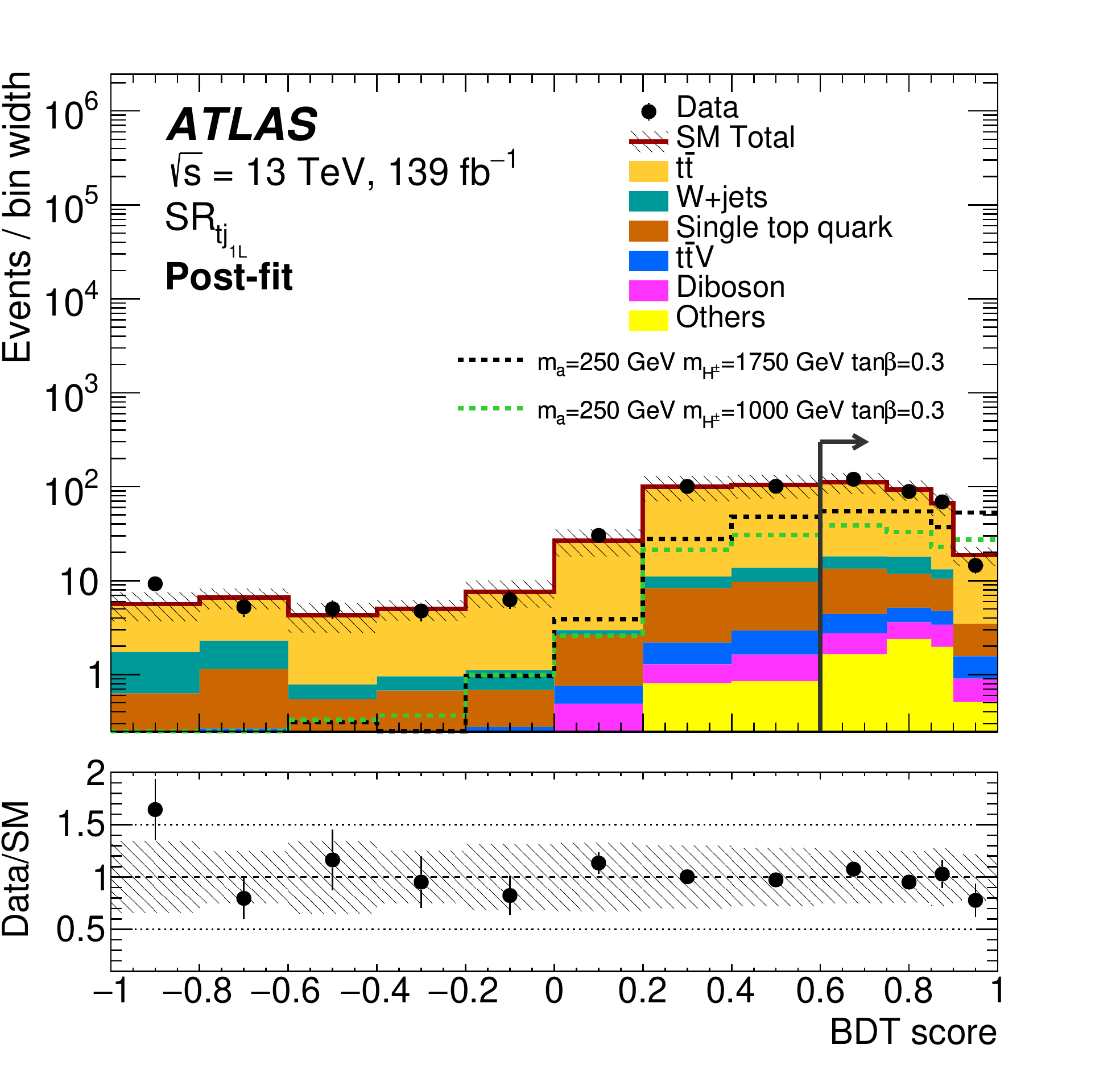}
\caption{\label{fig:tj1L_Nm1_BDT}}
\end{subfigure}
\begin{subfigure}[b]{0.31\textwidth}
\centering\includegraphics[width=\textwidth]{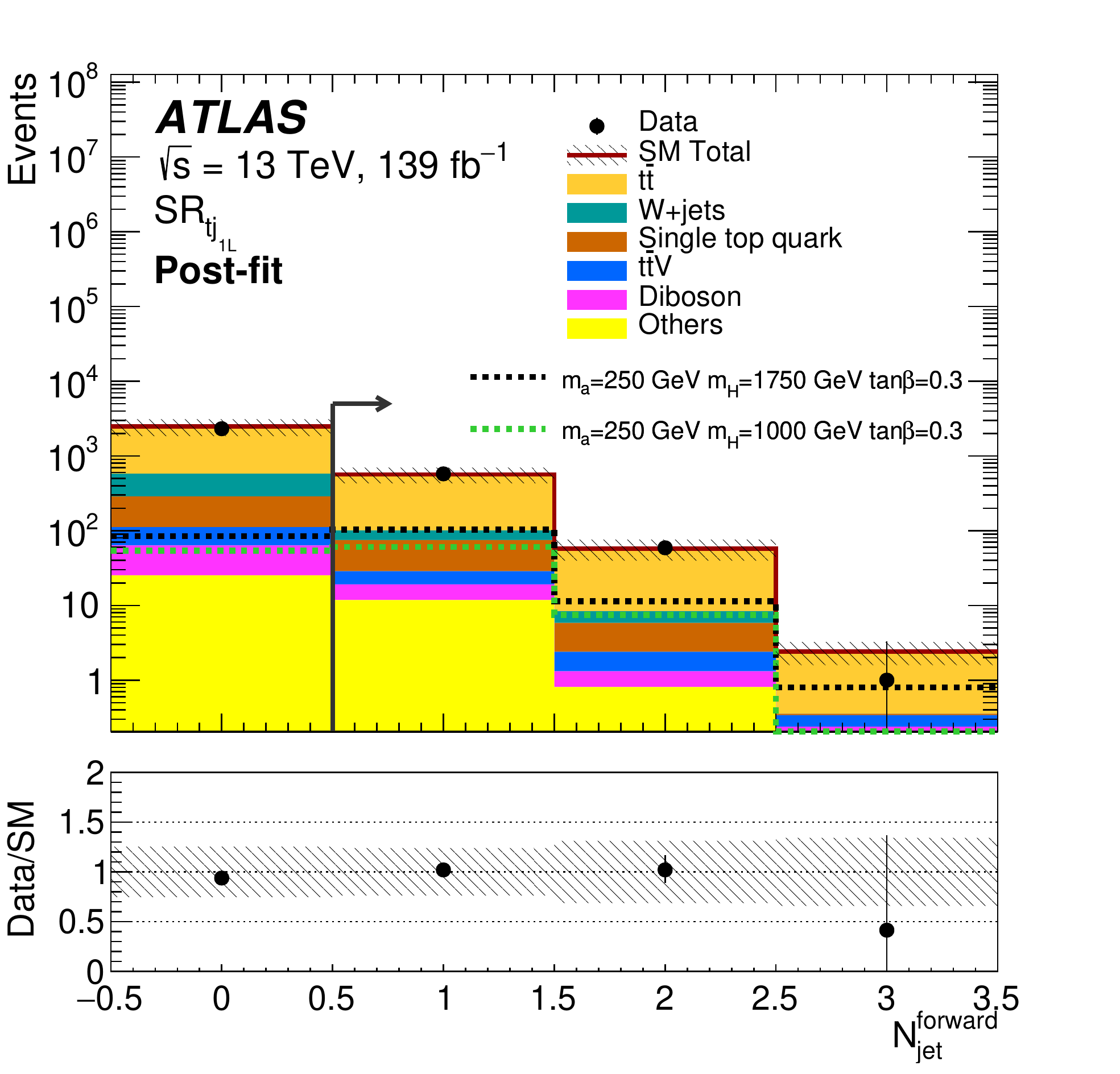}
\caption{\label{fig:tj1L_Nm1_Nj}}
\end{subfigure}
\begin{subfigure}[b]{0.31\textwidth}
\centering\includegraphics[width=\textwidth]{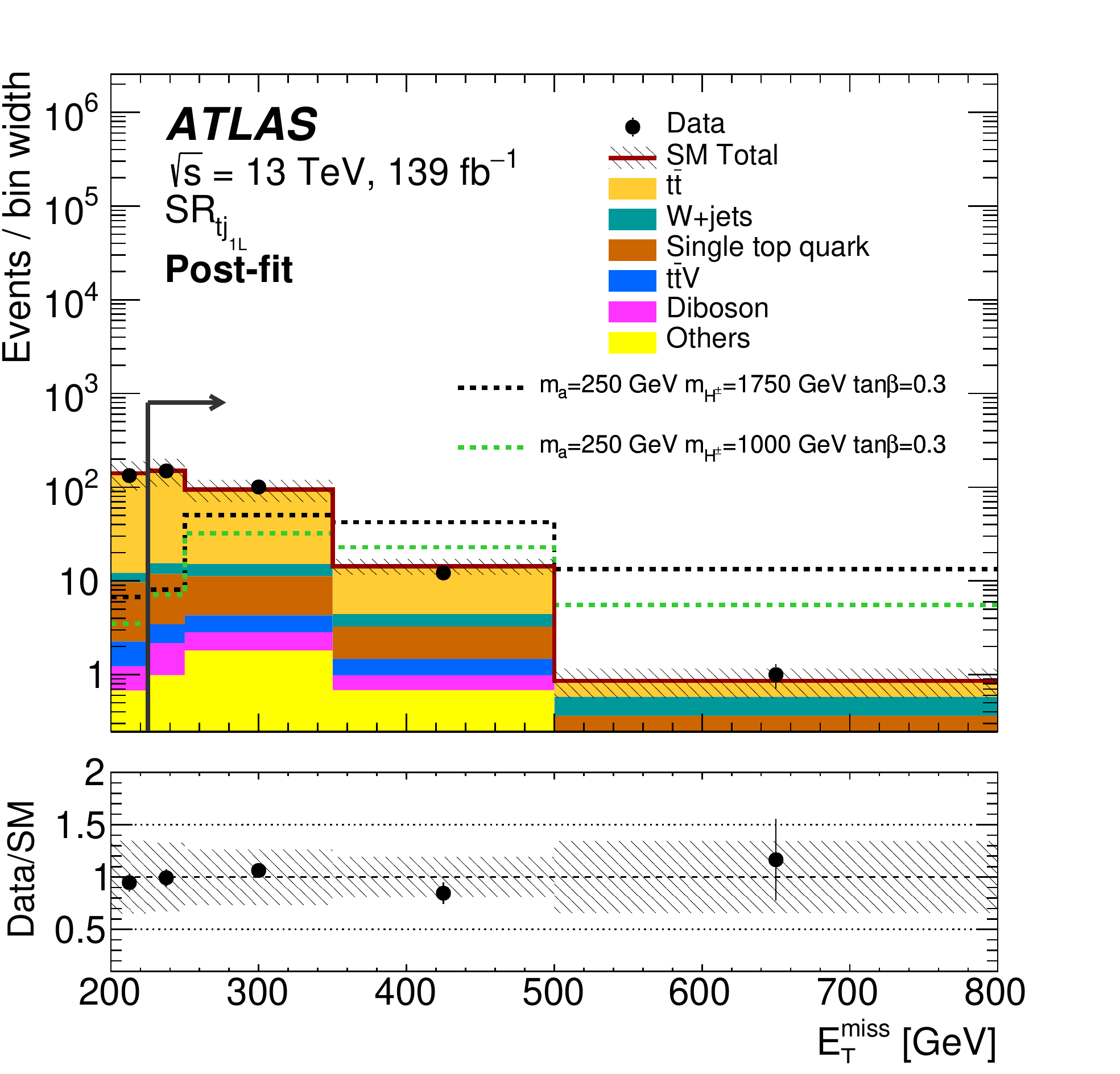}
\caption{\label{fig:tj1L_Nm1_met}}
\end{subfigure}
\end{center}
\caption{
Representative distributions of \subref{fig:tj1L_Nm1_BDT} BDT score, \subref{fig:tj1L_Nm1_Nj}~\njetforward\ and
\subref{fig:tj1L_Nm1_met}~\met in the \tjOneL\ channel.
Observed data are compared with the SM background predictions extrapolated from the background-only fit.
All SR selections except the one on the quantity shown are applied. The SR requirement is indicated by the arrow.
The `Others' category includes contributions from $Z$+jets and \tWZ\ production, and
rare processes such as triboson, \tttt, \ttWW, and Higgs boson production processes.
The expected distributions for representative scenarios with different $m_a$,  $m_{H^{\pm}}$, and $\tan\beta$ are
shown for illustrative purposes.
The overflow events, where present, are included in the last bin.
The lower panels show the ratio of data to the background prediction.
The hatched error bands indicate the combined experimental and MC statistical uncertainties on these background predictions.
}
\label{fig:tj1L_SROpt_Nminus1}
\end{figure}

Figures~\ref{fig:tW1L_SROpt_Nminus1} to~\ref{fig:tj1L_SROpt_Nminus1} show comparisons between
the observed data and the post-fit SM predictions for some relevant kinematic distributions in the
three analysis channels after applying all SR selection requirements except the one on the quantity shown
(except for Figures~\ref{fig:tW1L_SROpt_Nminus1}(c) and~\ref{fig:tW2L_SROpt_Nminus1}(b), where all events in the SR are shown).
For the \tWOneL\ analysis channel, the \mtw and \amt\ distributions are shown for all values of \met\ across the five bins.
Similarly, for the \tjOneL\ analysis channel, the number of forward jets and \met\ are shown for all BDT values above 0.6.
The expected distributions for representative scenarios with different $m_a$,  $m_{H^{\pm}}$, and $\tan\beta$
(depending on the analysis channel and SR considered) are shown for illustrative purposes.
Reasonable agreement is found between data and SM predictions in all distributions,
although a mild excess of data events is found in the \tWTwoL\ distributions, accounting for a discrepancy lower than 2$\sigma$
considering statistical and systematic uncertainties.
 
\subsection{Statistical combination of the \tWOneL\ and \tWTwoL\ analysis channels}
A statistical combination of results from the \tWOneL\ and \tWTwoL\ channels is performed to maximise the sensitivity to $tW$+DM models.
The simultaneous fit is performed such that the individual background normalisation factors, $\mu^{1L}_{\ttbar}$, $\mu^{2L}_{\ttbar}$,
$\mu^{}_{W+\mathrm{jets}}$ and $\mu^{}_{\ttbar V}$, are constrained in the same regions as the respective,
individual analyses to avoid extrapolations into a different phase space.
Experimental uncertainties in the background and signal are evaluated using the same methods as described in Section~\ref{sec:analysis}
and correlated across channels.
Modelling uncertainties from the same source for a given process are correlated, e.g.\ all modelling uncertainties for \ttbar
are correlated across the regions. Signal systematic uncertainties are also correlated for the exclusion fits described in the next section.
 
The predictions for SM backgrounds are, as expected, equivalent to those of the individual channels.
In particular, the values for the  $\mu^{}_{\ttbar}$ background normalisation factor are found to be
consistent for \tWOneL\ and \tWTwoL\ estimates, $\mu^{1L}_{\ttbar} = 0.97 \pm 0.08$ and
$\mu^{2L}_{\ttbar} = 1.00 \pm 0.03$, respectively.

\subsection{Model-independent limits}
The CL$_{\mathrm{s}}$ technique~\cite{Read:2002hq} is used to place 95\% confidence level (CL) upper
limits on event yields from physics BSM for each signal region (model-independent limits).
The profile-likelihood-ratio test statistic is used to exclude the signal-plus-background hypotheses for
specific signal models. When normalised to the integrated luminosity of the data sample, results can be interpreted
as corresponding upper limits on the visible cross section, $\sigma^{}_\textrm{vis}$, defined as the product of
the BSM production cross section, the acceptance and the selection efficiency of a BSM signal.
In the case of the \tWOneL\ analysis channel, the \etmiss\ bins are defined inclusively, i.e.\ all events above
the lowest bin-threshold in \etmiss are taken, to retain discovery potential. The SM predictions and their corresponding
uncertainties are reported in Table~\ref{tab:1Ldiscovery}.
In the case of the \tjOneL\ analysis channel, the last bin of the BDT score distribution, 0.9--1.0, is considered.
 
% The next lines are included from the .//tables/discoverytable.tex input file
\begin{table}[!h]
\caption{Background-only fit results for \tWOneL signal region bins for model-independent limits. Events with \etmiss above the lowest bin-threshold are retained in each bin.
This is indicated by value $X$ in SR$^{X}_{\tWOneL}$.
The backgrounds which contribute only a small amount (rare processes such as $tWZ$, triboson, \ttbar\ttbar, \ttbar$WW$ and Higgs boson production processes) are grouped and labelled as `Others'.
The quoted uncertainties of the fitted SM background include both the statistical and systematic uncertainties. }
\label{tab:1Ldiscovery}
\begin{center}
\setlength{\tabcolsep}{0.0pc}
{\small
 
\begin{tabular*}{\textwidth}{@{\extracolsep{\fill}}lccccc}
\toprule
& SR$^{250}_{\tWOneL}$  & SR$^{300}_{\tWOneL}$  & SR$^{400}_{\tWOneL}$ & SR$^{500}_{\tWOneL}$   & SR$^{600}_{\tWOneL}$      \\[-0.05cm]
\midrule
Observed events          & $469$              & $287$              & $96$              & $36$              & $12$                    \\
\midrule
Fitted SM bkg events     & $431 \pm 27$~~          & $262 \pm 20$~~          & $91 \pm 10$          & $36 \pm 5$~~          & $15.5 \pm 2.8$~~              \\
\midrule
\ttbar         & $213 \pm 25$~~          & $111 \pm 18$~~          & $28 \pm 7$~~          & $7.5 \pm 2.9$          & $2.3 \pm 1.5$              \\
Single top     & $43 \pm 15$          & $27 \pm 12$          & $9 \pm 7$          & $3.9 \pm 3.8$          & $1.7_{-1.7}^{+2.0}$\,~              \\
$W$+jets       & $91 \pm 8$~~          & $63 \pm 5$~~          & $26 \pm 3$~~          & $12.0 \pm 1.7$~~          & $5.9 \pm 1.1$              \\
$Z$+jets       & $3.8 \pm 1.0$          & $1.7 \pm 0.7$          & $0.6 \pm 0.1$          & $0.30 \pm 0.05$          & $0.15 \pm 0.02$              \\
Diboson        & $26 \pm 5$~~          & $19.0 \pm 3.9$~~          & $9 \pm 2$          & $4.8 \pm 1.0$          & $2.6 \pm 0.6$              \\
$\ttbar V$     & $47 \pm 3$~~          & $34.7 \pm 2.4$~~          & $15.3\pm 1.2$~~          & $6.6 \pm 0.5$          & $2.5 \pm 0.3$              \\
Others    & $7.5 \pm 0.5$          & $5.3 \pm 0.4$          & $2.3 \pm 0.1$          & $0.9 \pm 0.1$          & $0.40 \pm 0.05$              \\
\bottomrule
\end{tabular*}
}
\end{center}
\end{table}
 
% End of text imported from the .//tables/discoverytable.tex input file
 
\Tab{\ref{tab:upper}} summarises the observed ($S^{95}_\mathrm{obs}$)
and expected ($S^{95}_\mathrm{exp}$) 95\% CL upper limits on the number of BSM events and on $\sigma^{}_\textrm{vis}$
for all SRs. The $p_0$-values, which represent the probability of the SM background to
fluctuate to the observed number of events or higher, are also provided and are capped at $p_0 = 0.5$; the associated significance
is provided in parentheses.
 
% The next lines are included from the .//tables/tab-ModelInd.tex input file
\begin{table}
\caption{The 95\% CL upper limits on the visible cross section
($\langle\epsilon\sigma\rangle_\mathrm{obs}^{95}$) and on the number of
signal events ($S_\mathrm{obs}^{95}$ ) for all SRs and analysis channels as detailed in the text.
The fourth column ($S_\mathrm{exp}^{95}$) shows the 95\% CL upper limit on the number of
signal events, given the expected number (and $\pm 1\sigma$
exclusions of the expectation) of background events.
The last two columns indicate the $CL_B$ value, i.e.\ the confidence level observed for
the background-only hypothesis, and the discovery $p$-value ($p(s = 0)$). The associated significance is provided in parentheses.
}
\label{tab:upper}
\begin{center}
\setlength{\tabcolsep}{0.0pc}
{\small
\begin{tabular*}{\textwidth}{@{\extracolsep{\fill}}lccccc}
\toprule
{ Signal channel}                        & $\langle\epsilon{ \sigma}\rangle_\mathrm{obs}^{95}$[fb]  &  $S_\mathrm{obs}^{95}$  & $S_\mathrm{exp}^{95}$ & $CL_{B}$ & $p(s=0)$ ($Z$)  \\
\midrule
SR$^{250}_{\tWOneL}$     & $0.72$ &  $100.6$~~ & $ { 67 }^{ +33 }_{ -16 }$ & $0.85$ & $ 0.12$~$(1.16)$ \\[0.1ex]
SR$^{300}_{\tWOneL}$   & $0.51$ &  $70.8$ & $ { 54 }\pm{16}$ & $0.85$ & $ 0.15$~$(1.02)$ \\[0.1ex]
SR$^{400}_{\tWOneL}$   & $0.24$ &  $32.9$ & $ { 29 }^{ +10 }_{ -6 }$ & $0.64$ & $ 0.30$~$(0.52)$ \\[0.1ex]
SR$^{500}_{\tWOneL}$    & $0.14$ &  $18.9$ & $ { 19 }^{ +8 }_{ -5 }$~~ & $0.52$ & $ 0.45$~$(0.13)$ \\[0.1ex]
SR$^{600}_{\tWOneL}$   & $0.08$ &  $10.6$ & $ { 12 }^{ +3 }_{ -4 }$ & $0.24$ & $ 0.94$~$(-1.54)$ \\[0.1ex]
\midrule
SR$^{}_{\tWTwoL}$ & $0.10$ &  $13.8$ & $ { 7.3 }^{ +2.9 }_{ -1.1 }$ & $0.97$ & $ 0.02$~$(2.07)$ \\[0.1ex]
SR$^{}_{\tjOneL}$ (BDT$>$0.9)   & $0.10$ &  $14.4$ & $ { 19 }^{ +6 }_{ -5 }$~~ & $0.24$ & $ 0.50$~$(0.00)$ \\%
\bottomrule
\end{tabular*}}
\end{center}
\end{table}
 
% End of text imported from the .//tables/tab-ModelInd.tex input file
 
\subsection{Model-dependent limits}
Model-dependent exclusion limits are placed on the common signal parameters $m_a$,  $m_{H^{\pm}}$, and $\tan\beta$
in the 2HDM+$a$ models considered in the analysis.
Following the prescriptions in Ref.~\cite{Abe:2018bpo}, the masses of the bosons $H$, $H^\pm$ and $A$ are set to be equal.
The three quartic couplings between the scalar doublets and the $a$ boson ($\lambda_{P1}, \lambda_{P2}$ and $\lambda_3$) are all set equal to $3$,
in order to reduce the number of parameters and evade the constraints from electroweak precision measurements~\cite{Haisch:2016gry}.
To further reduce the parameter space, unitary couplings between the $a$ boson mediator and the DM particle $\chi$ ($g_\chi = 1$) are considered,
with the DM particle mass set to $m_\chi=10\;\GeV$.
The mixing angle $\theta$ is fixed at $\sin\theta=1/\sqrt{2}$, yielding full mixing between the $a$ and $A$ bosons
and the largest cross sections for the processes of interest.
Two sets of samples are considered,\footnote{As pointed out in Ref.~\cite{Abe:2018bpo}, for the two considered sets of samples, only values
$m_{H^{\pm}} < 600$~\GeV\ provide a bounded-from-below scalar potential~\cite{Gunion:2002zf} for the 2HDM+$a$ model.
This constraint can be relaxed up to a factor 2 if the quartic coupling $\lambda_3$ assumes a value closer
to the perturbativity limit and they can be relaxed further in more general 2HDMs containing
additional quartic couplings~\cite{Bauer:2017fsw}, as discussed in Ref.~\cite{Haisch:2018djm}.}
varying either the $(m_a, m_{H^{\pm}})$ parameters and setting $\tan\beta$ to unity,
or varying the $(m_{H^{\pm}}, \tan\beta)$ parameters and setting $m_a=250$~\GeV.
The fit procedure takes into account correlations in the yield predictions between
control and signal regions due to common background normalisation parameters and systematic uncertainties.
The experimental systematic uncertainties in the signal are taken into account for the calculation
and are assumed to be fully correlated with those in the SM background.
The results of the combined fit for the \tWOneL and \tWTwoL\ channels are interpreted using the
sum of the respective signal yield estimates for each generated sample, with overlap between the samples removed
according to the procedure illustrated in Ref.~\cite{EXOT-2017-32}.
 
\begin{figure}[h!]
\begin{center}
\begin{subfigure}[b]{0.75\textwidth}
\centering\includegraphics[width=\textwidth]{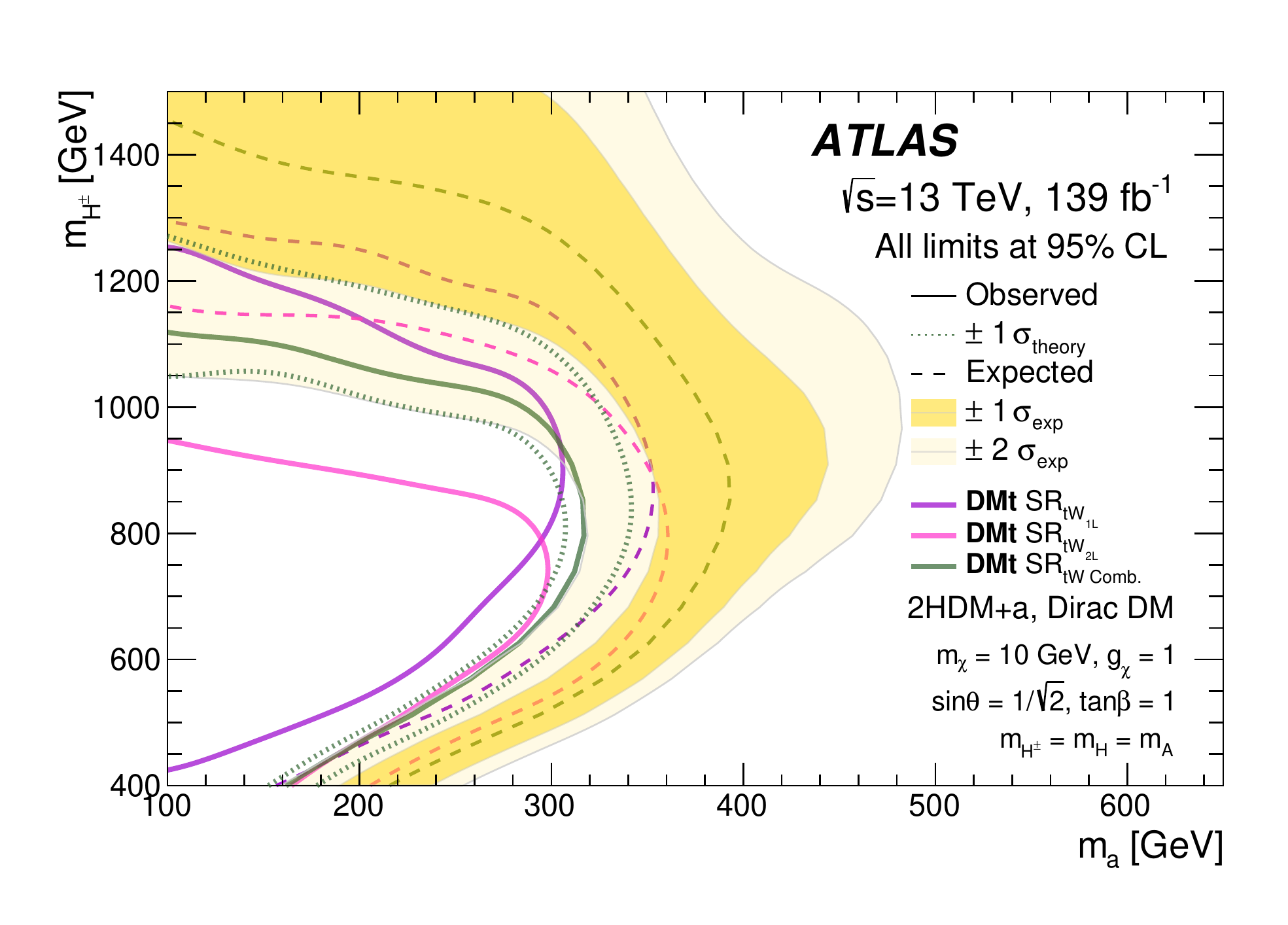}
\caption{DM$t$, $m_a$--$m_{H^{\pm}}$ plane \label{fig:comb_nominal_ma}}
\end{subfigure}
\begin{subfigure}[b]{0.75\textwidth}
\centering\includegraphics[width=\textwidth]{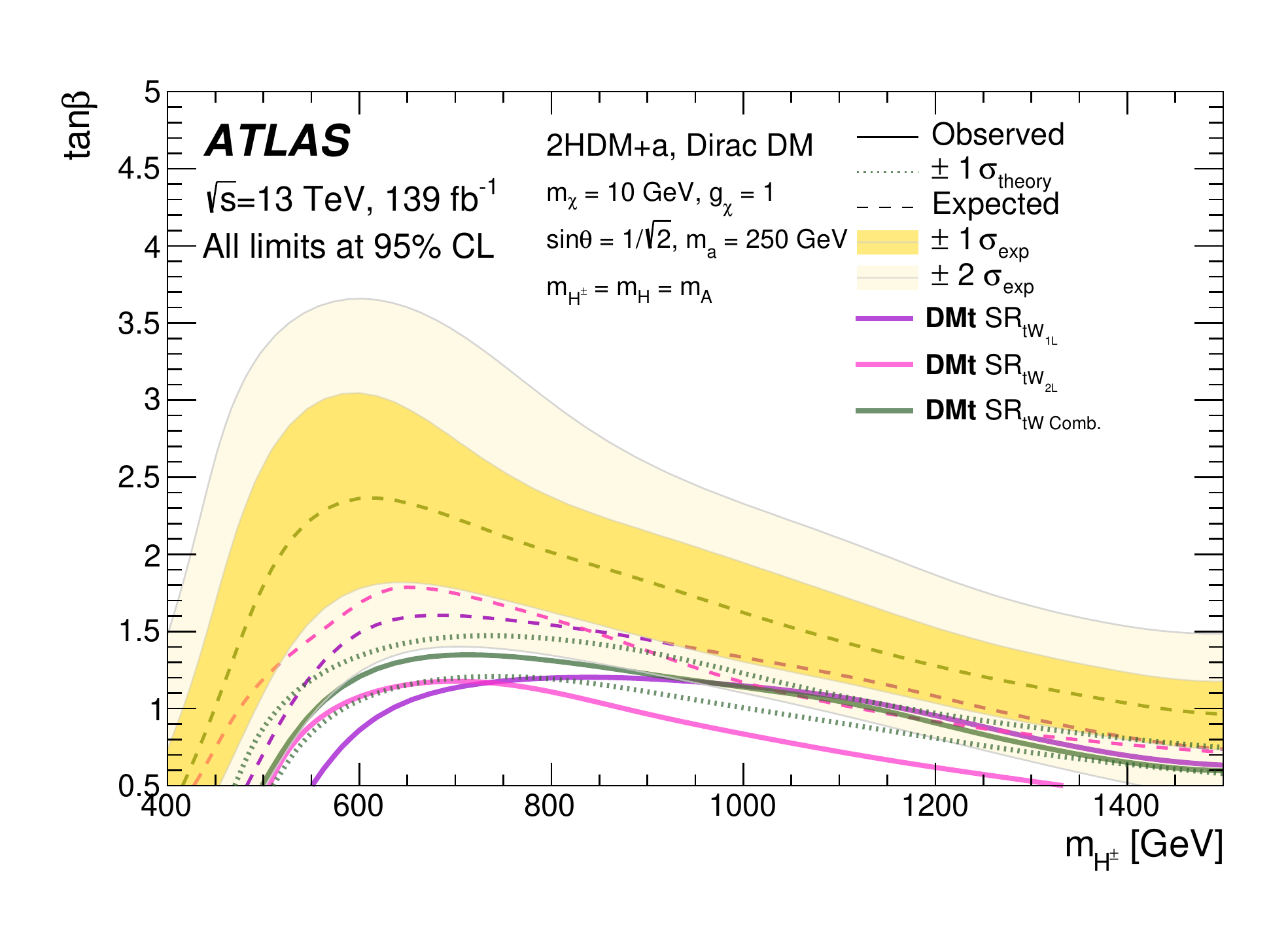}
\caption{DM$t$, $m_{H^{\pm}}$--$\tan\beta$ plane \label{fig:comb_nominal_tanb}}
\end{subfigure}
\end{center}
\caption{The expected and observed exclusion contours as a function of $(m_a, m_{H^{\pm}})$ (top) and $(m_{H^{\pm}}, \tan\beta)$ (bottom), assuming only $tW$+DM contributions, for the individual \tWOneL\ (purple line) and \tWTwoL\ (pink line) analysis channels, and for their statistical combination (green line).
Experimental and theoretical systematic uncertainties, as described in Section~\ref{sec:sys},
are applied to background and signal samples and illustrated by the $\pm 1$ standard-deviation and $\pm 2$ standard-deviation yellow bands and the green dotted contour lines, respectively, for the statistical combination.}
\label{fig:results-lim}
\end{figure}

\begin{figure}[tp]
\begin{center}
\begin{subfigure}[b]{0.75\textwidth}
\centering\includegraphics[width=\textwidth]{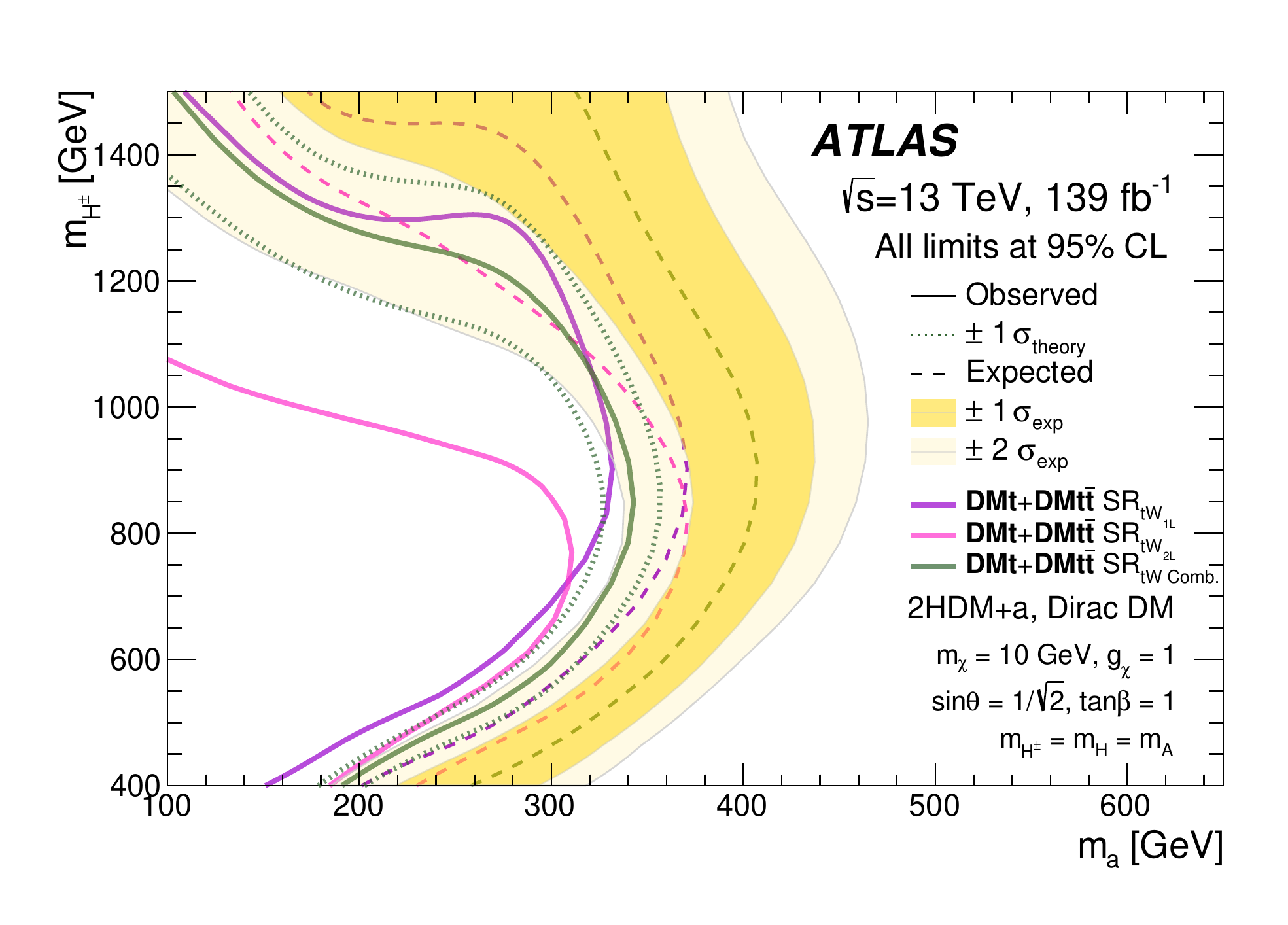}
\caption{DM$t$+DM$\ttbar$, $m_a$--$m_{H^{\pm}}$ plane \label{fig:comb_nominal_ma2}}
\end{subfigure}
\begin{subfigure}[b]{0.75\textwidth}
\centering\includegraphics[width=\textwidth]{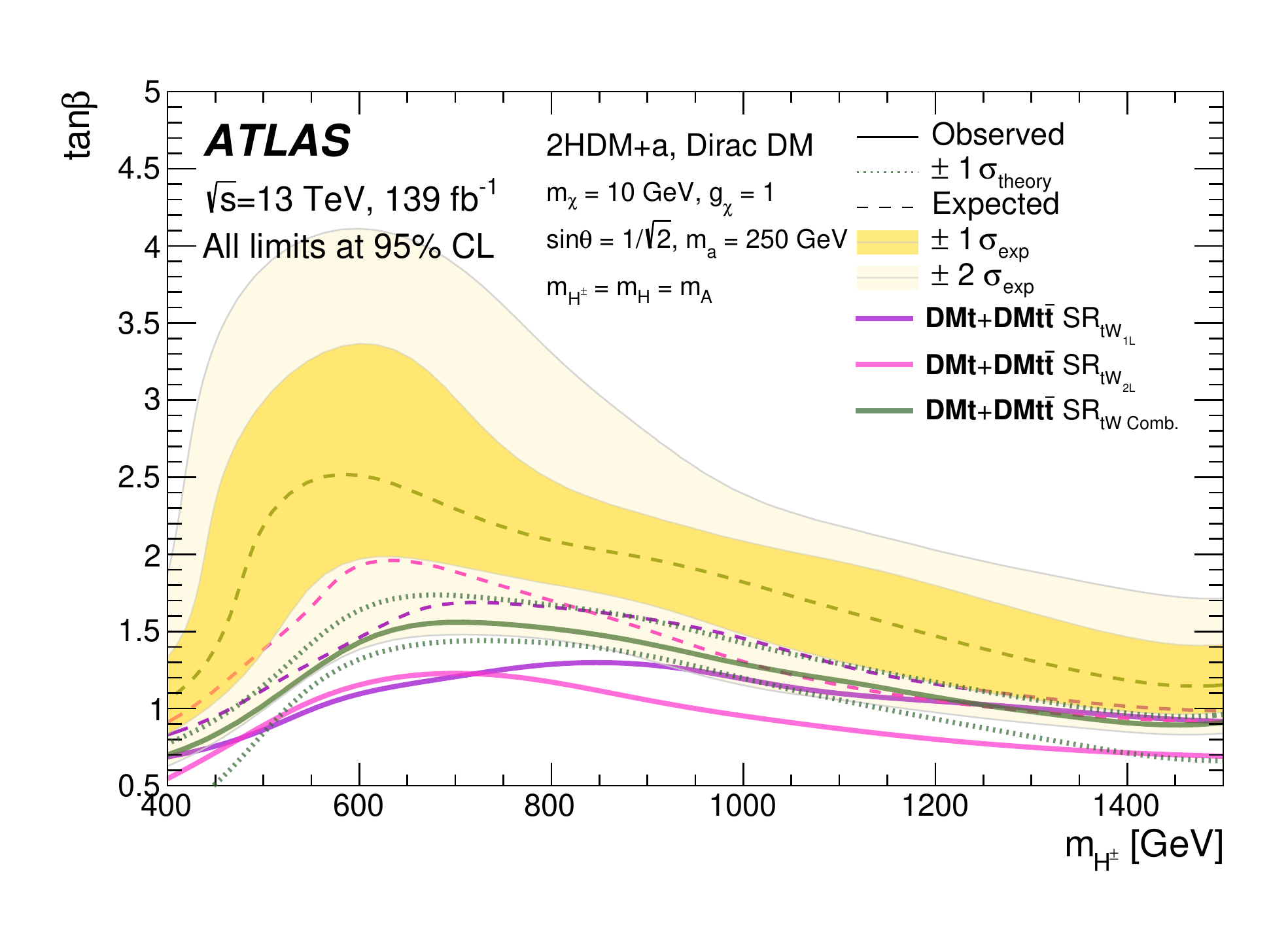}
\caption{DM$t$+DM$\ttbar$, $m_{H^{\pm}}$--$\tan\beta$ plane \label{fig:comb_nominal_tanb2}}
\end{subfigure}
\end{center}
\caption{The expected and observed exclusion contours as a function of $(m_a, m_{H^{\pm}})$ (top) and $(m_{H^{\pm}}, \tan\beta)$ (bottom), assuming DM$\ttbar$ and DM$t$ contributions, for the individual \tWOneL\ (purple line) and \tWTwoL\ (pink line) analysis channels, and for their statistical combination (green line).
Experimental and theoretical systematic uncertainties, as described in Section~\ref{sec:sys},
are applied to background and signal samples and illustrated by the $\pm 1$ standard-deviation and $\pm 2$ standard-deviation yellow bands and the green dotted contour lines, respectively, for the statistical combination.}
\label{fig:results-lim2}
\end{figure}

Figures~\ref{fig:comb_nominal_ma} and \ref{fig:comb_nominal_tanb} show
the observed and expected exclusion contours as functions of $(m_a, m_{H^{\pm}})$ and
$(m_{H^{\pm}}, \tan\beta)$, respectively, for the \tWOneL and \tWTwoL\ channels, presented both  individually and statistically combined.
In this case, only the DM$t$ contribution of the signal is taken into account to better illustrate the sensitivity to
single-top signatures.
Figures~\ref{fig:comb_nominal_ma2} and \ref{fig:comb_nominal_tanb2} show the observed and expected exclusion contours for the same
models, but also include the expected contributions from the DM$\ttbar$ process.
Figures~\ref{fig:results-lim} and \ref{fig:results-lim2} also report the 1$\sigma$ and 2$\sigma$ uncertainty bands around the observed limit contour, as well as
the variations obtained by changing the theoretical cross-section predictions for signal to be 15\% above or below the nominal value (as this is
expected to be largest uncertainty in the signal yields across the plane).
For low $H^{\pm}$ masses, DM$t$ production generally dominates DM\ttbar production,
due to the contribution from the resonant $H^{\pm}$ diagrams, except when the mass difference $m_{H^{\pm}}-m_a$ is
small enough to suppress the branching fraction of $H^{\pm} \to Wa$ decay relative to  $H^{\pm} \to t\bar{b}$.
On the other hand, DM\ttbar contributions are dominant at high $m_{H^{\pm}}$.
The width of $H^{\pm}$ also increases at high  $m_{H^{\pm}}$, and it is about 20\% of its mass for  $m_{H^{\pm}}=1$~\TeV.
Moreover, as studied in Ref.~\cite{EXOT-2017-32}, the DM\ttbar cross section is proportional to $1/\tan^2\beta$, whereas the $H^{\pm}$ production cross section has a more complex dependence, with a minimum for $\tan\beta\sim10$ and an enhancement for high values of $\tan\beta$. 
For $\tan\beta=1$ and $m_{H^{\pm}}\sim m_a+m_W$, the DM\ttbar cross section also dominates the DM$t$
cross section.
Assuming $m_\chi = 10$~\GeV\ and $g_\chi = 1$, masses of $a$ below 190~\GeV\ are
excluded at 95\% CL for all values of $m_{H^{\pm}}$ in the range 400--1400~\GeV, and up to 330~\GeV\
for $m_{H^{\pm}}$ around 800~\GeV. When only DM$t$ contributions are taken into account,
the constraints on $m_a$ decrease by 20--50~\GeV. In the case where $m_a = 250$~\GeV,
all values of $m_{H^{\pm}}$ between 450~\GeV\ and 1.5~\TeV\ are excluded for $\tan\beta$ around and
below unity, and scenarios with $\tan\beta$ below 1.5 are excluded for masses of $H^{\pm}$
around 800~\GeV.
 
\begin{figure}[!ht]
\centering
\begin{subfigure}[b]{0.48\textwidth}
\centering\includegraphics[width=\textwidth]{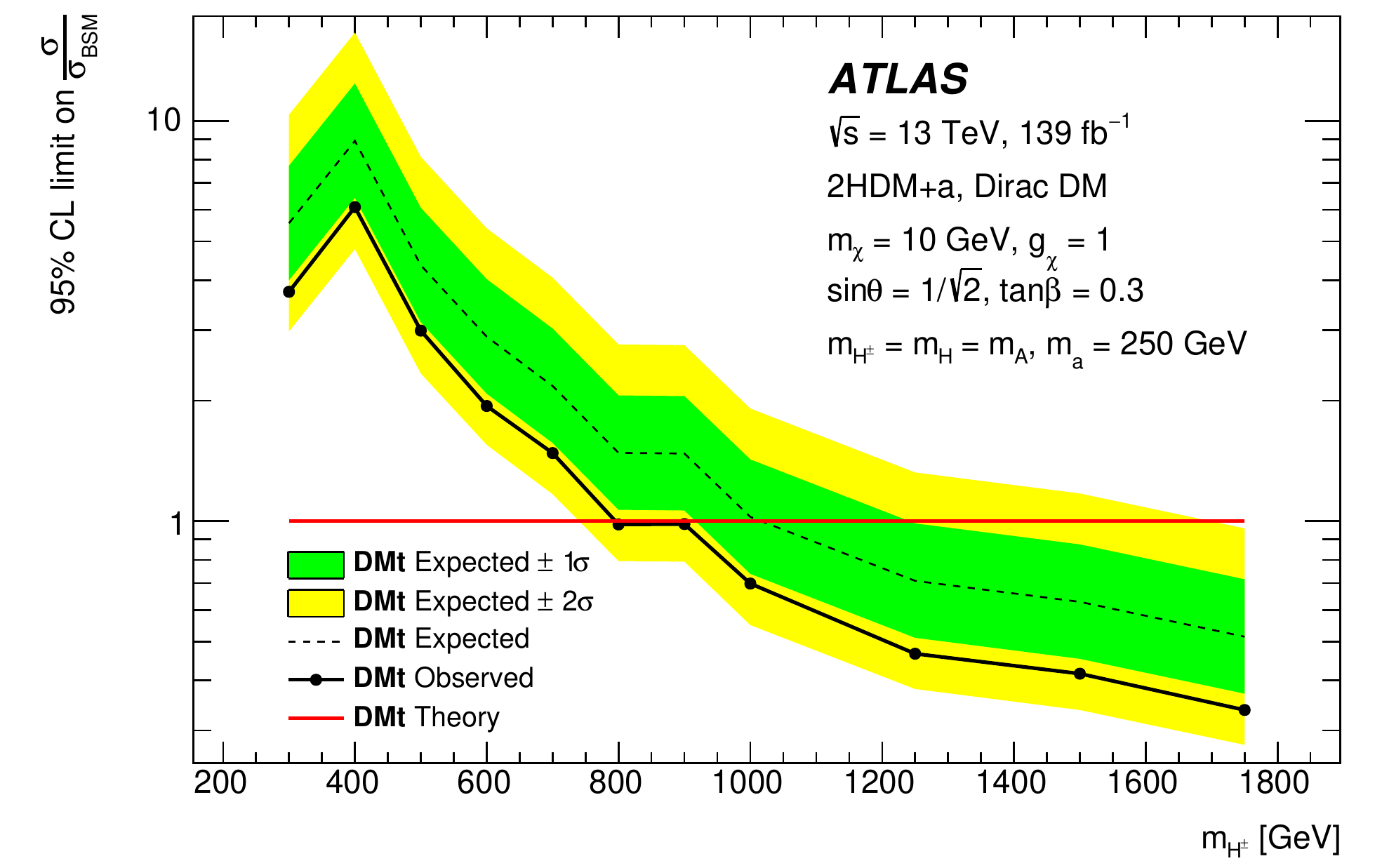}
\caption{$\tan\beta=0.3$\label{fig:limit_mH_tj1L_tb30}}
\end{subfigure}
\begin{subfigure}[b]{0.48\textwidth}
\centering\includegraphics[width=\textwidth]{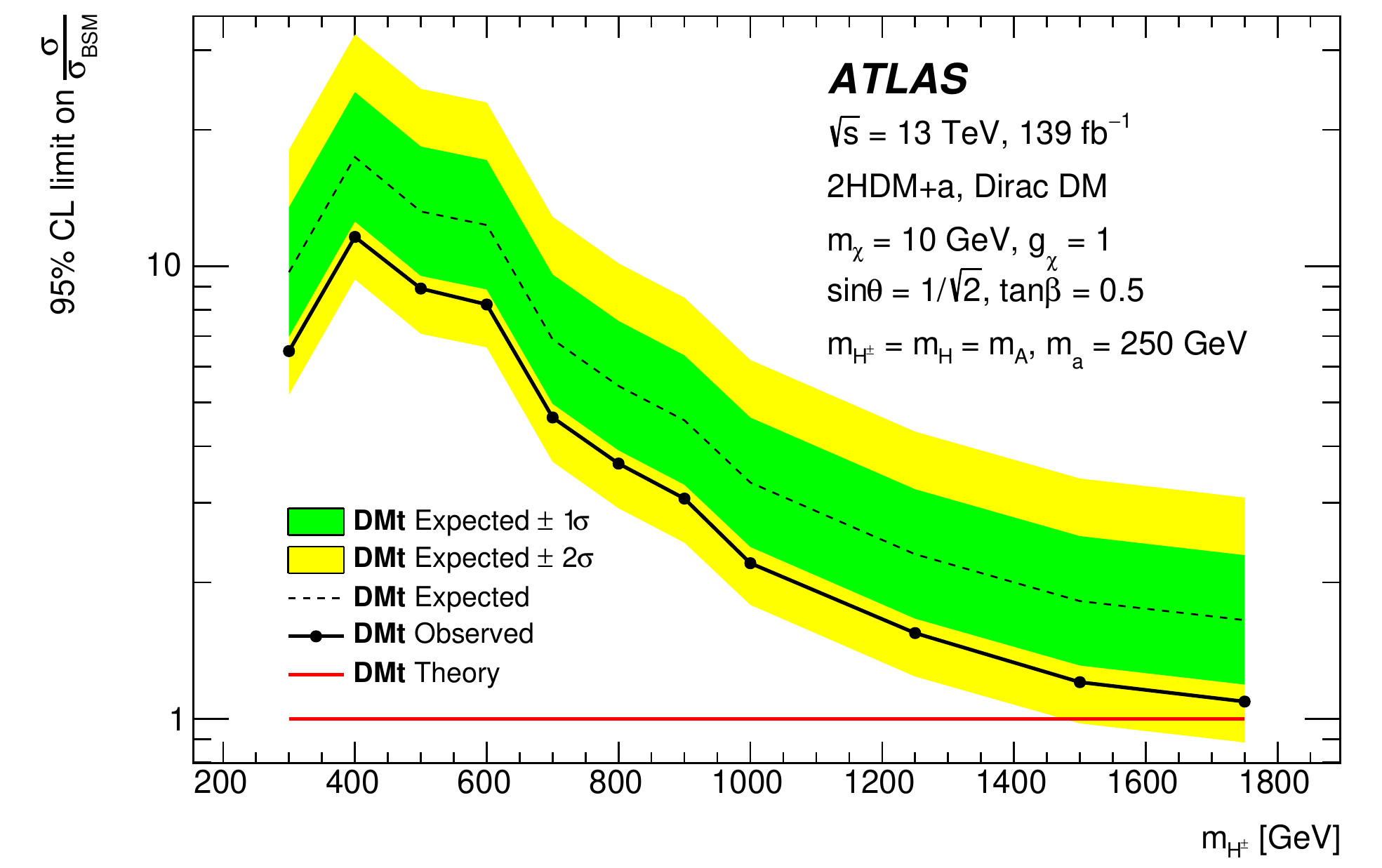}
\caption{$\tan\beta=0.5$\label{fig:limit_mH_tj1L_tb50}}
\end{subfigure}
\caption{
The expected and observed cross-section exclusion limits as a function of $m_{H^{\pm}}$
in the \tjOneL\ analysis channel for signal models with $m_a = 250~\GeV$, and
\subref{fig:limit_mH_tj1L_tb30}~$\tan\beta=0.3$ and \subref{fig:limit_mH_tj1L_tb50}~$\tan\beta=0.5$.
The theoretical cross section of the $t$-channel DM production process is denoted by `Theory'.
}
\label{fig:limit_mH_tj1L}
\end{figure}

The sensitivity of the \tjOneL\ channel is small compared to the other analysis channels.
It targets the $t$-channel production component of the DM$t$ signal, which has a smaller
cross section with respect to the $tW$+DM process.
The observed and expected cross-section limits at 95\% CL
as a function of $m_{H^{\pm}}$ for two representative values of $\tan\beta$ are shown in
Figure~\ref{fig:limit_mH_tj1L} assuming a fixed value of $m_a=250$~\GeV.
The limits are shown as a multiple of $\sigma^{}_\mathrm{BSM}$, the theoretical cross section of the $t$-channel DM production process.
For $\tan\beta=0.3$, $H^{\pm}$ masses above 900~\GeV\ are excluded under these hypotheses, whilst no exclusion
is obtained for $\tan\beta=0.5$.
 
% End of text imported from the .//sections/results.tex input file
\FloatBarrier

\section{Conclusion}
\label{sec:conclusion}
% The next lines are included from the .//sections/conclusion.tex input file
A search for dark matter has been performed in the context of a two-Higgs-doublet model together with an additional pseudoscalar mediator, $a$, which decays
into the dark-matter particles. Processes where the pseudoscalar mediator is produced in association with a single top quark in the 2HDM+$a$ model are explored for the first time at the LHC.
Several final states which include either one or two leptons (electrons or muons) and a significant amount of missing transverse momentum are considered.
The analysis makes use of proton--proton collision data at $\sqrt{s} = 13$~\TeV\ collected by the ATLAS experiment during LHC Run 2 (2015--2018),
corresponding to an integrated luminosity of 139~fb$^{-1}$.
No significant excess above the Standard Model predictions is found.
The results are expressed as 95\% confidence-level limits on the 2HDM+$a$ signal models considered.
Assuming dark-matter particles with mass $m_\chi = 10$~\GeV\ and coupling $g_\chi = 1$ to the mediator,
and full mixing between the $a$ and $A$ bosons, masses of $a$ below 200~\GeV\ are
excluded at 95\% CL for all values of $m_{H^{\pm}}$ in the range 400--1400~\GeV, and up to 330~\GeV\
for $m_{H^{\pm}}$ around 900~\GeV. For $m_a = 250$~\GeV, all values of $m_{H^{\pm}}$ below 1.5~\TeV\
are excluded for $\tan\beta$ below unity, and scenarios with $\tan\beta$ below 1.5
are excluded for masses of $H^{\pm}$ around 800~\GeV.

% End of text imported from the .//sections/conclusion.tex input file

\section*{Acknowledgements}
 
% The next lines are included from the .//acknowledgements/Acknowledgements.tex input file

We thank CERN for the very successful operation of the LHC, as well as the
support staff from our institutions without whom ATLAS could not be
operated efficiently.
 
We acknowledge the support of ANPCyT, Argentina; YerPhI, Armenia; ARC, Australia; BMWFW and FWF, Austria; ANAS, Azerbaijan; SSTC, Belarus; CNPq and FAPESP, Brazil; NSERC, NRC and CFI, Canada; CERN; ANID, Chile; CAS, MOST and NSFC, China; COLCIENCIAS, Colombia; MSMT CR, MPO CR and VSC CR, Czech Republic; DNRF and DNSRC, Denmark; IN2P3-CNRS and CEA-DRF/IRFU, France; SRNSFG, Georgia; BMBF, HGF and MPG, Germany; GSRT, Greece; RGC and Hong Kong SAR, China; ISF and Benoziyo Center, Israel; INFN, Italy; MEXT and JSPS, Japan; CNRST, Morocco; NWO, Netherlands; RCN, Norway; MNiSW and NCN, Poland; FCT, Portugal; MNE/IFA, Romania; JINR; MES of Russia and NRC KI, Russian Federation; MESTD, Serbia; MSSR, Slovakia; ARRS and MIZ\v{S}, Slovenia; DST/NRF, South Africa; MICINN, Spain; SRC and Wallenberg Foundation, Sweden; SERI, SNSF and Cantons of Bern and Geneva, Switzerland; MOST, Taiwan; TAEK, Turkey; STFC, United Kingdom; DOE and NSF, United States of America. In addition, individual groups and members have received support from BCKDF, CANARIE, Compute Canada, CRC and IVADO, Canada; Beijing Municipal Science \& Technology Commission, China; COST, ERC, ERDF, Horizon 2020 and Marie Sk{\l}odowska-Curie Actions, European Union; Investissements d'Avenir Labex, Investissements d'Avenir Idex and ANR, France; DFG and AvH Foundation, Germany; Herakleitos, Thales and Aristeia programmes co-financed by EU-ESF and the Greek NSRF, Greece; BSF-NSF and GIF, Israel; La Caixa Banking Foundation, CERCA Programme Generalitat de Catalunya and PROMETEO and GenT Programmes Generalitat Valenciana, Spain; G\"{o}ran Gustafssons Stiftelse, Sweden; The Royal Society and Leverhulme Trust, United Kingdom.
 
The crucial computing support from all WLCG partners is acknowledged gratefully, in particular from CERN, the ATLAS Tier-1 facilities at TRIUMF (Canada), NDGF (Denmark, Norway, Sweden), CC-IN2P3 (France), KIT/GridKA (Germany), INFN-CNAF (Italy), NL-T1 (Netherlands), PIC (Spain), ASGC (Taiwan), RAL (UK) and BNL (USA), the Tier-2 facilities worldwide and large non-WLCG resource providers. Major contributors of computing resources are listed in Ref.~\cite{ATL-SOFT-PUB-2020-001}.
 
% End of text imported from the .//acknowledgements/Acknowledgements.tex input file

\printbibliography

\clearpage% ATLAS Collaboration author list
% Reference date of EXOT-2018-43 is 2020-07-08
% Author list last updated on date 18-NOV-20
% Data extracted on 18-Nov-2020 for paper reference EXOT-2018-43
% at 9:18am
 
\begin{flushleft}
\hypersetup{urlcolor=black}
{\Large The ATLAS Collaboration}

\bigskip

\AtlasOrcid[0000-0002-6665-4934]{G.~Aad}$^\textrm{\scriptsize 102}$,    
\AtlasOrcid[0000-0002-5888-2734]{B.~Abbott}$^\textrm{\scriptsize 128}$,    
\AtlasOrcid{D.C.~Abbott}$^\textrm{\scriptsize 103}$,    
\AtlasOrcid[0000-0002-2788-3822]{A.~Abed~Abud}$^\textrm{\scriptsize 36}$,    
\AtlasOrcid[0000-0002-1002-1652]{K.~Abeling}$^\textrm{\scriptsize 53}$,    
\AtlasOrcid[0000-0002-2987-4006]{D.K.~Abhayasinghe}$^\textrm{\scriptsize 94}$,    
\AtlasOrcid[0000-0002-8496-9294]{S.H.~Abidi}$^\textrm{\scriptsize 167}$,    
\AtlasOrcid[0000-0002-8279-9324]{O.S.~AbouZeid}$^\textrm{\scriptsize 40}$,    
\AtlasOrcid{N.L.~Abraham}$^\textrm{\scriptsize 156}$,    
\AtlasOrcid[0000-0001-5329-6640]{H.~Abramowicz}$^\textrm{\scriptsize 161}$
% ...
\bigskip
\\

% ...
$^{36}$CERN, Geneva; Switzerland.\\
% ...
$^{40}$Niels Bohr Institute, University of Copenhagen, Copenhagen; Denmark.\\
% ...
$^{53}$II. Physikalisches Institut, Georg-August-Universit\"{a}t G\"ottingen, G\"ottingen; Germany.\\
% ...
$^{94}$Department of Physics, Royal Holloway University of London, Egham; United Kingdom.\\
% ...
$^{102}$CPPM, Aix-Marseille Universit\'e, CNRS/IN2P3, Marseille; France.\\
$^{103}$Department of Physics, University of Massachusetts, Amherst MA; United States of America.\\
% ...
$^{128}$Homer L. Dodge Department of Physics and Astronomy, University of Oklahoma, Norman OK; United States of America.\\
% ...
$^{156}$Department of Physics and Astronomy, University of Sussex, Brighton; United Kingdom.\\
% ...
$^{161}$Raymond and Beverly Sackler School of Physics and Astronomy, Tel Aviv University, Tel Aviv; Israel.\\
% ...
$^{167}$Department of Physics, University of Toronto, Toronto ON; Canada.\\
% ...

\end{flushleft}

% Created with Glance <Atlas.Glance@cern.ch>

\end{document}